\documentclass[aps,prb,showpacs,a4paper,superscriptaddress,groupedaddress]{revtex4}
\usepackage[english]{babel}
\usepackage[utf8]{inputenc}
\usepackage{amsmath}
\usepackage{graphicx}
\usepackage[toc,page]{appendix}
\usepackage{multirow}
\usepackage{nonfloat}
\usepackage[colorinlistoftodos]{todonotes}
\usepackage{verbatim}

\newcommand\scalemath[2]{\scalebox{#1}{\mbox{\ensuremath{\displaystyle #2}}}}

\begin{document}
\title{Coupled spatial multi-mode solitons in microcavity wires}

\author{G. Slavcheva}\email{g.slavcheva@bath.ac.uk}
\affiliation{Department of Physics, University of Bath, Bath, BA2 7AY, United Kingdom}
\author{A. V. Gorbach}
\affiliation{Department of Physics, University of Bath, Bath, BA2 7AY, United Kingdom}
\author{A. Pimenov}
\affiliation{Weierstrass Institute, Mohrenstrasse 39, D-10117 Berlin, Germany}

\date{\today}

\begin{abstract}
A modal expansion approach is developed and employed to investigate and elucidate the nonlinear mechanism behind the multistability and formation of coupled multi-mode polariton solitons in microcavity wires. With pump switched on and realistic dissipation parameters, truncating the expansion up to the second-order wire mode, our model predicts two distinct coupled soliton branches: stable and ustable. Modulational stability of the homogeneous solution and soliton branches stability are studied. Our simplified 1D model is in remarkably good agreement with the full 2D mean-field Gross-Pitaevskii model, reproducing correctly the soliton existence domain upon variation of pump amplitude and the onset of multistability.

\end{abstract}

\maketitle
\section{Introduction}
\label{sec1}
Nonlinear propagation in multimode systems, such as optical fibres and nonlinear optical waveguides, has recently become a topic of considerable interest in view of pushing the inherent limits for information transmission of the communication technologies by space-division multiplexing, exploiting multiple spatial transverse mode channels, and high-speed applications, such as all-optical switching using ultrashort high peak power pulses. Several experiments have demonstrated that there is also significant potential for new spatial and spectral nonlinear effects in multimode fibers\cite{Wright}$^,$\cite{Russell} and nonlinear waveguides \cite{Modotto}. Recently, non-classical light generation has been demonstrated using two-photon interference in a multi-mode nonlinear waveguide as a parametric down-conversion source, controlling the spatial characteristics of the down-conversion process via intermodal dispersion \cite{Karpinski}.

Self-localisation optical phenomena in multimode systems are possible as a result of counter-balancing of a combination of dispersive effects: (i) material dispersion, due to frequency-dependent dielectric response; (ii) waveguide modes dispersion; (iii) variation of the group velocity of each waveguide mode, and nonlinearity. For instance in optical fibres, complex 'envelope' multimode solitons have been theoretically predicted in the early $80s$ \cite{Hasegawa}$^-$\cite{Crosignani&Cutolo&DiPorto} and only very recently experimentally studied \cite{Renninger}$^,$\cite{Wright&Wise}. Nonlinear localisation effects such as 'soliton trapping'\cite{Agrawal} has recently been theoretically demonstrated whereby two solitons in different modes shift their spectra and travel at the same speed in spite of considerable intermodal differential group delay between them. For third-order nonlinear processes such as four-photon mixing, which are not automatically phase matched, it is possible to use the dispersion of the waveguide modes to compensate for material group dispersion in such a way as to achieve phase matching. This has been demonstrated by the observation of four-photon mixing in a multimode fibre \cite{Stolen&Ashkin}. In multimode systems the dominant dispersive effect originates from the difference in the group velocity of each excited mode. Under suitable conditions, the different modes interact among themselves in such a way as to give rise to a self-localisation mechanism, due to non-resonant (intermodal cross-phase modulation through Kerr nonlinearity) and/or resonant (four-wave mixing) nonlinearities\cite{Agrawal}, that prevents the pulse from broadening as a consequence of modal dispersion\cite{Crosignani&Cutolo&DiPorto}.

The concept of a multi-component or vector soliton has been introduced by Christodoulides et al. \cite{Christodoulides} in the context of nonlinear optical wave propagation in birefringent Kerr media. It has been shown that birefringent media support solitons that consist of a bound state of two distinct orthogonally polarised solitons. Conversely, bound states of solitons can exist without birefringence and four-wave mixing simply on the basis of mutual trapping induced by cross-phase modulation (incoherent coupling) between the circularly polarised light components in Kerr media \cite{Haelterman1, Haelterman2, Snyder}. In the spatial domain these solitons can be viewed as a superposition of the fundamental and the higher-order waveguide modes induced by the modes themselves through self- and cross-phase modulations.

Recently a number of nonlinear self-localisation phenomena with light-matter wave packets, cavity polaritons, in strongly-coupled planar quantum well semiconductor microcavities \cite{Kavokin} have been demonstrated such as bright polariton solitons \cite{Sich_NaturePhotonics}, superfluidity \cite{Amo1,Amo2}, pattern formation \cite{Philippe, Kwong}, vortices \cite{Nardin,Lagoudakis}. Here we show that similar nonlinear self-confinement mechanism takes place when a light-matter wave polariton soliton, rather than an optical soliton, propagates in a microcavity wire. Self-localised light-matter wave packets in multimode polariton systems result from compensation of the polariton modes dispersion and group velocity dispersion of each cavity mode with nonlinearity. A new nonlinearity component is the intermodal nonlinear coupling that arises from intermodal cross-phase modulation (through Kerr nonlinearity) and polariton parametric scattering (polariton four-wave mixing). Hence, multi-mode polariton solitons can be viewed as resulting from distribution of the excitation energy over multiple spatial modes and consisting of synchronized, non-dispersive pulses in multiple spatial modes, interacting via parametric polariton nonlinearity. Here, the dominant dispersive effect originates from the difference in the group velocity of each excited cavity polariton mode.

For the purpose of theoretical description of the ultrashort pulse propagation in multimode waveguides a set of multimode generalised nonlinear Schr\"{o}dinger equations (MM-GNLSE), including a range of intermodal nonlinear effects has been derived \cite{Poletti}$^,$\cite{Horak}. A tractable simplified form of the MM-GNLSE has been recently proposed in \cite{Mafi}. The nonlinear polariton dynamics in microcavity wires is generally described by Gross-Pitaevskii type equations which are obtained as a special case of the MM-GNLSE equations when Raman, shock and dispersion orders greater than the second contributions are neglected. In fact, resonant coherent interactions among different modes through Brillouin (involving acoustical phonons) and Raman (involving optical phonons) scattering are unlikely to occur because the matching conditions are not generally satisfied and can be neglected \cite{Hasegawa}. Hence the resonant nonlinearities will be dominated by four-photon processes, such as polariton parametric scattering (polariton four-wave mixing).

In a recent work \cite{our_OL} we found composite 'multi-mode' polariton soliton solutions in 1D microcavity wires that result from the superposition of the fundamental and multiple higher-order co-existing transverse cavity modes. Most recently, similar spatial multi-stability behaviour has been experimentally observed in laterally confined microcavity exciton-polaritons \cite{Deveaud}. Unlike single-mode polariton soliton solutions previously found in planar semiconductor microcavities that are stable within the bistability domain, the multimode solitons exhibit more complex multi-stable behaviour. We have previously found inconsistencies between the calculated domains of soliton existence and a peculiar non-monotonous soliton velocity behaviour for different wire widths \cite{JOSAB}, which we were unable to explain within the framework of our full mean-field model. In order to get a deeper insight into the soliton solutions we develop a modal expansion method, expanding the nonlinear polariton modes in the basis of free polariton modes. The multi-mode analysis helps to investigate in detail conditions, dynamics and stability of coupled soliton formation and identify reliably ranges of soliton existence, which holds benefits for the fabrication technologies targeting novel polaritonic integrated devices based on structured microcavities \cite{Wertz_APL}$^,$\cite{Wertz_Nature}.

Polariton propagation in multi-mode systems is an interplay of complex nonlinear spatiotemporal phenomena and waveguide imperfections: the pulse effective coherence length is reduced from the strictly infinite coherence length of perfect phase matching by waveguide imperfections. In a parametric scattering process the pump can be either redistributed between several different polariton modes and Stokes signal appears in one of these modes while anti-Stokes appears in a different mode, or the pump photons can be in the same mode. These two cases are referred to as "mixed-mode pump" and "single-mode pump" processes. It has been demonstrated that mixed-mode pump processes result in pulses that have much longer coherence lengths than single-mode pump processes \cite{Stolen}. This is a key reason for the interest in multi-mode polariton solitons, as they are expected to be more robust and able to propagate over much longer distances without being destroyed in a realistic waveguide with imperfections. Furthermore, as has been pointed out in \cite{Wright&Wise}, they are expected to exhibit novel spatiotemporal dynamics and, like single-mode solitons, may provide a convenient framework for understanding more complex nonlinear phenomena in confined multi-mode polariton systems.

\section{Modal equations}
\label{sec2}
\begin{figure}
\resizebox{.9\textwidth}{!}{%
\includegraphics[height=5cm]{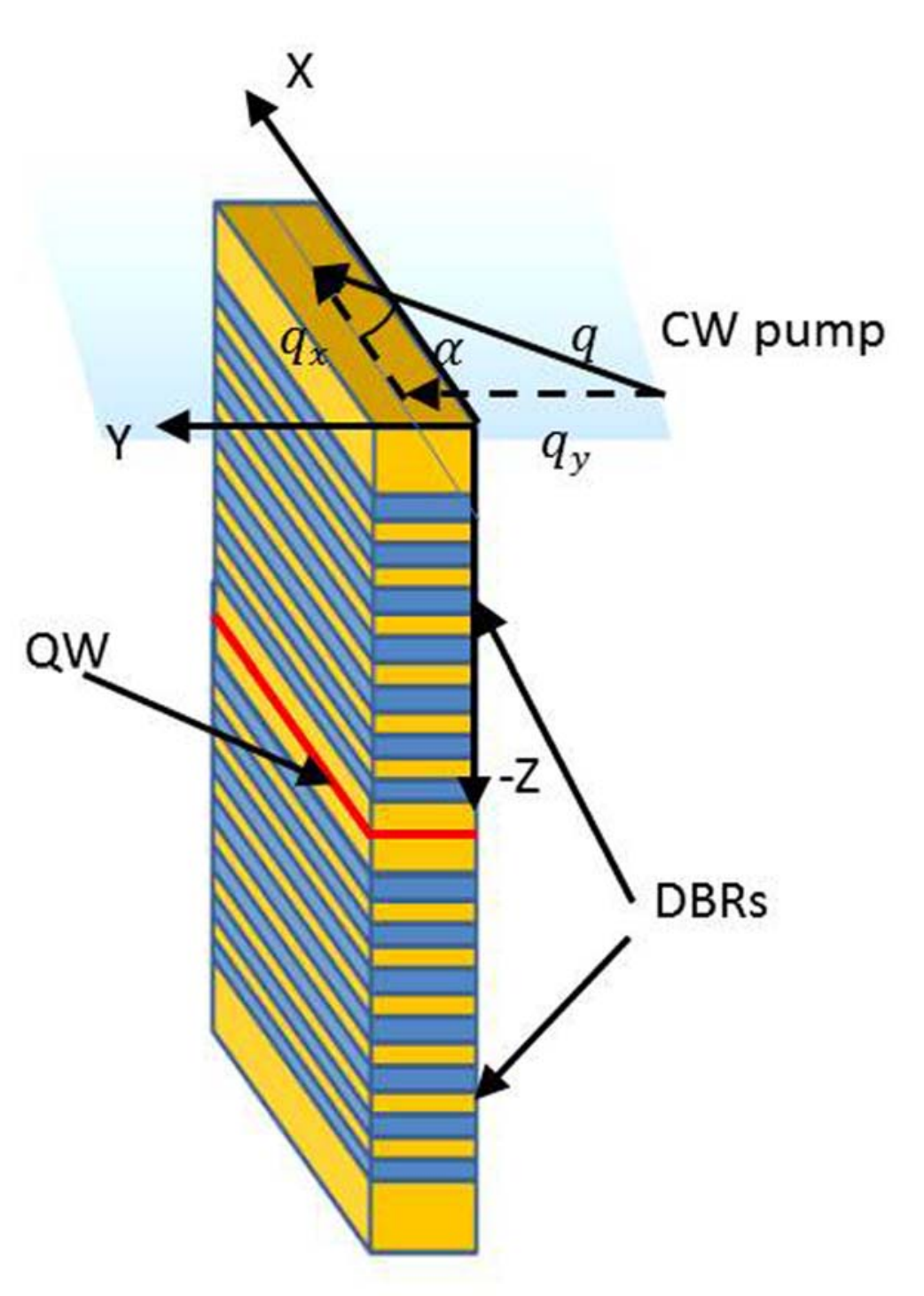}
\quad
\includegraphics[height=5cm]{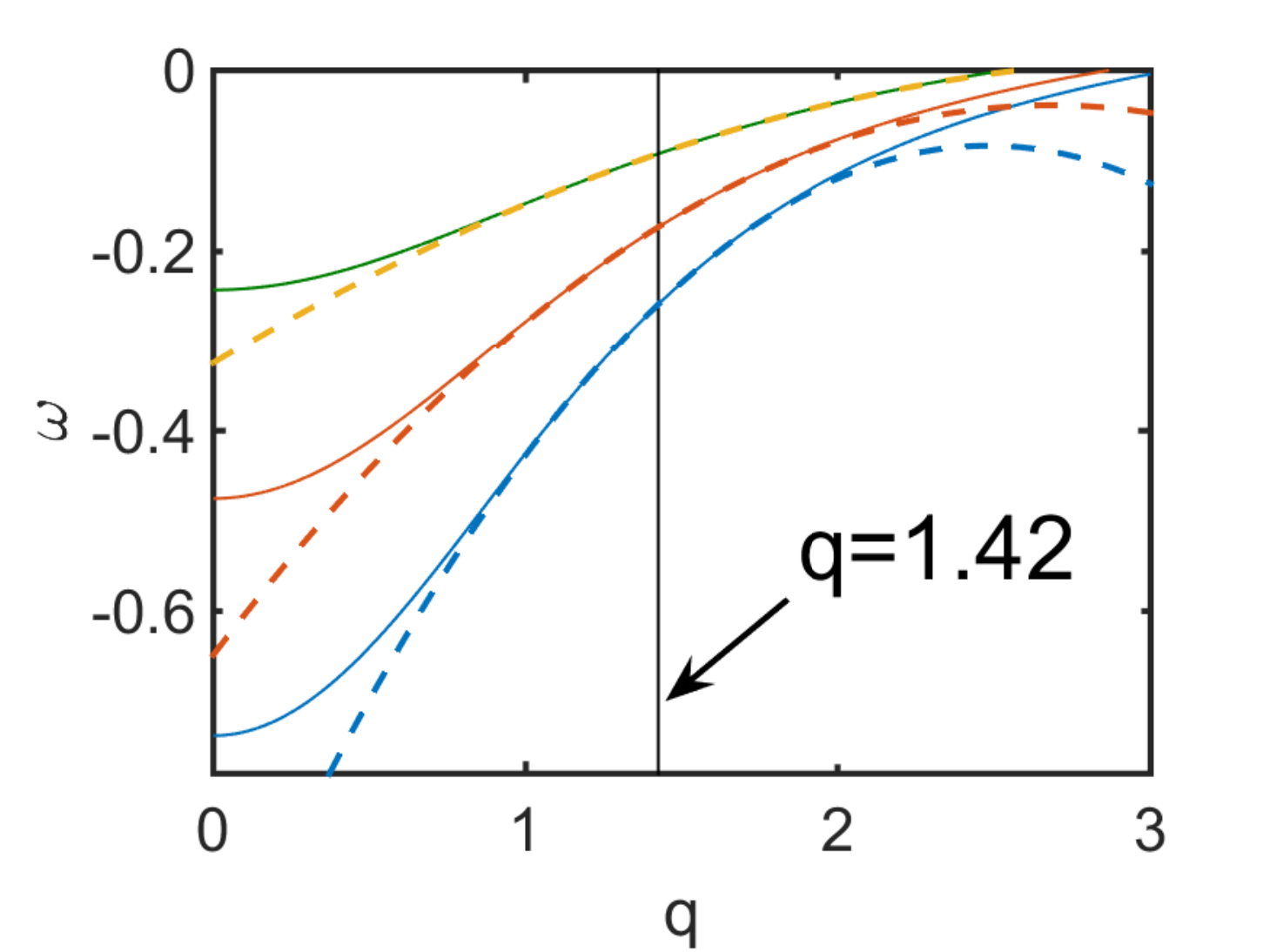}
\quad
\includegraphics[height=5cm]{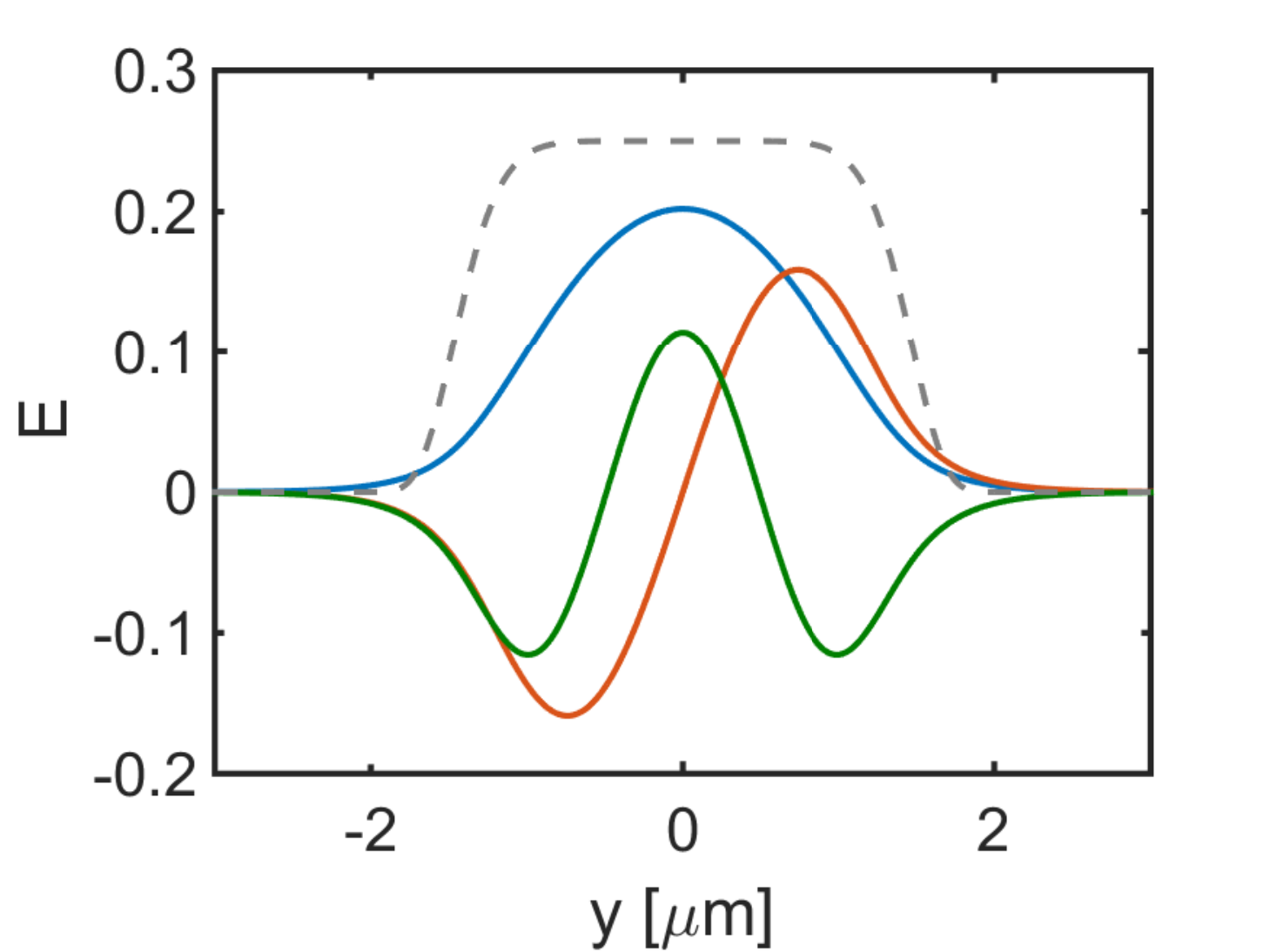}
}
\caption{(a) Scheme of the microcavity polaritonic wire structure with a tilted in-plane cw pump; (b) Linear polariton dispersion with parabolic fit (dashed curves): all parameters are as in Ref. \cite{our_OL}, coefficients for parabolic fits are listed in Eqs.~(\ref{eq:omega_parabolic0}-\ref{eq:omega_parabolic2}). (c) Modal profiles for $q=1.4237$ (which corresponds to pump inclination at 20 degrees). All modes are normalized such that $N_j=1$ in Eq.~(\ref{eq:norm_my}). Dashed curve indicate scaled profile of the potential $\Omega_R(y)$}
\label{fig:geometry_dispersion}
\end{figure}

The starting point is our mean-field driven-dissipative Gross-Pitaevskii model \cite{our_OL} in a tilted along the wire reference frame, in which, for the sake of generality, we introduce inclined at an angle $\alpha$ to the channel pump with in-plane wave vector components, $q_x=q\cos(\alpha)\;q_y=q\sin(\alpha)\;$ (Fig. \ref{fig:geometry_dispersion}a):
\begin{eqnarray}
\nonumber
i\partial_t E +\left[(\partial_x+iq_x)^2+\partial_y^2\right]E+\left[i\gamma_c+\delta_c+U(y)\right]E\\
+\Omega_R(y)\psi=iE_pe^{iq_y y-i\Delta t}\;,
\label{eq:eqE}\\
i\partial_t \Psi +\left(i\gamma_e+\delta_e\right)\psi
+\Omega_R(y)E=|\psi|^2\psi\;,
\label{eq:eqPsi}
\end{eqnarray}
where $\delta_e, \delta_c$, and the pump frequency, $\Delta$ are detunings from a reference frequency.


We treat nonlinearity, pump and dissipation as perturbations, and expand the solutions in terms of slowly time-varying amplitudes, $\vec{A}_j,\vec{B}_j$ of the $j^{th}$ free polariton mode.
\begin{eqnarray}
[E,\psi]^T=\vec{x}=\sum_j \int \left[s^{1/2}\vec{A}_j+s^{3/2}\vec{B}_j+O(s^{5/2})\right]e^{ikx-i\omega_j(k)t} dk
\label{eq:modal_decomposition}
\end{eqnarray}

Here $s$ is a dummy small parameter, we assume $\partial_t E \sim s E\sim \Delta E \sim \gamma_c E\ll \delta_c E$, $|\psi|^2 \sim s$, $\partial_t\psi\sim s \psi \sim \gamma_e \psi$, $E_p\sim s^{3/2}$. The sum is performed over a discrete set of polariton modes, specified below.
In the lowest order ($s^{1/2}$) and pump parallel to the wire, i.e. $\alpha=0$, the system Eqs.~(\ref{eq:eqE}), (\ref{eq:eqPsi}) is reduced to an eigenvalue problem from which the mode dispersion, $\omega=\omega_j(k)$ , of each discrete polariton mode is obtained (Fig. \ref{fig:geometry_dispersion}b) and we seek the solution of the form:
\begin{equation}
\vec{A}_j=\frac{1}{N_j}F_j(t,k)\vec{x}_j(y,k)\;, \qquad \vec{x}_j=\left[x_{je}, x_{j\psi}\right]^T
\label{eq:0s12_solution}
\end{equation}
 where $\vec{x}_j(y,k)$ is the corresponding eigenvector, with $x_{je}(y)$ and $x_{j\psi}(y)$ its photonic and excitonic components, respectively; $N_j$ is a normalisation factor and $F_j(t,k)$ are the modal amplitudes. In the next order ($s^{3/2}$), introducing new modal amplitudes by the transformation  $F_j(k,t)e^{-i\omega_j(k)t+i\Delta t}=Q_j(k,t)$, taking into account the phase-matching condition for polariton four-wave mixing, $k_3=k-k_1+k_2$, projecting the equations onto  $\vec{x}_j$ and expanding in Taylor series the mode frequencies $\omega_j(k)$ around $k=0$ and retaining up to second-order terms and performing inverse Fourier transform, we obtain the modal equations for $Q_j(x,t)$:
\begin{eqnarray}
\nonumber
i\partial_t Q_j +(\Delta- \omega_{j0}+i\gamma_{0}) Q_j + i \omega_{j1}\partial_x Q_j +\frac{\omega_{j2}}{2}\partial^2_x Q_j = \\
=\sum_{l m n}\Gamma_{lmn,j}Q_l Q_m^* Q_n
+ih_j \;.
\label{eq:fin_modal_gen}
\end{eqnarray}

where the following set of parameters have been introduced:
\begin{eqnarray}
\label{eq:hj}
h_j&=&N_j\frac{\int E_p e^{iq_y y} x_{je}^*(y)dy}
{\int \left(|x_{je}|^2+|x_{j\psi}|^2\right) dy}\;,\\
\label{eq:damping}
\gamma_j&=&\frac{\int \left[\gamma_c|{x}_{je}|^2+\gamma_e|{x}_{j\psi}|^2\right]dy}
{\int \left(|x_{je}|^2+|x_{j\psi}|^2\right) dy}\\
\label{eq:gam_lmnj}
\Gamma_{lmn,j}&=&\frac{N_j}{N_l N_m N_n}
\frac{\int x_{l\psi} x_{m\psi}^* x_{n\psi} x_{j\psi}^* dy}
{\int \left(|x_{je}|^2+|x_{j\psi}|^2\right) dy}\;,
\end{eqnarray}
The physical meaning of the above parameters is as follows: $h_j$ is the pump projection onto $j^{th}$ eigenmode; $\gamma_j$ are normalised dissipation parameters and $\Gamma_{lmn,j}$ are the intermodal nonlinear coupling coefficients. Making the reasonable assumption $\gamma_c=\gamma_e=\gamma_0$, leads to $\gamma_j=\gamma_0\;\; \forall j$. We choose the normalisation:
\begin{equation}
N_j^2=\int \left(|x_{je}|^2+|x_{j\psi}|^2\right) dy
\label{eq:norm_my}
\end{equation}
since it introduces additional symmetries with respect to permutation of indices for $\Gamma$-coefficients, and thus reduces the number of nonlinear coefficients to be computed.

\subsection{Three coupled modes}
\label{sssec:2.1}
Consider three coupled modes ($j=0,1,2$) and introduce linear operators:$\hat{\mathcal{L}}_j=\Delta- \omega_{j0}+i\gamma_{0} + i \omega_{j1}\partial_x  +\frac{\omega_{j2}}{2}\partial^2_x$. Owing to the mode symmetry there are only a few non-vansihing non-linear coefficients,$\Gamma_{lmn,j}$ and with the chosen normalisation, $\Gamma_{jk}=\Gamma_{kj}$. The full system can be written as:

\begin{eqnarray}
\nonumber
i\partial_t Q_0 + \hat{\mathcal{L}}_0 Q_0&=&ih_0+\left(\Gamma_{00} |Q_0|^2
+2\Gamma_{10} |Q_1|^2+2\Gamma_{20} |Q_2|^2\right)Q_0\\
\nonumber
&&+\Gamma_{002,0}\left(2|Q_0|^2Q_2+Q_0^2 Q_2^*\right)\\
\nonumber
&&+\Gamma_{112,0}\left(2|Q_1|^2Q_2+Q_1^2 Q_2^*\right)\\
\label{eq:cm3_Q0}
&&+\Gamma_{222,0}|Q_2|^2 Q_2\\
\label{eq:cm3_Q1}
i\partial_t Q_1 + \hat{\mathcal{L}}_1 Q_1&=&ih_1+\left(\Gamma_{11} |Q_1|^2
+2\Gamma_{01} |Q_0|^2+2\Gamma_{21} |Q_2|^2\right)Q_1\\
\label{eq:cm3_Q2}
i\partial_t Q_2 + \hat{\mathcal{L}}_2 Q_2&=&ih_2+\left(\Gamma_{22} |Q_2|^2
+2\Gamma_{02} |Q_0|^2+2\Gamma_{12} |Q_1|^2\right)Q_2\\
\nonumber
&&+\Gamma_{220,2}\left(2|Q_2|^2Q_0+Q_2^2 Q_0^*\right)\\
\nonumber
&&+\Gamma_{110,2}\left(2|Q_1|^2Q_0+Q_1^2 Q_0^*\right)\\
\nonumber
&&+\Gamma_{000,2}|Q_0|^2 Q_0
\end{eqnarray}

Rescaling $x$-coordinate and transforming into moving frame: $\xi=\frac{1}{\sqrt{|\omega_{02}|}}\left(x-\omega_{01}t\right)$, the linear operators become:
\begin{eqnarray}
\hat{\mathcal{L}}_0&=&\Delta- \omega_{00}+i\gamma_{0} +\frac{d_0}{2}\partial^2_\xi\;,\\
\hat{\mathcal{L}}_1&=&\Delta- \omega_{10}+i\gamma_{0} +iv_1 \partial_\xi
+\frac{d_1}{2}\partial^2_\xi\;,\\
\hat{\mathcal{L}}_2&=&\Delta- \omega_{20}+i\gamma_{0} +iv_2 \partial_\xi
+\frac{d_2}{2}\partial^2_\xi\;,
\end{eqnarray}
where
\begin{eqnarray}
\label{eq:modal_group_velocity}
v_j=\frac{\omega_{j1}-\omega_{01}}{\sqrt{|\omega_{02}|}}\;,\\
d_j=\frac{\omega_{j2}}{|\omega_{02}|}\, \qquad (d_0=\pm 1)\;.
\end{eqnarray}

where $v_j$ is the relative group velocity of the $j^{th}$ mode with respect to the fundamental mode velocity.

\subsection{Parameter set for the microcavity wire}
\label{sssec:2.2}
We choose all parameters as in \cite{our_OL}.
In Fig.~\ref{fig:geometry_dispersion}b dispersions of the energetically lowest-lying three lower polariton branches are plotted, along with their parabolic fits, giving:
\begin{eqnarray}
\label{eq:omega_parabolic0}
\omega_{00}=-0.2583, \qquad \omega_{01}=0.3311, \qquad \omega_{02}=-0.3129\;, \\
\label{eq:omega_parabolic1}
\omega_{10}=-0.1731, \qquad \omega_{11}=0.2134, \qquad \omega_{12}=-0.1693, \\
\label{eq:omega_parabolic2}
\omega_{20}=-0.0922, \qquad \omega_{21}=0.1172, \qquad \omega_{22}=-0.0647,
\end{eqnarray}
and the higher-order dispersion coefficients:
\begin{eqnarray}
v_1=-0.2103, \qquad v_2=-0.3823,\\
d_0=-1, \qquad d_1=-0.5409, \qquad d_2=-0.2067\;.
\end{eqnarray}

The corresponding modes are displayed in Fig.~\ref{fig:geometry_dispersion}c.
Some nonlinear coefficients (for two coupled modes) are listed below:
\begin{eqnarray}
\label{eq:self_gammas_OL}
\Gamma_{00}=0.2111\;, \qquad \Gamma_{11}=0.2312\;, \qquad \Gamma_{22}=0.2443\;,\\
\label{eq:xpm01_gammas_OL}
\Gamma_{01}=\Gamma_{10}=0.1481\;,\\
\label{eq:xpm02_gammas_OL}
\Gamma_{02}=\Gamma_{20}=0.1543\;,\\
\label{eq:xpm12_gammas_OL}
\Gamma_{12}=\Gamma_{21}=0.1568\\\
\label{eq:strange_gammas}
\Gamma_{000,2}=\Gamma_{002,0}=0.0750\;,\qquad  \Gamma_{222,0}=\Gamma_{220,2}=0.0054\;,\\
\label{eq:strange_gammas_part2}
\Gamma_{112,0}=\Gamma_{110,2}=-0.0764\;,
\end{eqnarray}

And the pump coefficients are (for $q_y=0$):
\begin{eqnarray}
\label{eq:pump_coeffs_0}
h_0&=&E_p\cdot 0.7709\;,\\
h_1&=&0\;,\\
\label{eq:pump_coeffs_1}
h_2&=&E_p\cdot (-0.2208)\;,
\label{eq:pump_coeffs_2}
\end{eqnarray}

\begin{figure}
\includegraphics[width=0.7\textwidth]{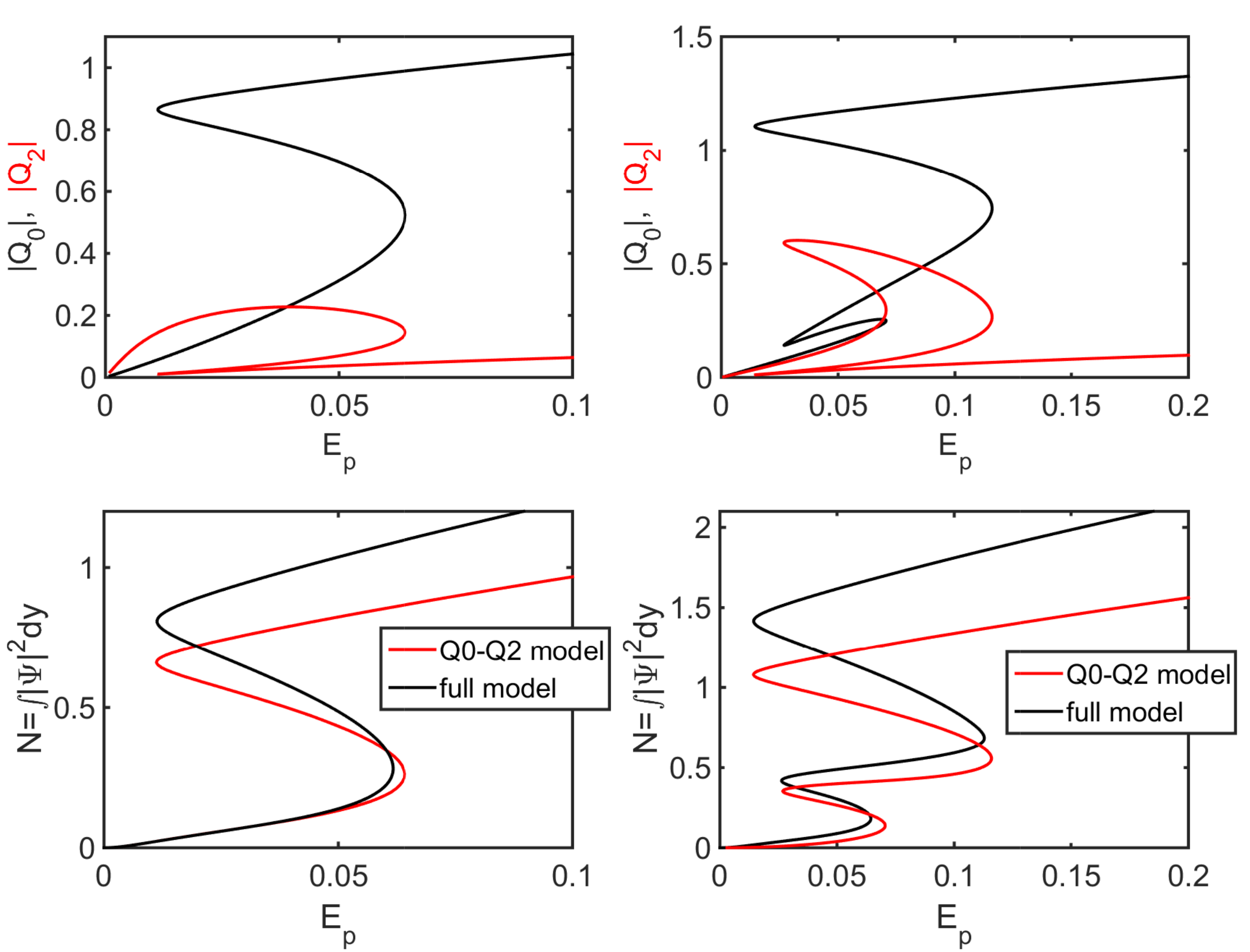}
\caption{Bi- and multistability of the homogeneous solution for $\gamma=0.01$, $\Delta=-0.1$ (left column) and $\Delta=0$ (right column), cf.  Figs. 2b and 2c in Ref. \cite{our_OL}. Solution in terms of $Q_0$ and $Q_2$ is shown in the top row, and the corresponding conversion in terms of the full $\Psi$ field is illustrated in the bottom row ($\Psi$-norm, $N = \int {\left| \Psi  \right|} ^2 dy$).}
\label{fig:compare_with_full_model}
\end{figure}
To check the validity of our approach, in Fig.~\ref{fig:compare_with_full_model} the multistability of the homogeneous solution is illustrated for a different set of parameters (as in \cite{our_OL}). First, we select a realistic value for the dissipation parameter, e.g. $\gamma=0.01$ and detunings $\Delta=-0.1$ and $\Delta=0$ for comparison with the full model. The stationary homogeneous solution of Eqs. (\ref{eq:cm3_Q0}-\ref{eq:cm3_Q2}) was computed (see top row in Fig.~\ref{fig:compare_with_full_model}) . It was then converted to the full $E$- and $\Psi$ fields according to Eqs.~(\ref{eq:modal_decomposition},\ref{eq:0s12_solution}) and compared against stationary solutions of the full model, given by Eqs.~(\ref{eq:eqE},\ref{eq:eqPsi}).
Transition from bi- to multi-stable behaviour upon variation of the detuning parameter from $\Delta=-0.1$ to $\Delta=0$ is observed as in Figs. 2b and 2c in Ref. \cite{our_OL}, confirming the validity of our approach restricted to three coupled modes, of which, effectively only even modes survive (see pump coefficients in Eq.\ref{eq:pump_coeffs_0}-\ref{eq:pump_coeffs_2}).
Discrepancy between the reduced and the full model is noticeable at high amplitudes. This is most likely due to the self-focusing of the nonlinear waveguide mode in a transverse to propagation direction (in our modal expansion scheme the $y$-profile of the mode is fixed).

\subsection{Stability analysis of the homogenous solution}
\label{sssec:3.1}
We perform first stability analysis of the homogeneous solution multistability curves, adding small perturbations to the modal profiles: $Q_0=A+\epsilon_fe^{iqx}e^{(\lambda-i\delta)t}+\epsilon_b^*e^{-iqx}e^{(\lambda+i\delta)t}$, $Q_2=B+p_fe^{iqx}e^{(\lambda-i\delta) t}+p_b^*e^{-iqx}e^{(\lambda+i\delta) t}$with $q,\delta,\lambda$ all real. Introducing linear operators,  $\hat{\mathcal{L}}_{0,2}$ without spatial derivatives, and ${\vec{\bf{x}}} = \left[ {\varepsilon _f ,\varepsilon _b ,p_f ,p_b } \right]^T$, the resulting eigenvalue problem reads:

\begin{eqnarray}
&& (\delta+i\lambda)\vec{x}=
\left[\scalemath{0.8}{
\begin{array}{cccc}
-\hat{\mathcal{L}}_0 & \Gamma_{00} A^2 & 2\Gamma_{20}AB^* & 2\Gamma_{20}AB \\
-\Gamma_{00}(A^*)^2 & \hat{\mathcal{L}}_0^* & -2\Gamma_{20}A^*B^* & -2\Gamma_{20}A^*B  \\
2\Gamma_{02}A^*B & 2\Gamma_{02}AB & -\hat{\mathcal{L}}_2+v_2q+2(\Gamma_{22}|B|^2+\Gamma_{02}|A|^2) &  \Gamma_{22}B^2 \\
-2\Gamma_{02}A^*B^* & -2\Gamma_{02}AB^* & -\Gamma_{22}(B^*)^2 & \hat{\mathcal{L}}_2^*+v_2q-2(\Gamma_{22}|B|^2+\Gamma_{02}|A|^2)
\label{eq:eigenvalue_problem}
\end{array}}
\right]\vec{x}\;,
\end{eqnarray}

\begin{figure}

\resizebox{.9\textwidth}{!}{%
\includegraphics[scale=0.5]{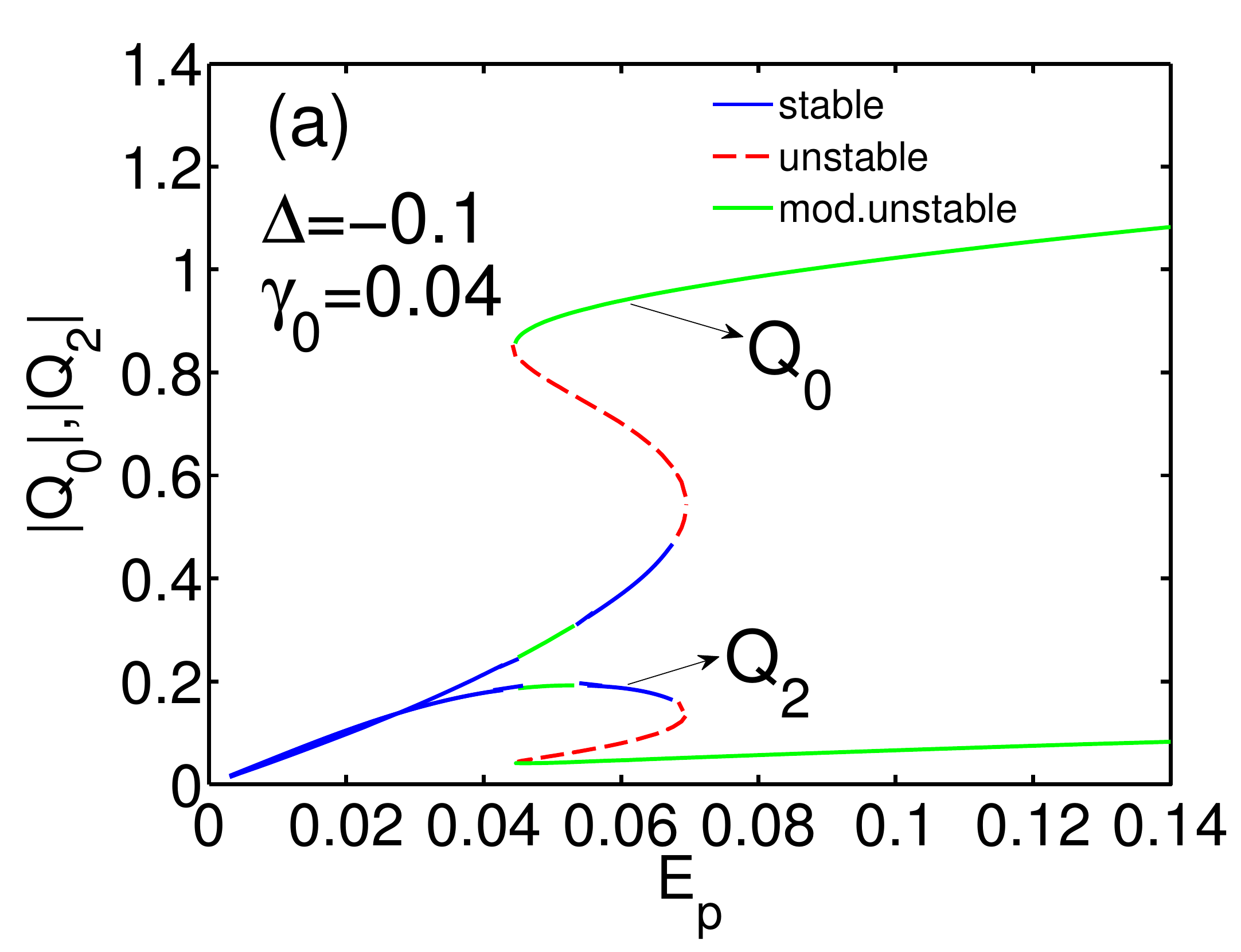}
\includegraphics[scale=0.5]{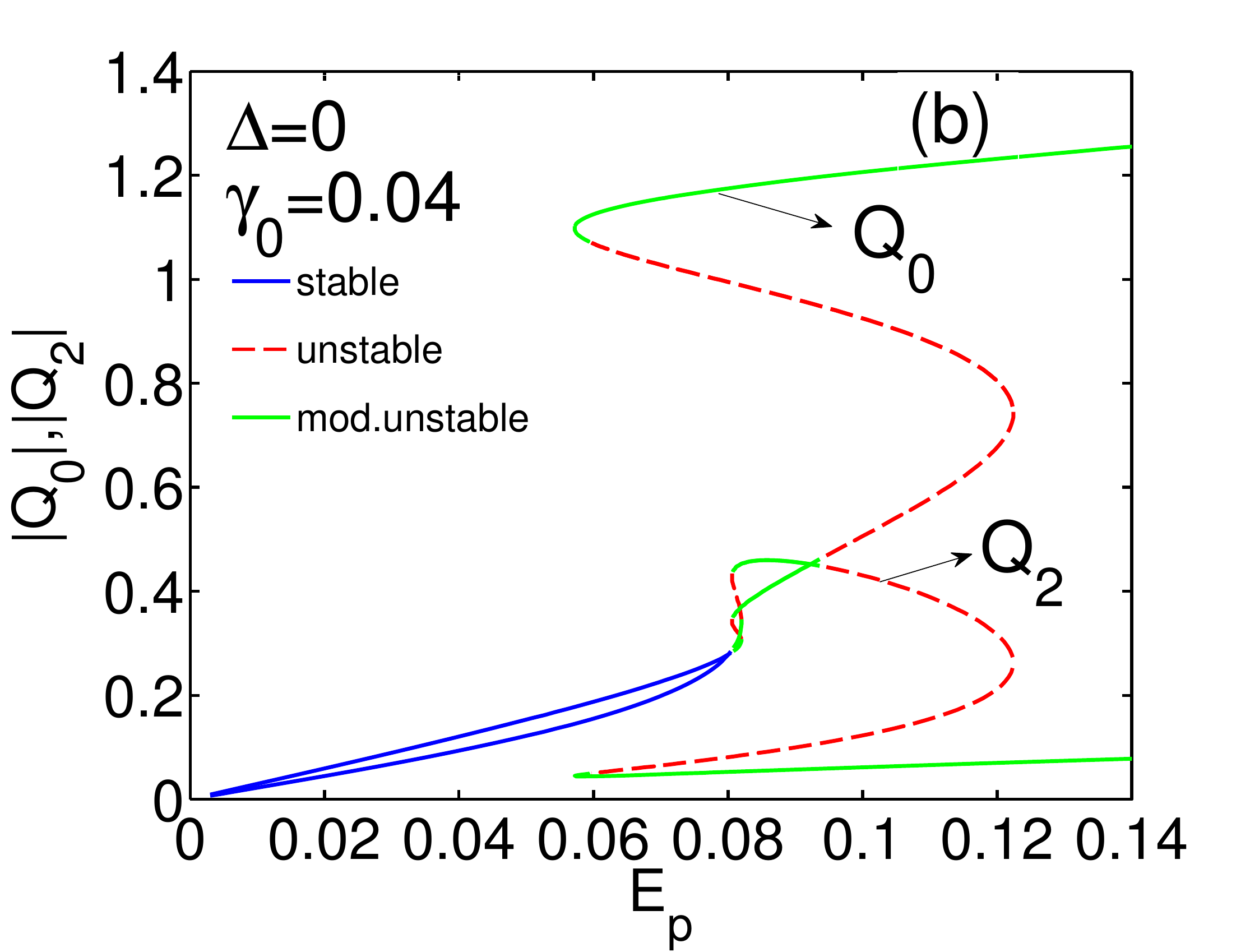}
}
\caption{Stability analysis of the homogeneous solution for (a) $\Delta=-0.1, \gamma_0=0.04$; (b) $\Delta=0, \gamma_0=0.04$. Solid blue, dashed red and solid green curves correspond to stable, unstable and modulationally unstable branches, respectively.}
\label{Fig:Stability_analysis_homogeneous_Delta_0_m0p1}
\end{figure}

The eigenvalues and engenvectors are found numerically and the stable ($\lambda<0$), unstable($\lambda(q=0)>0$) and modulationally unstable ($\lambda>0$ only for some $q\ne 0$) branches of the multi-stability curves for $Q_0$ and $Q_2$ homogeneous solutions are plotted against pump amplitude $E_p$ for $\Delta=0, -0.1$ and $\gamma_0=0.04$ in Fig. \ref{Fig:Stability_analysis_homogeneous_Delta_0_m0p1}. The stability of the homogeneous nonlinear solution provides background for analysis of the soliton solutions, which we shall compute in the following section.

\section{Coupled soliton families with non-zero pump and dissipation}
\label{sec3}
In this section we calculate the coupled soliton branches as a function of the pump amplitude for a dissipation parameter $\gamma=0.04$ and detunings $\Delta=-0.1$ and $\Delta=0$. We solve self-consistently Eqs.(\ref{eq:cm3_Q0}), (\ref{eq:cm3_Q2}) for the ($Q_0, Q_2$) coupled soliton in a moving with the soliton reference frame, introducing additional unknown parameter: the soliton velocity, $u$. Two types stable and unstable solitons are found, which we will refer to as type 1 and 2 (soliton branch stability investigated below).

\begin{figure}
\includegraphics[scale=0.5]{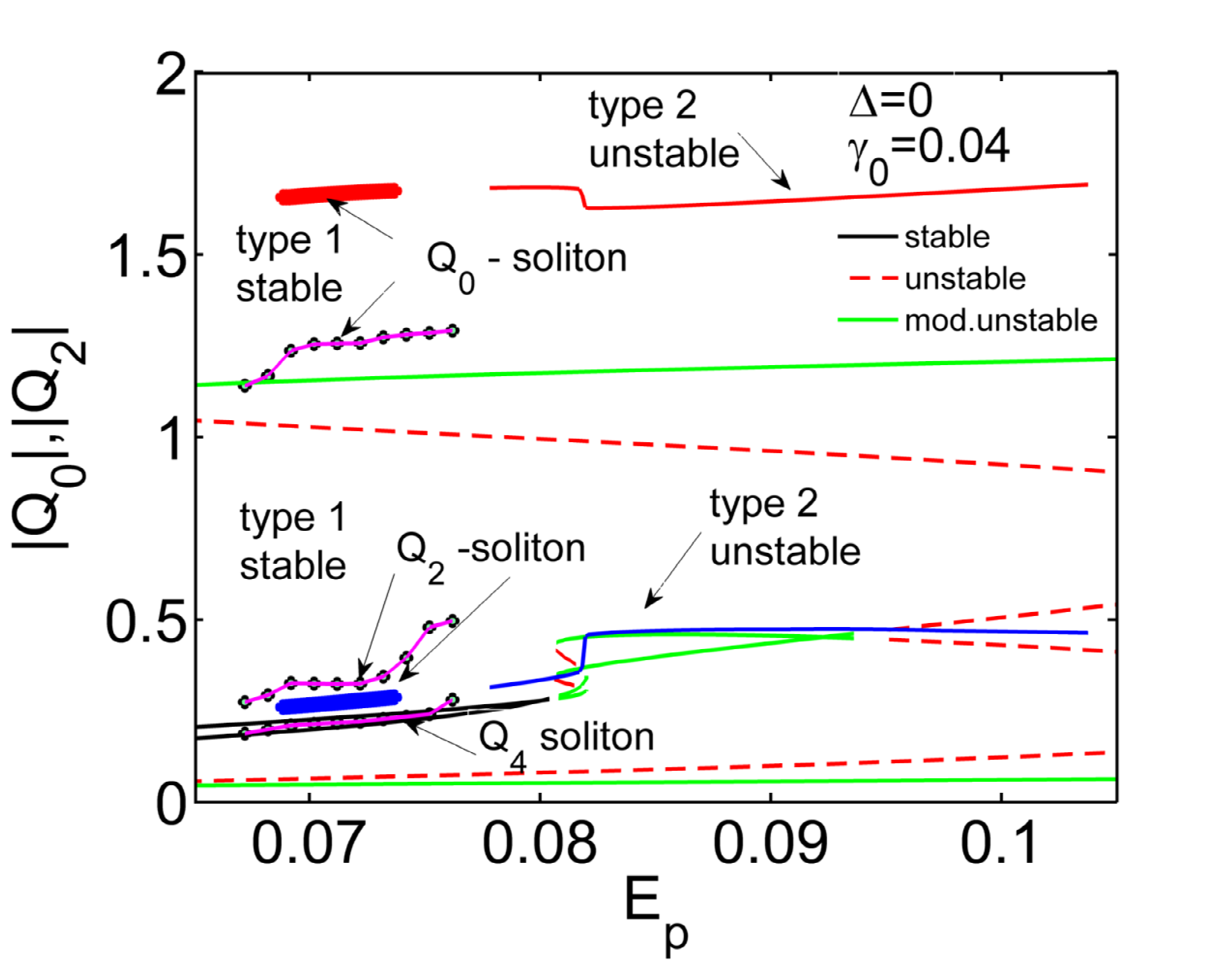}
\resizebox{.9\textwidth}{!}{%
\includegraphics[scale=0.3]{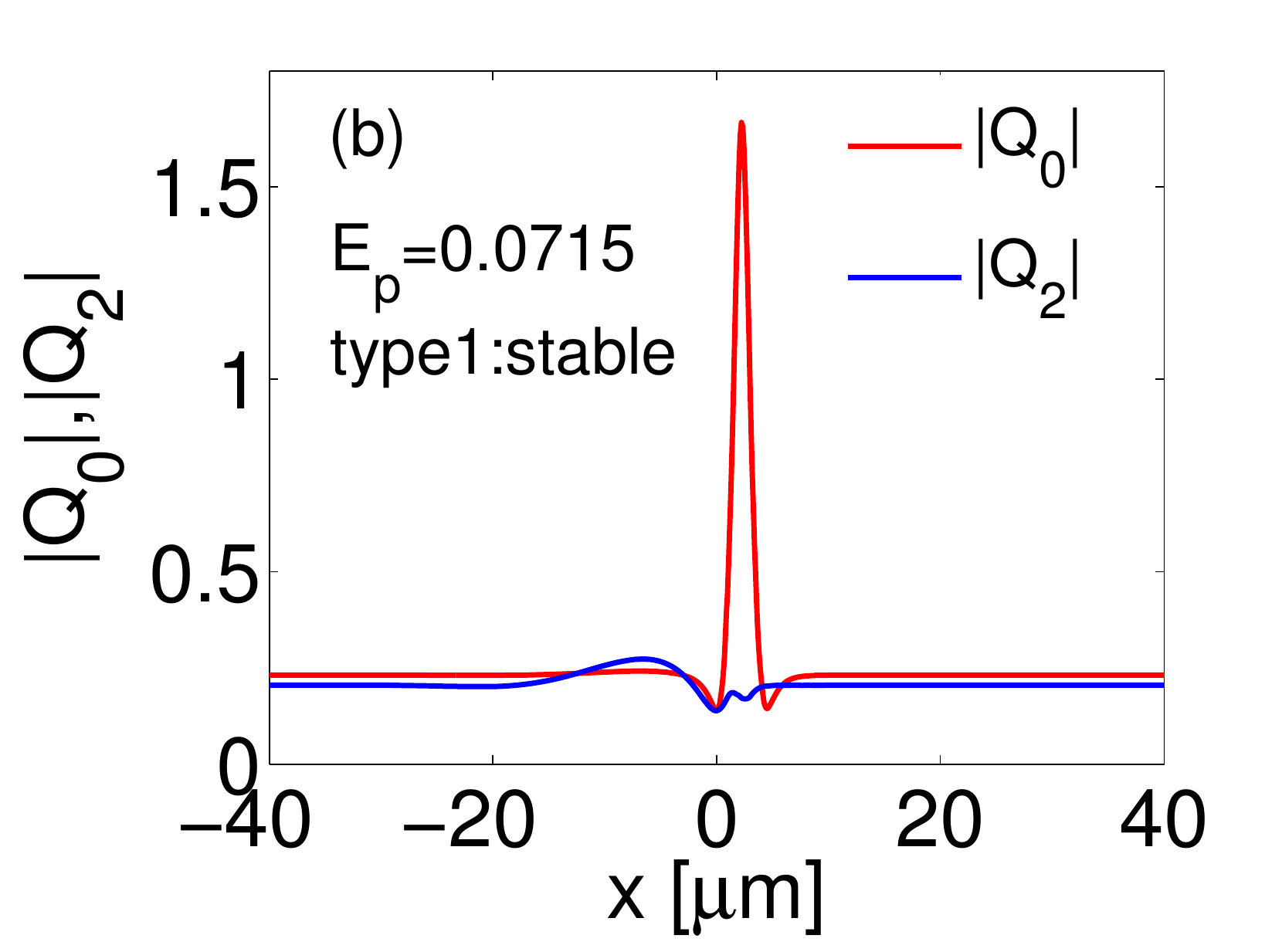}
\includegraphics[scale=0.3]{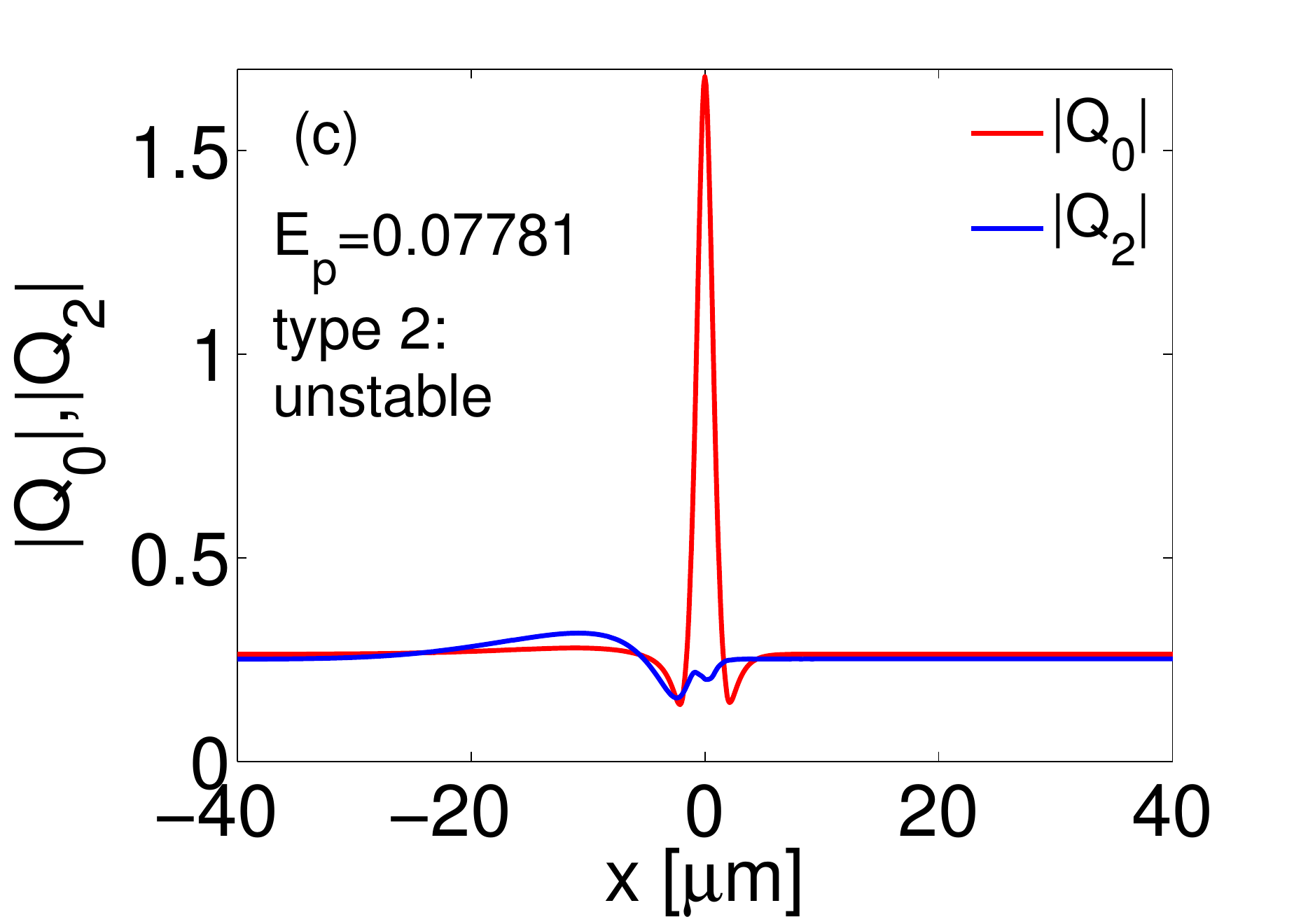}
\includegraphics[scale=0.3]{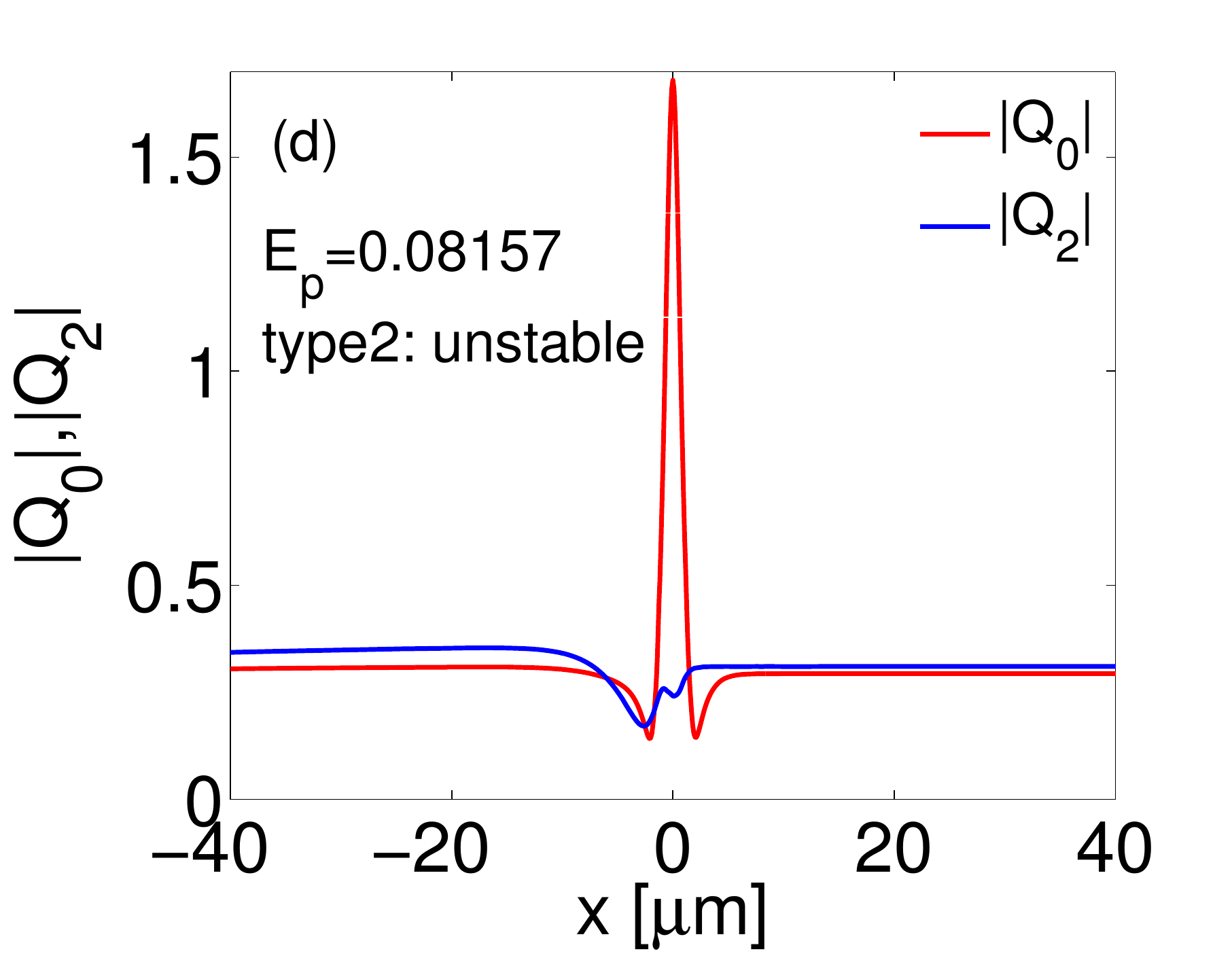}
}
\resizebox{.9\textwidth}{!}{%
\includegraphics[scale=0.3]{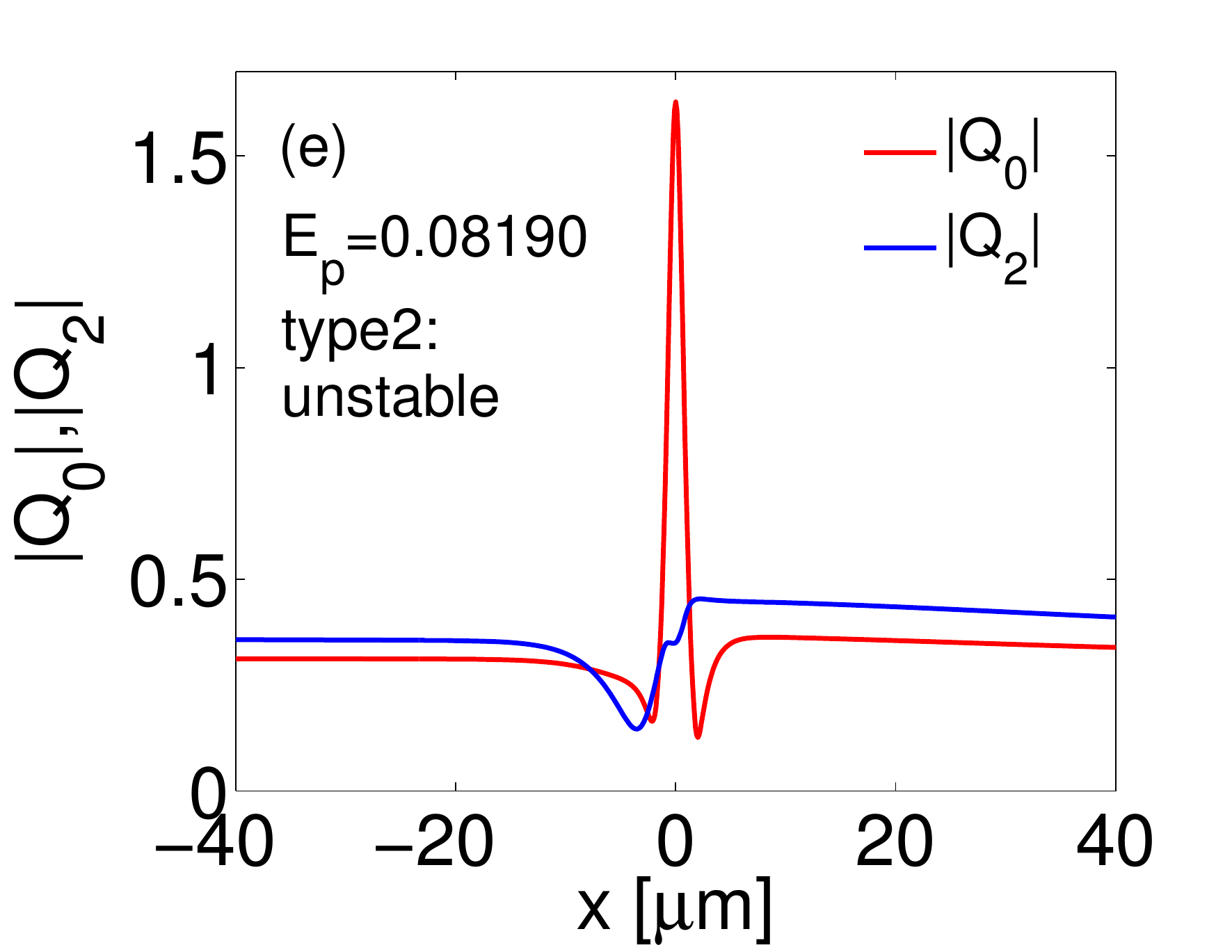}
\includegraphics[scale=0.3]{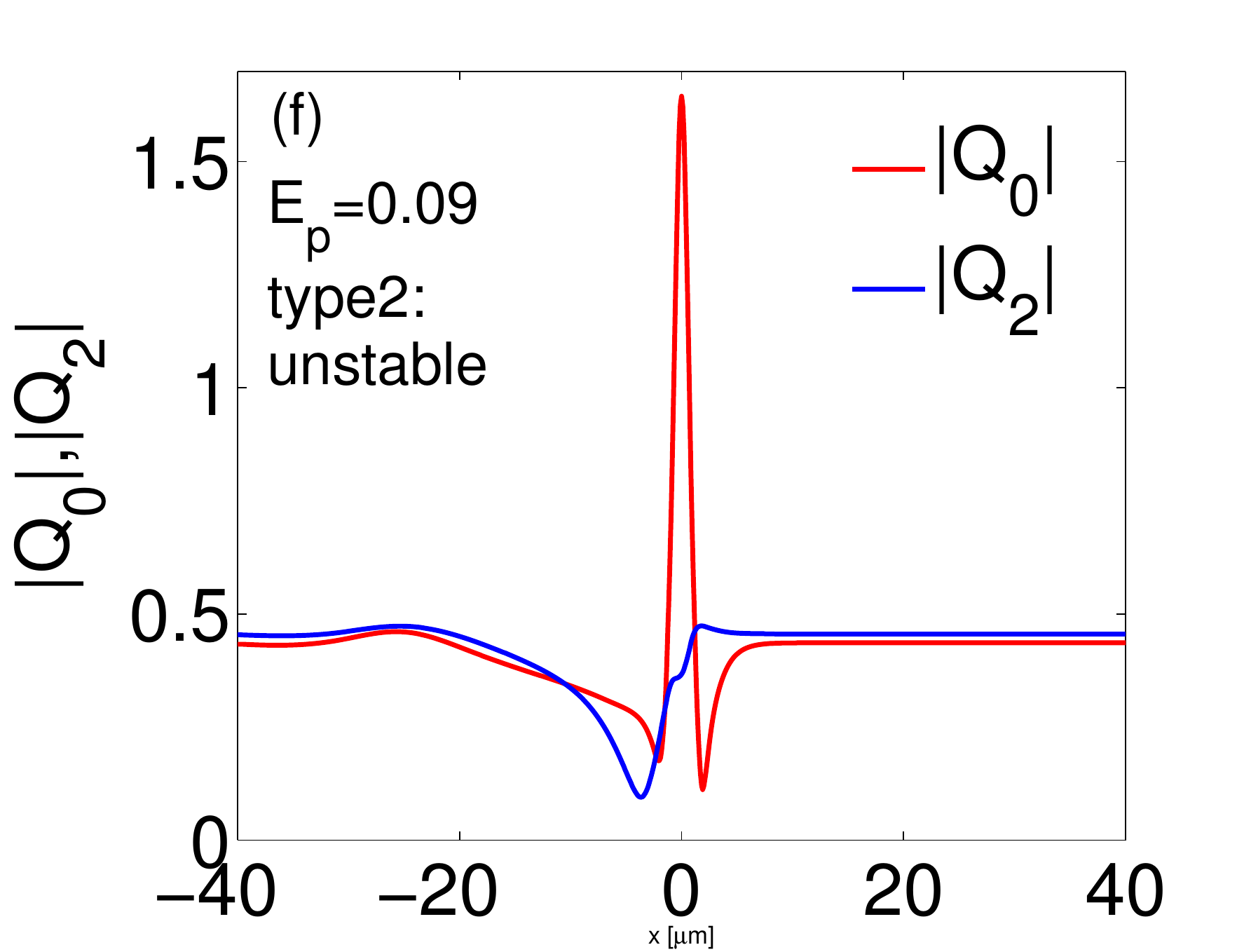}
\includegraphics[scale=0.3]{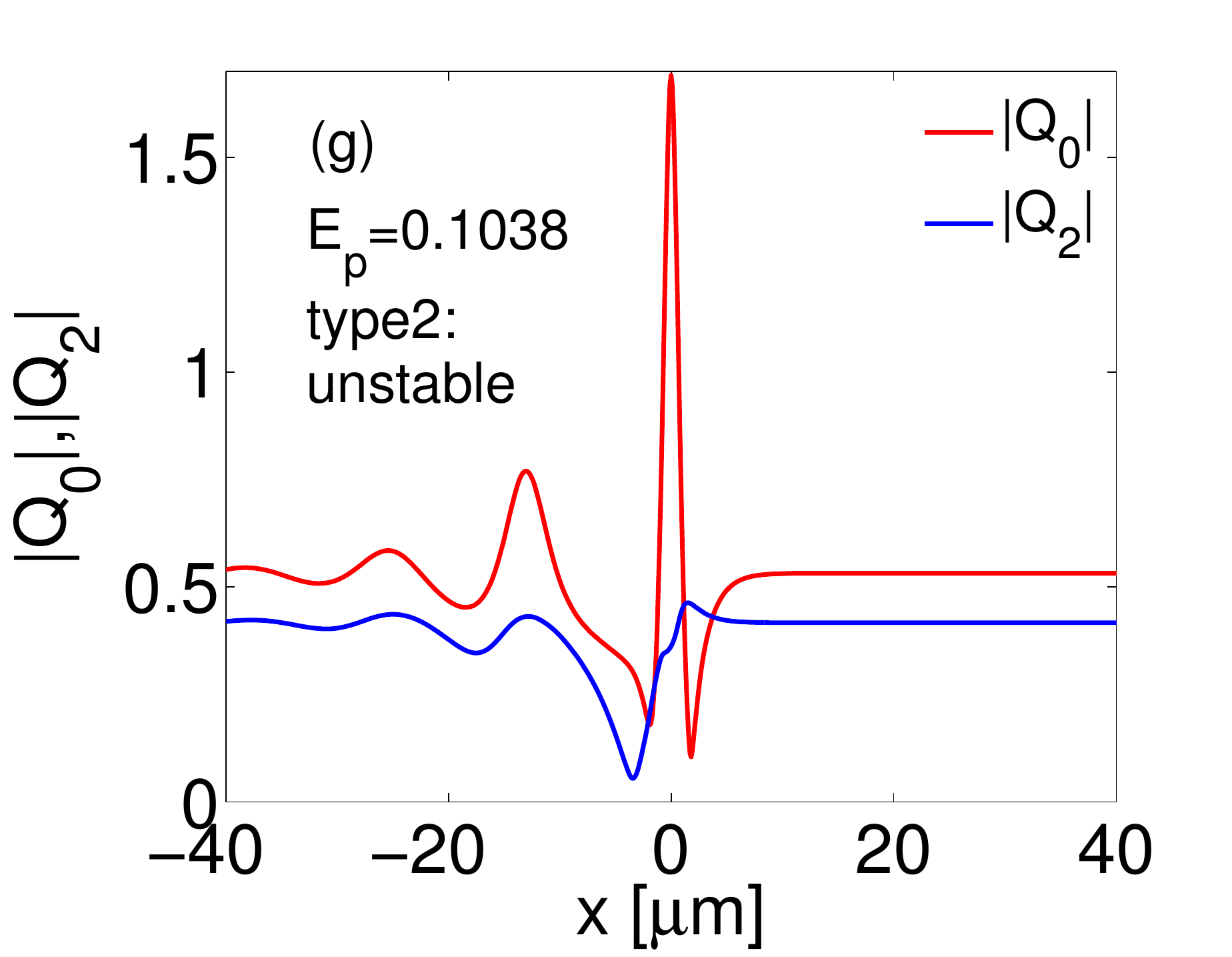}
}
\caption{(a) Coupled stable (type 1)  and unstable (type2) soliton branches, superimposed on the multistability curves of the stationary nonlinear coupled ($Q_0,Q_2$)-modes vs $E_p$ (homogeneous solution stabilty indicated) at $\Delta=0$ and $\gamma_0=0.04$; note that the stable soliton branches sit on a stable background (black curve), while the unstable soliton branches sit on modulationally unstable background (green curve); Soliton branches corresponding to $Q_0$, $Q_2$ and $Q_4$ soliton components, inferred from the inverse transform (Eq.\ref{eq:modal_decomposition}) of the full-model 2D solitons, computed by a dynamical (split-step) model from (Eq.\ref{eq:eqE}, \ref{eq:eqPsi}), are shown with open circles connected by magenta line; (b) Stable type1 $|Q_0|$ and $|Q_2|$-soliton profiles in the middle of the stable soliton branch at $E_p=0.0715$; this soliton profile remains unchanged from the left stable soliton branch edge at $E_p=0.06886$ to the right edge at $E_p=0.07366$; Unstable type2 profiles at the (c) left edge of unstable soliton branch at $E_p=0.07781$; (d) right edge before kink at $E_p=0.08157$; (e) left edge after kink at $E_p=0.08190$; (f) at $E_p=0.09$; (g) right edge of unstable soliton branch at $E_p=0.1038$.}
\label{Fig:Stability_analysis_homogeneous_type1_solitons_Delta_0}
\end{figure}

The stable soliton type 1 branch is found from the final evolved profiles of the time-dependent equations (Eqs. (\ref{eq:eqE}),(\ref{eq:eqPsi})), solved by Fourier split-step method (see \cite{our_OL}), taken as initial guess for the Newton-Raphson method. The unstable type 2 soliton branch is obtained numerically from the coupled ($Q_0,Q_2$) stationary equations setting initially $Q_2=0$ with non-zero pump terms, $h_0,h_2$. Both type 1 and 2 soliton branches are shown in Fig.\ref{Fig:Stability_analysis_homogeneous_type1_solitons_Delta_0}(a) for zero detuning, $\Delta=0$, superimposed on the homogeneous solution multistability curves. Type 1 $Q_0$- and $Q_2$ soliton branches are displayed by a thick red/blue lines, respectively. In what follows we shall show that the solitons along these branches are stable. The corresponding soliton type 2 branches are denoted by solid red/blue lines. The transition between the unstable and stable branches is clearly visible by the kink in the curve. We note that there is also a gap between these two types soliton branches discussion of which we will postpone to Sec. \ref{sssec:4.3}.

The soliton profiles for the $Q_0$ and $Q_2$ components along the branches are shown in Fig. \ref{Fig:Stability_analysis_homogeneous_type1_solitons_Delta_0} (b-g) for different pump amplitudes, sweeping the curve from the left edge of the stable type 1 soliton branch up to the right edge of the unstable (type 2) branch. We note that the stable solitons (type 1) sit on a stable background of the homogeneous solution (see black portions of the curves in Fig.\ref{Fig:Stability_analysis_homogeneous_type1_solitons_Delta_0} (a) and the soliton profile remains unchanged as in Fig. \ref{Fig:Stability_analysis_homogeneous_type1_solitons_Delta_0} (b) along the stable branch. By contrast, the unstable soliton type 2 sits on a modulationally unstable homogeneous solution background and this soliton profile significantly changes when sweeping the unstable branch from the left edge (Fig. \ref{Fig:Stability_analysis_homogeneous_type1_solitons_Delta_0}(c)) to the right edge (Fig. \ref{Fig:Stability_analysis_homogeneous_type1_solitons_Delta_0}(g)) where higher-amplitude oscillations appear in the soliton tail.

\begin{figure}
\includegraphics[scale=0.5]{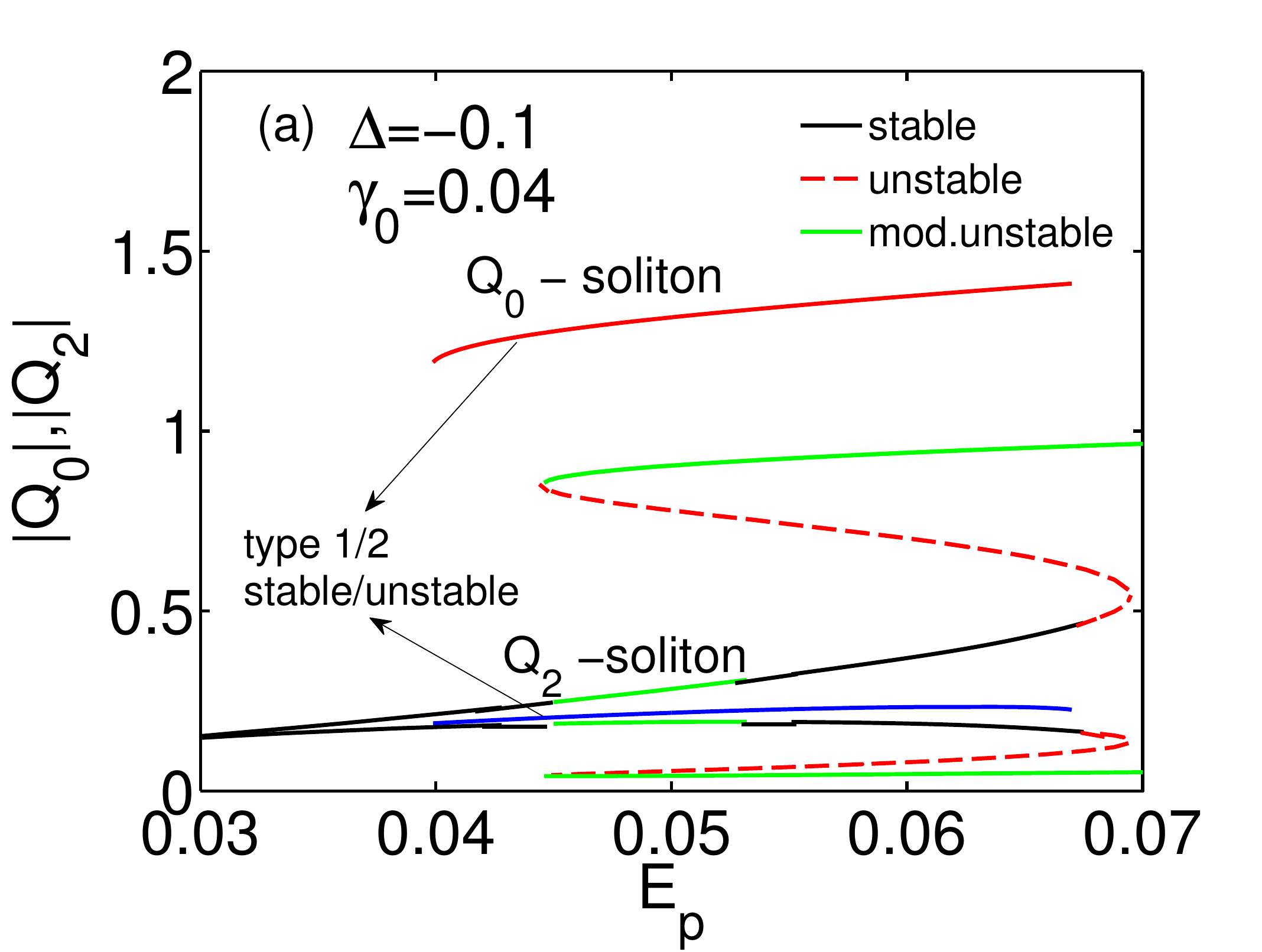}
\resizebox{.9\textwidth}{!}{%
\includegraphics[scale=0.3]{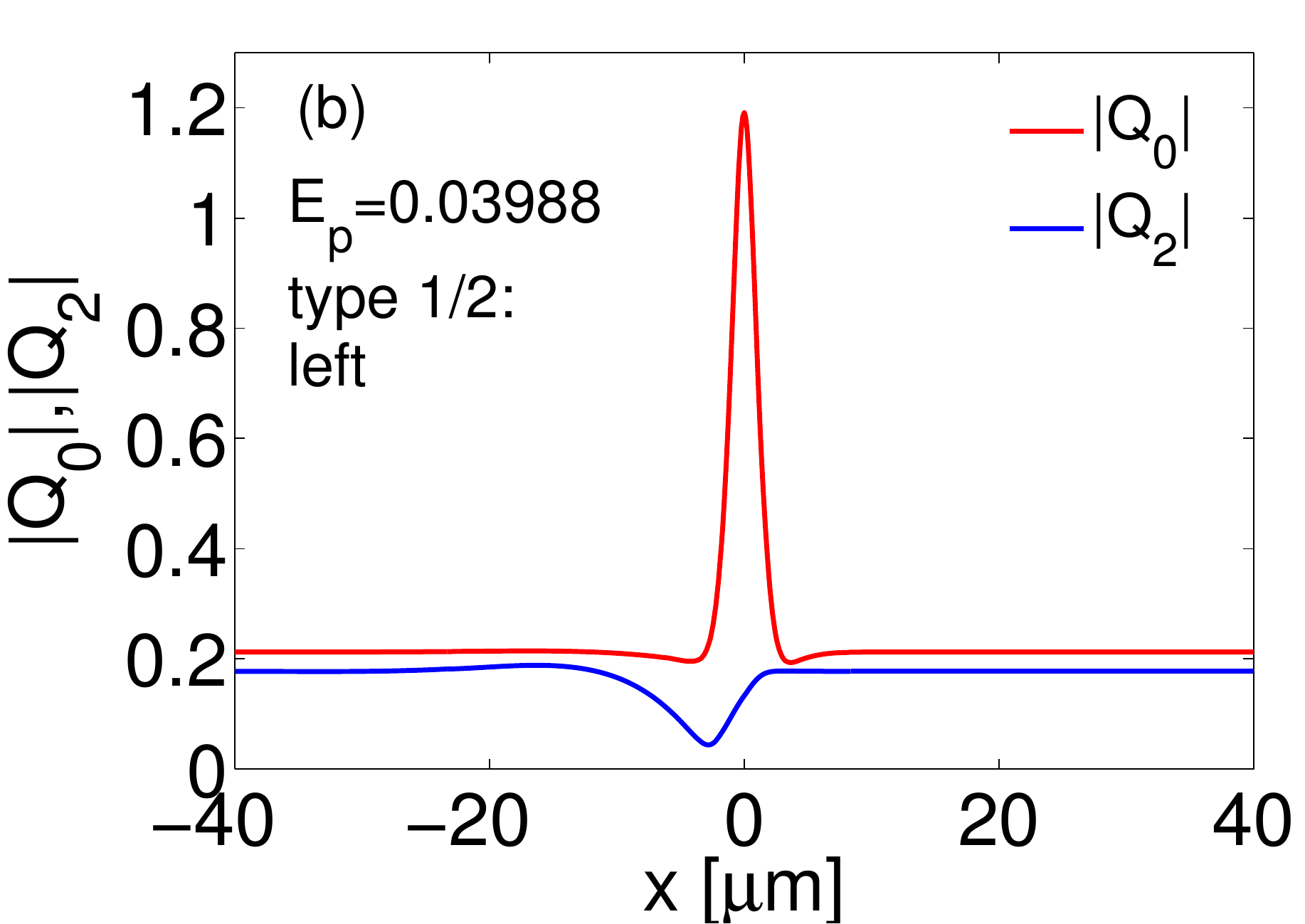}
\includegraphics[scale=0.3]{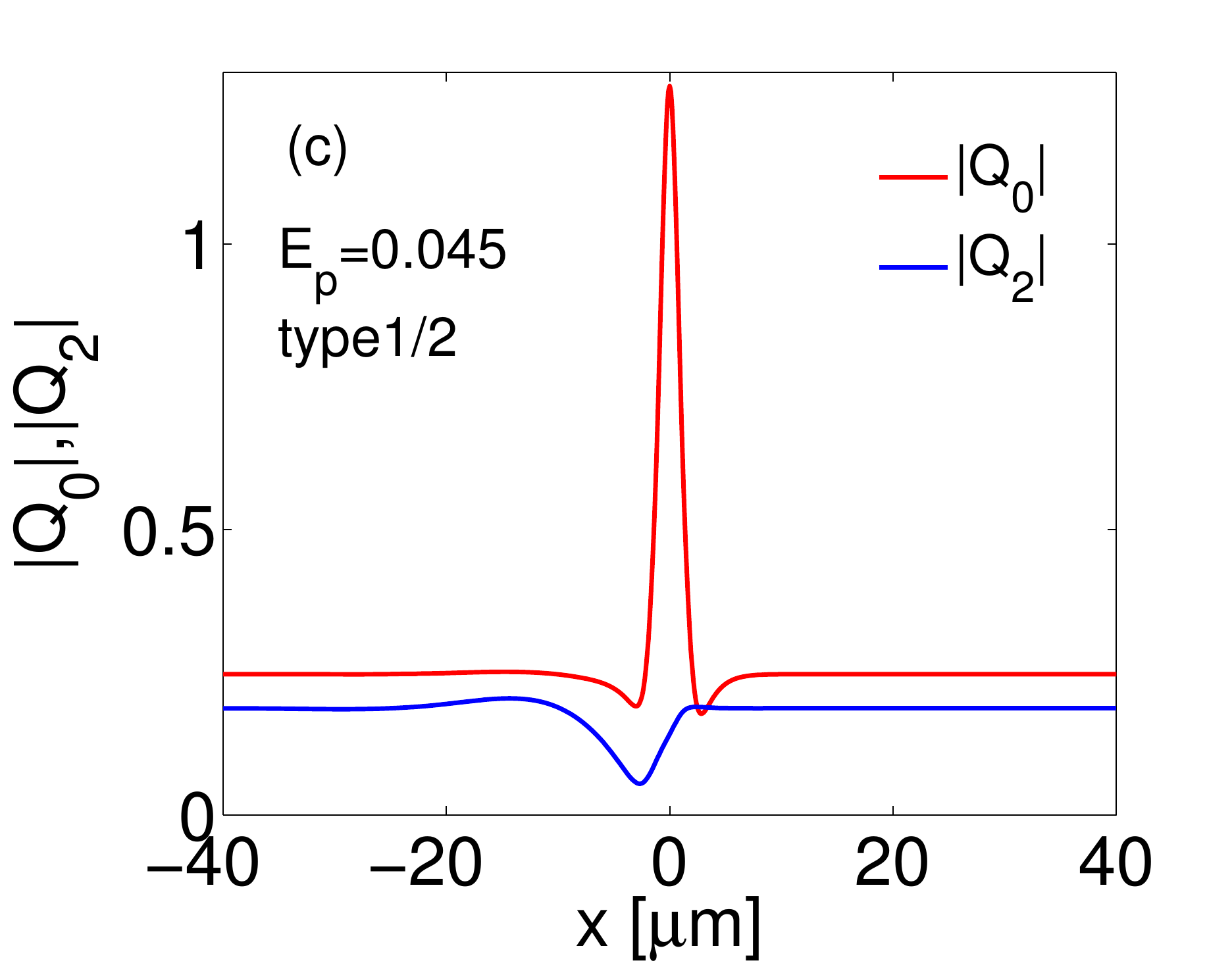}
}
\resizebox{.9\textwidth}{!}{%
\includegraphics[scale=0.3]{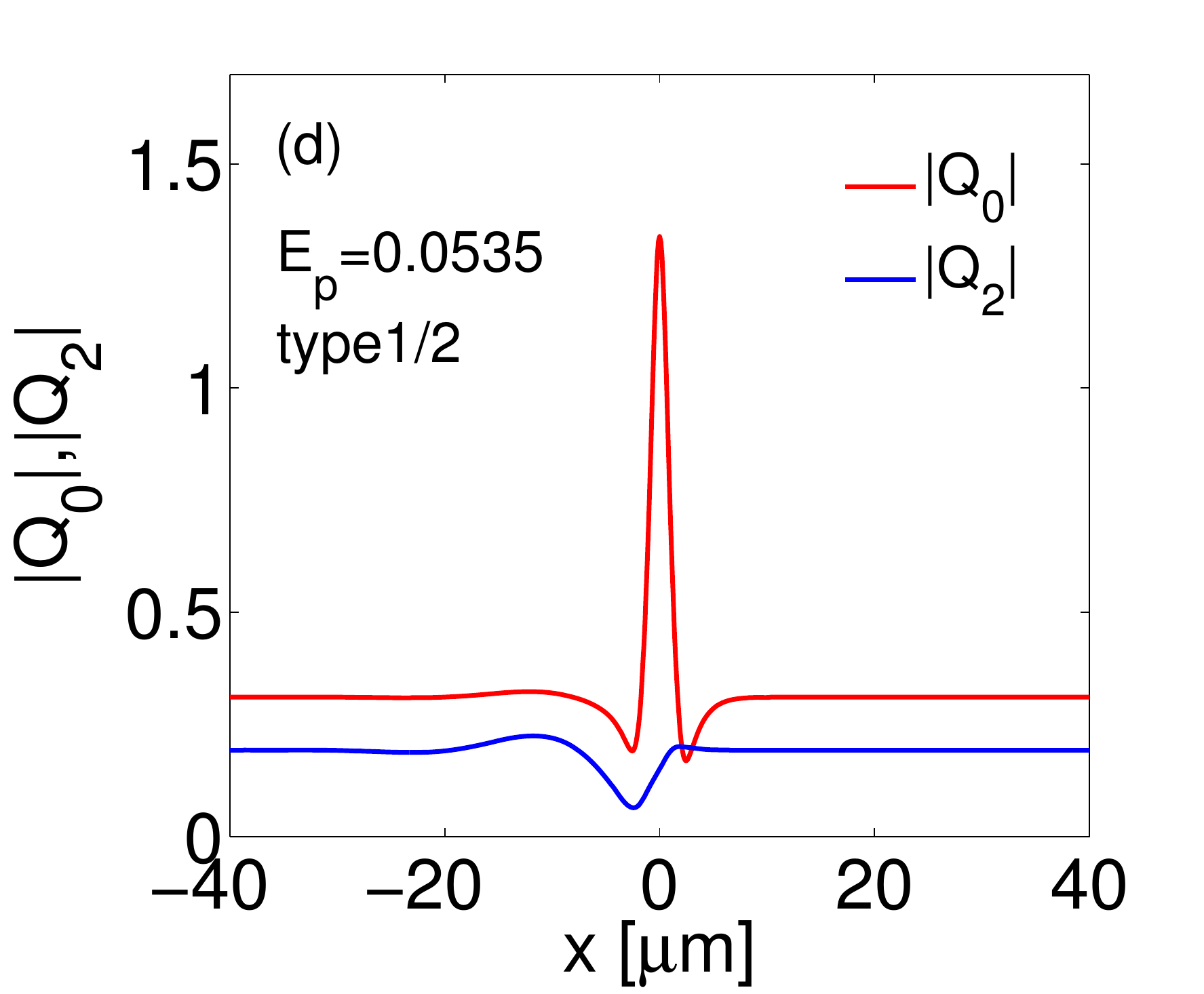}
\includegraphics[scale=0.3]{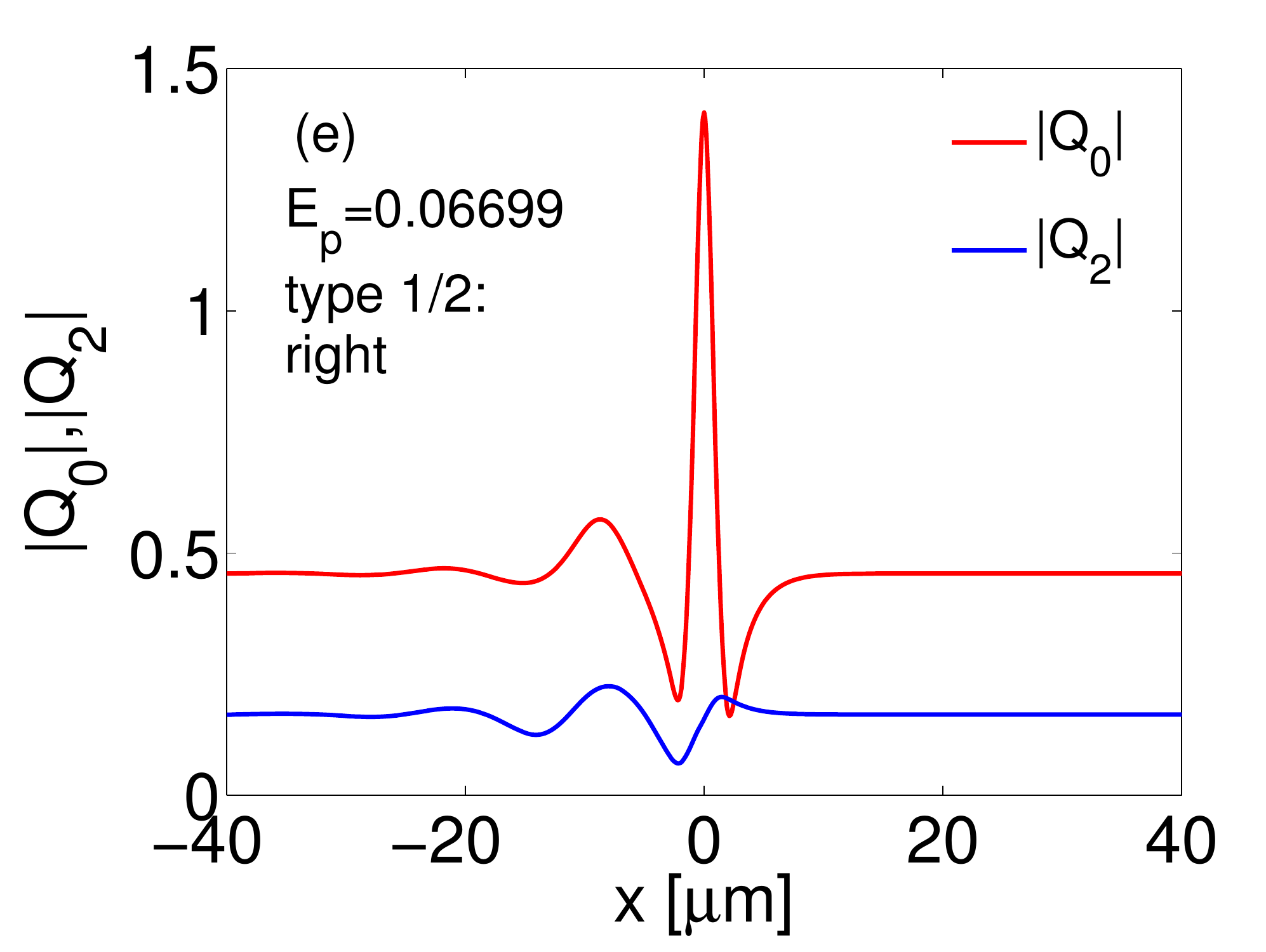}
}
\caption{(a) Coupled (type 1) unstable soliton branches vs $E_p$, superimposed on the homogeneous stationary solution multistability curves (stabilty indicated) at $\Delta=-0.1$ and $\gamma_0=0.04$; (b) Stable/unstable type1 soliton profiles at the left edge of the soliton branch at $E_p=0.03988$; (c) at $E_p=0.045$ at the edge of the homogeneous solution stable region (black curve); (d) at $E_p=0.0535$; (e) Stable/unstable type1 profiles at the right edge of the soliton branch at $E_p=0.06699$.}
\label{Fig:Stability_analysis_homogeneous_type1_solitons_Delta_-0.1}
\end{figure}

The soliton branches for the stable (type 1) and unstable solitons (type 2) merge for $\Delta=-0.1$ and $\gamma_0=0.04$ and are plotted in Fig.\ref{Fig:Stability_analysis_homogeneous_type1_solitons_Delta_-0.1}. The solitons are very close to stable at the left edge of the soliton branch and become more unstable towards the right edge, where the background is modulationally unstable. This is confirmed by our stability analysis of the soliton branches below.

We should note that the $Q_2$ soliton component of the stable soliton in Fig.\ref{Fig:Stability_analysis_homogeneous_type1_solitons_Delta_0}(b) at $\Delta=0, \gamma_0=0.04$ intersects the $Q_0$ one, thereby contributing to the specific spatial dynamics of the reconstructed 2D soliton shown in Fig. \ref{Fig:Reconstructed_2D_soliton_Delta_0} (a-c), namely $Q_0$-component dominates the soliton core, while the $Q_2$-component dominates the tail behaviour, leading to the characteristic split double-lobe tail in transverse to the propagation direction. By contrast, in the case $\Delta=-0.1, \gamma_0=0.04$, $Q_2$-component contributes to the soliton core but hardly has any influence on the soliton tail, thus leading to the single-lobe 2D spatial profile of the soliton tail, observed in Fig. \ref{Fig:Reconstructed_2D_soliton_Delta_m0p1}(a-c).

We perform stability analysis of the soliton solutions in a moving with the soliton frame. Introducing a new variable, $\eta=\xi-ut$, the $Q_0$ and $Q_2$-components of the coupled soliton can be written as:
\begin{equation}
\begin{array}{l}
 Q_0 \left( \eta  \right) = A_0 \left( {\xi  - ut} \right) = A_0 \left( \eta  \right) \\
 Q_2 \left( \eta  \right) = A_2 \left( {\xi  - ut} \right) = A_2 \left( \eta  \right) \\
 \label{soliton_profiles}
 \end{array}
\end{equation}
Adding small perturbations to the soliton profiles Eqs.~(\ref{soliton_profiles}), according to:
\begin{equation}
\begin{array}{l}
 Q_0 \left( \eta  \right) = A_0 \left( \eta  \right) + \varepsilon _f \left( \eta  \right)e^{\left( {\lambda  - i\delta } \right)t}  + \varepsilon _b^* \left( \eta  \right)e^{\left( {\lambda  + i\delta } \right)t}  \\
 Q_2 \left( \eta  \right) = A_2 \left( \eta  \right) + p_f \left( \eta  \right)e^{\left( {\lambda  - i\delta } \right)t}  + p_b^* \left( \eta  \right)e^{\left( {\lambda  + i\delta } \right)t}
 \label{soliton_perturbations}
 \end{array}
\end{equation}
and introducing linear operators: $\hat{\mathcal{L}}_j=\Delta- \omega_{j0}+i\gamma_{0} + i (v_j-u) \partial_\xi  +\frac{d_j}{2}\partial^2_\xi$ for $j=0,2$, and $v_0=0$, we obtain the following eigenvalue problem:
\begin{eqnarray}
\label{eq:soliton_eigenvalue_problem}
&& (\delta+i\lambda)\vec{x}=
\left[\scalemath{0.7}{
\begin{array}{cccc}
-\hat{\mathcal{L}}_0 +2(\Gamma_{00}|A_0|^2+\Gamma_{20}|A_2|^2)& \Gamma_{00} A_0^2 & 2\Gamma_{20}A_0A_2^* & 2\Gamma_{20}A_0A_2 \\
-\Gamma_{00}(A_0^*)^2 & \hat{\mathcal{L}}_0^* -2(\Gamma_{00}|A_0|^2+\Gamma_{20}|A_2|^2)& -2\Gamma_{20}A_0^*A_2^* & -2\Gamma_{20}A_0^*A_2)  \\
2\Gamma_{02}A_0^*A_2 & 2\Gamma_{02}A_0A_2 & -\hat{\mathcal{L}}_2+2(\Gamma_{22}|A_2|^2+\Gamma_{02}|A_0|^2) &  \Gamma_{22}A_2^2 \\
-2\Gamma_{02}A_0^*A_2^* & -2\Gamma_{02}A_0A_2^* & -\Gamma_{22}(A_2^*)^2 & \hat{\mathcal{L}}_2^*-2(\Gamma_{22}|A_2|^2+\Gamma_{02}|A_0|^2)
\end{array}}
\right]\vec{x}\;,
\end{eqnarray}

\begin{figure}
\resizebox{.9\textwidth}{!}{%
\includegraphics[height=5cm]{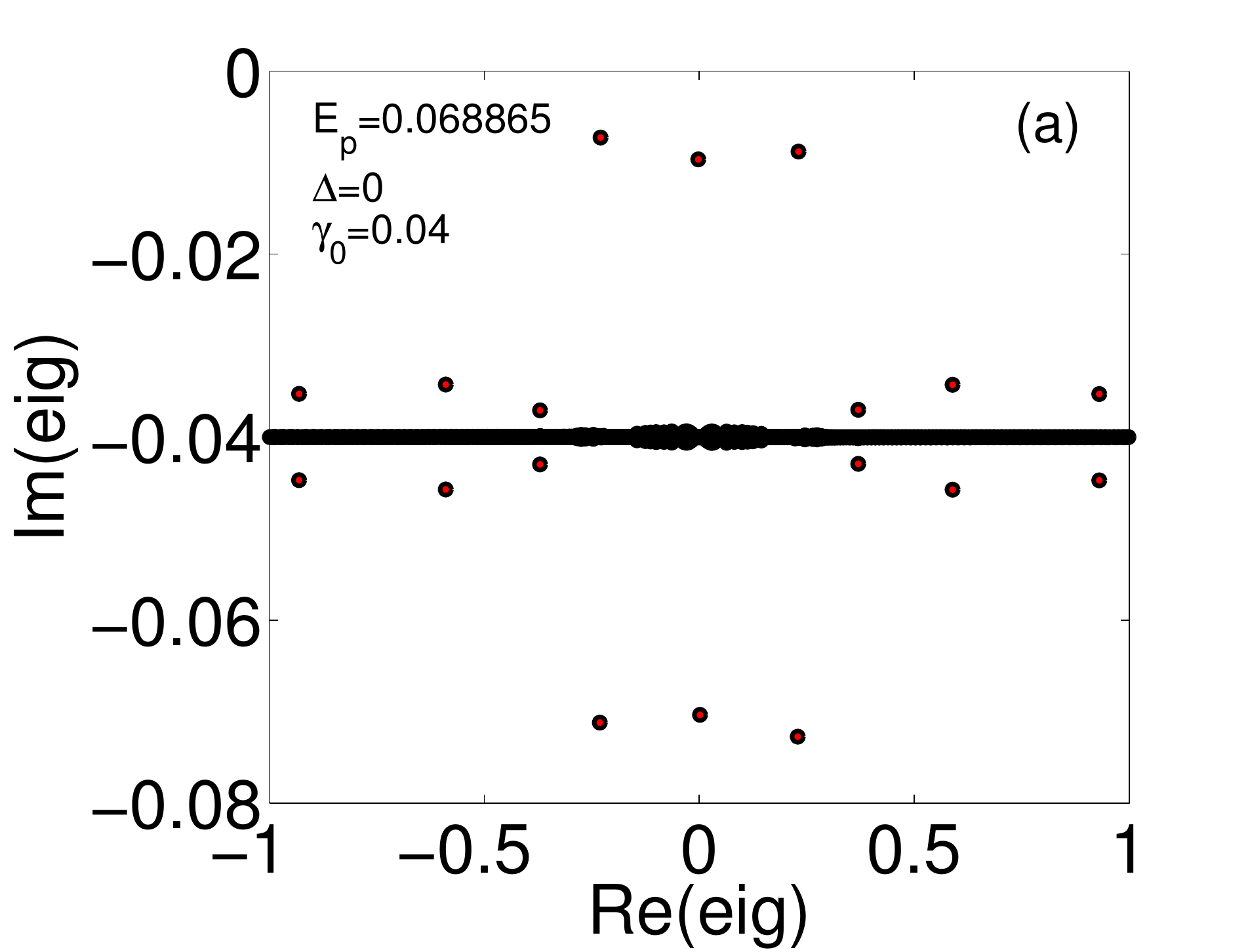}
\includegraphics[height=5cm]{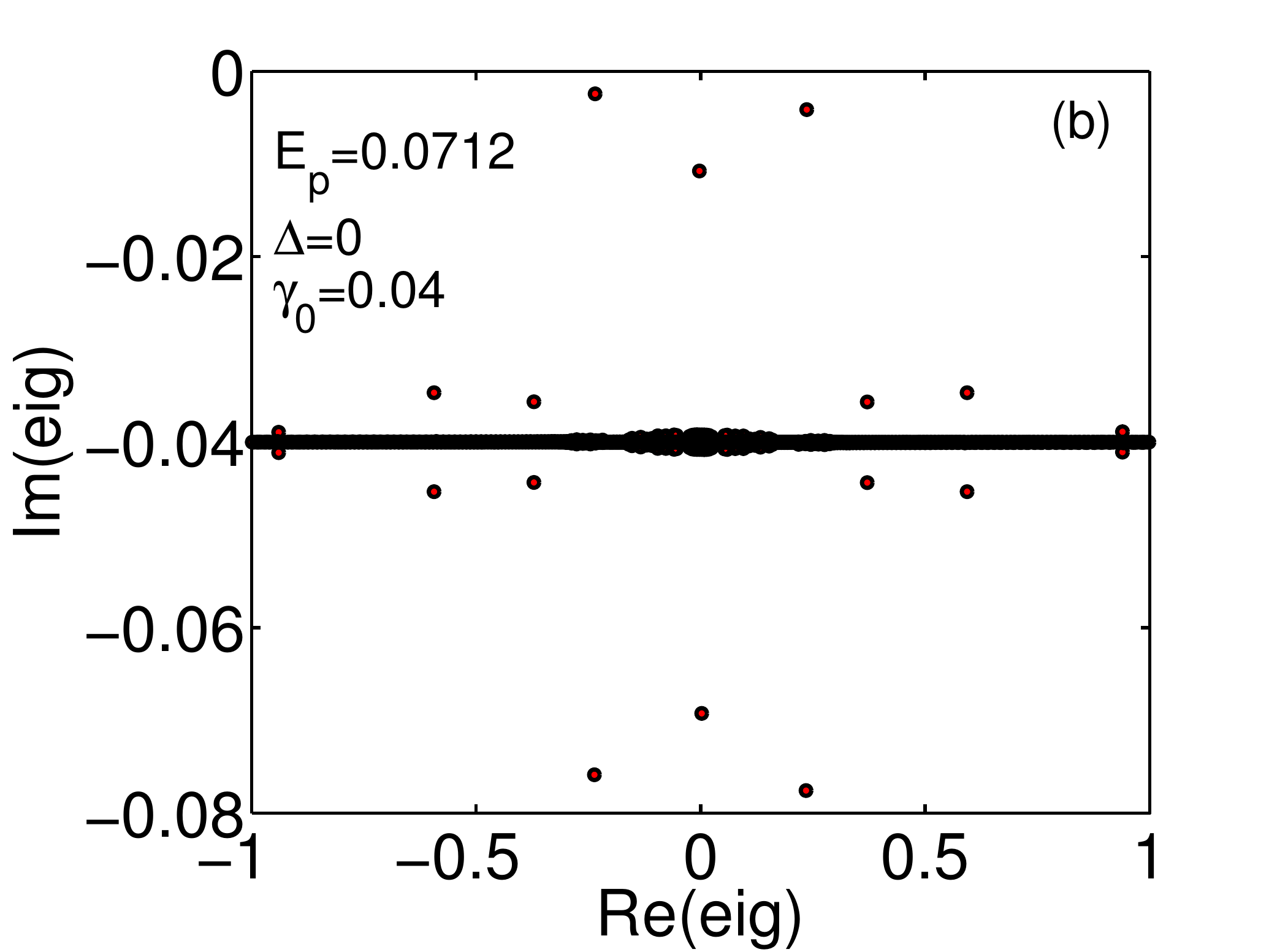}
}
\resizebox{.9\textwidth}{!}{%
\includegraphics[height=5cm]{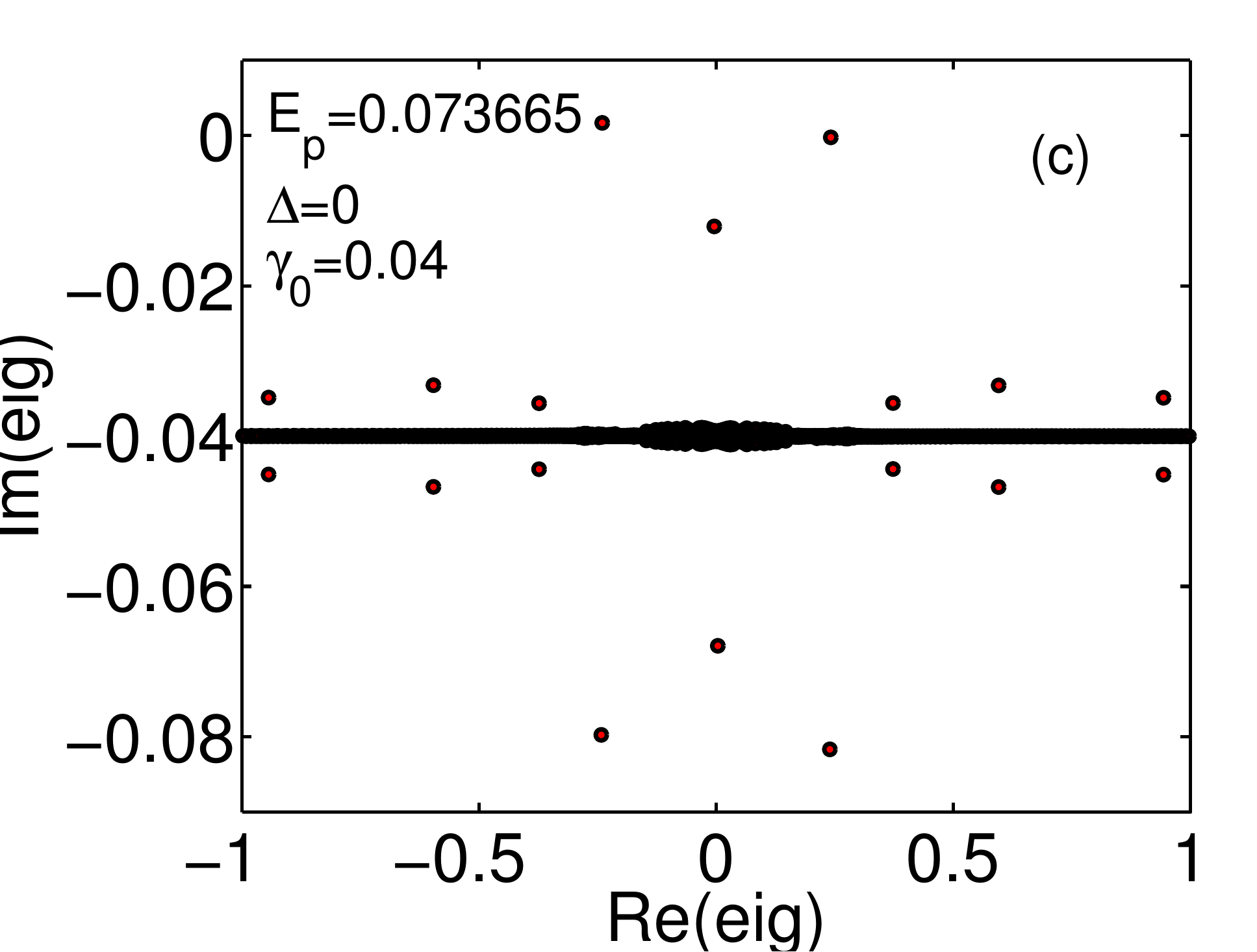}
\includegraphics[height=5cm]{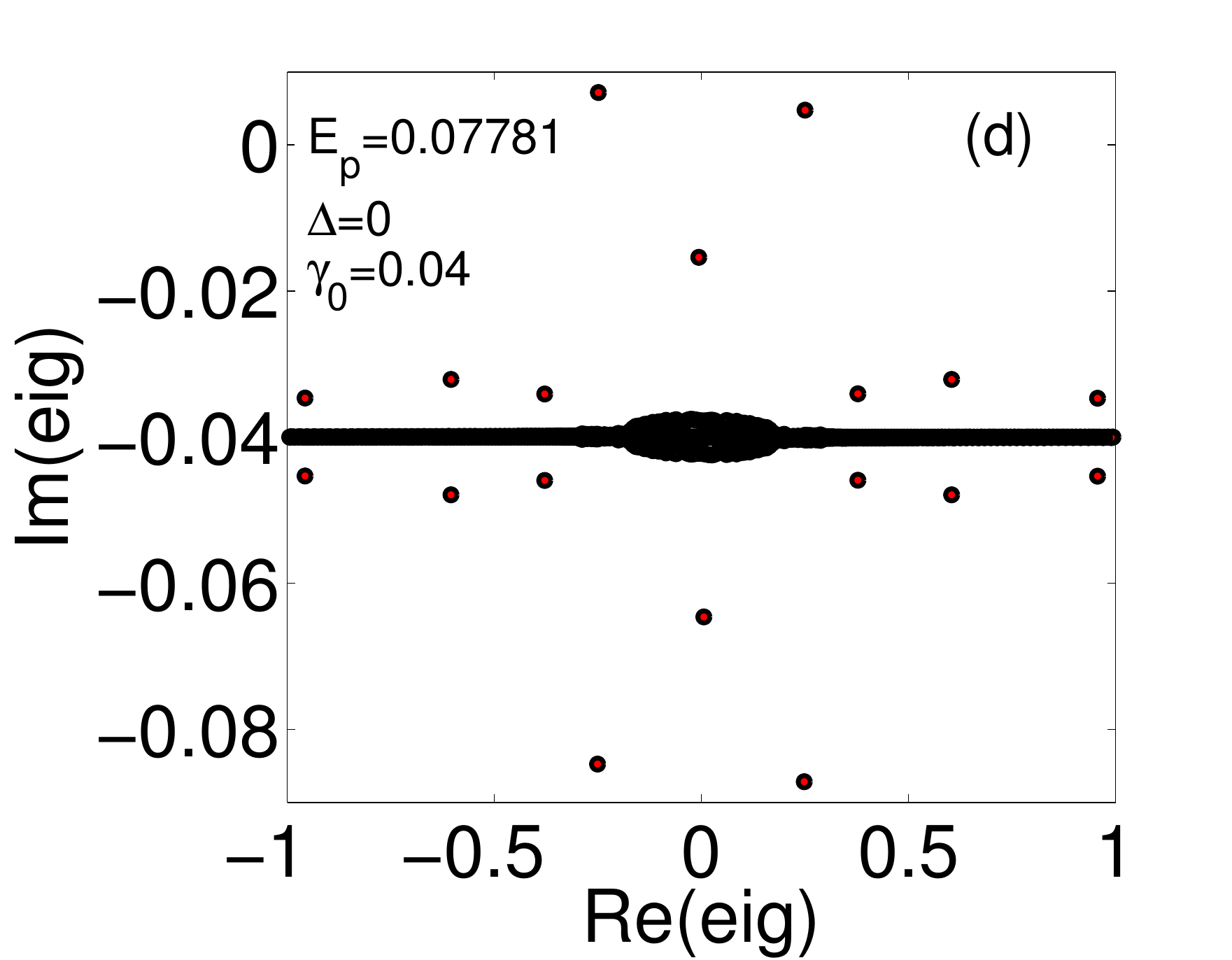}
}
\resizebox{.9\textwidth}{!}{%
\includegraphics[height=5cm]{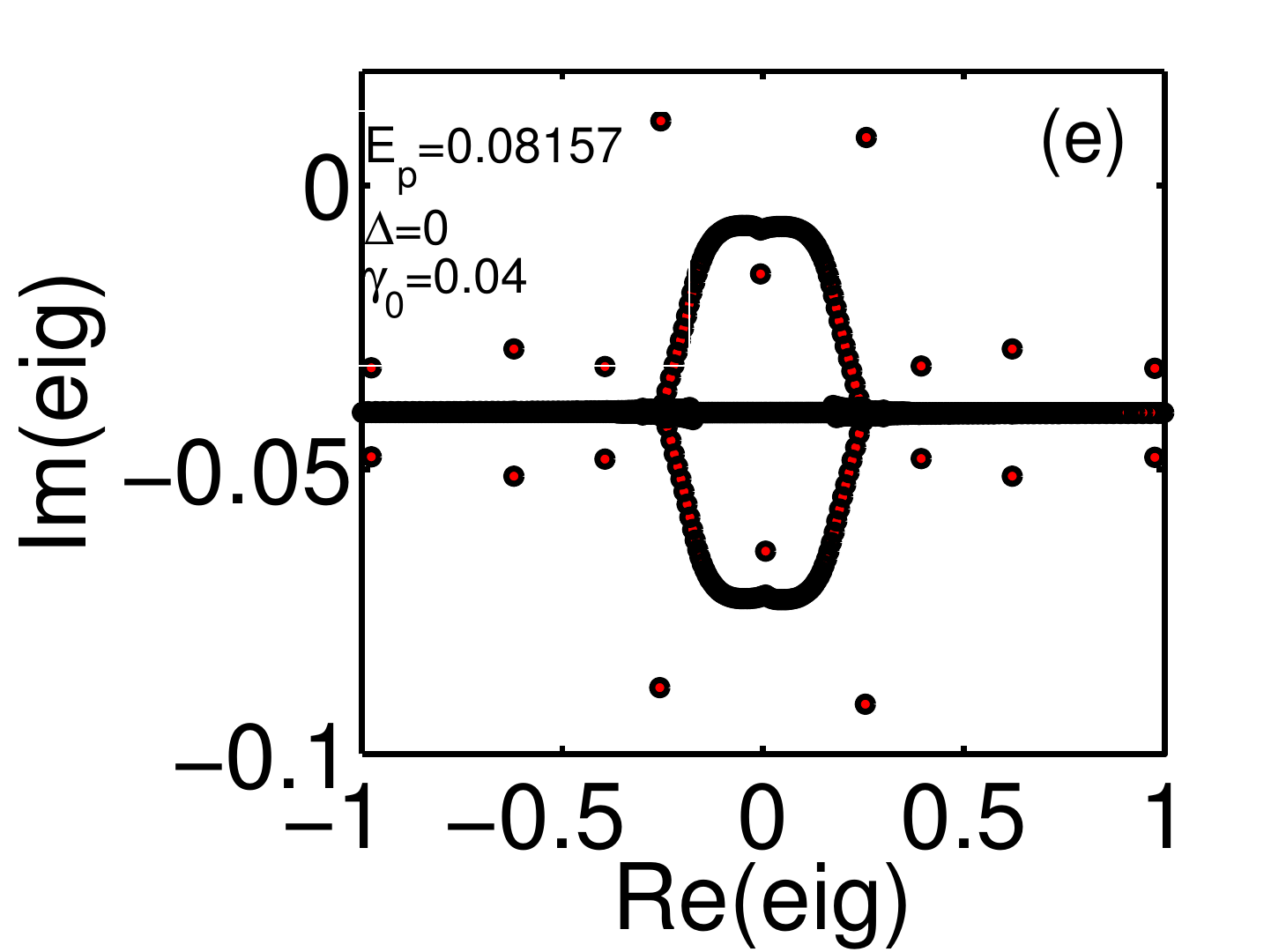}
\includegraphics[height=5cm]{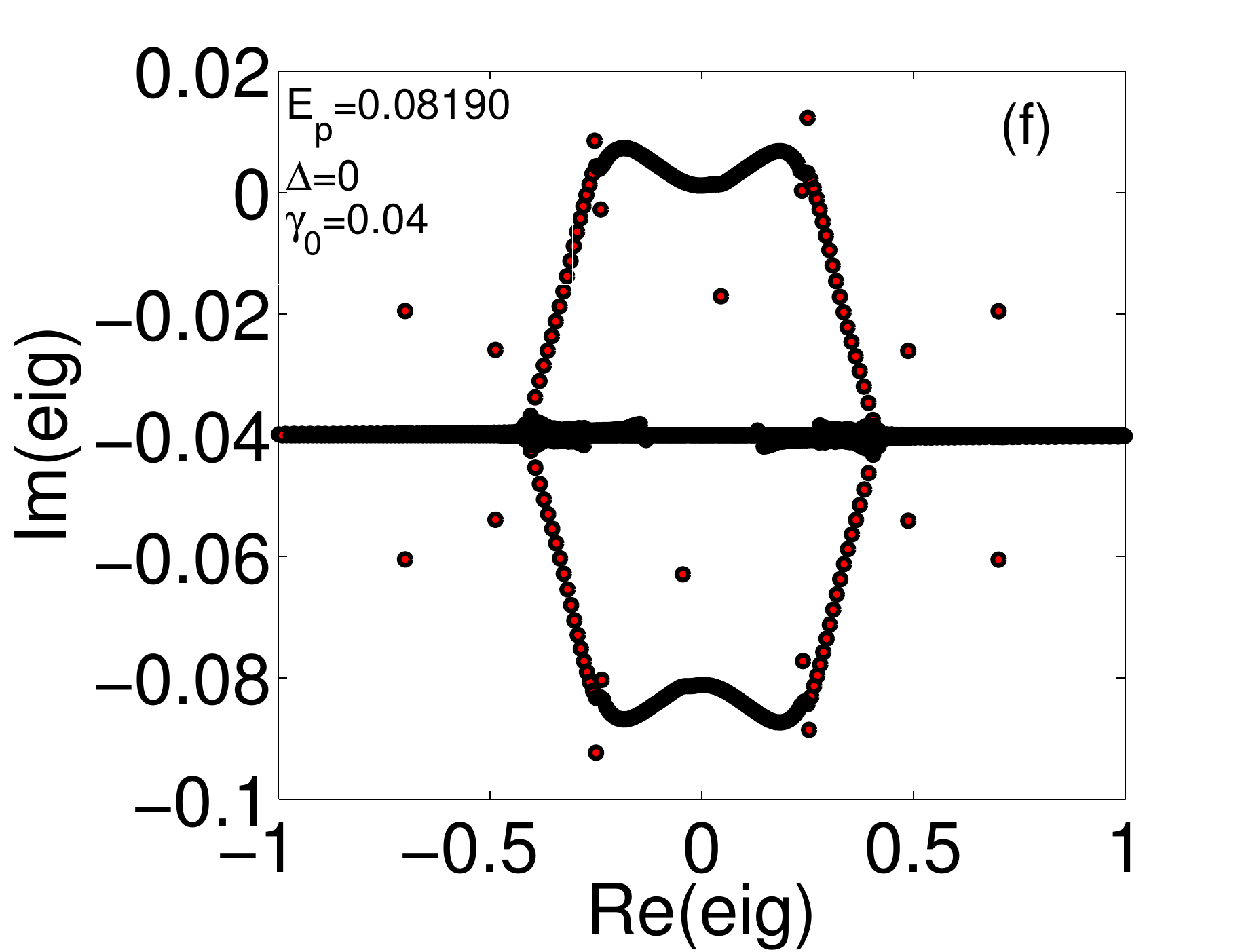}
}
\resizebox{.9\textwidth}{!}{%
\includegraphics[height=5cm]{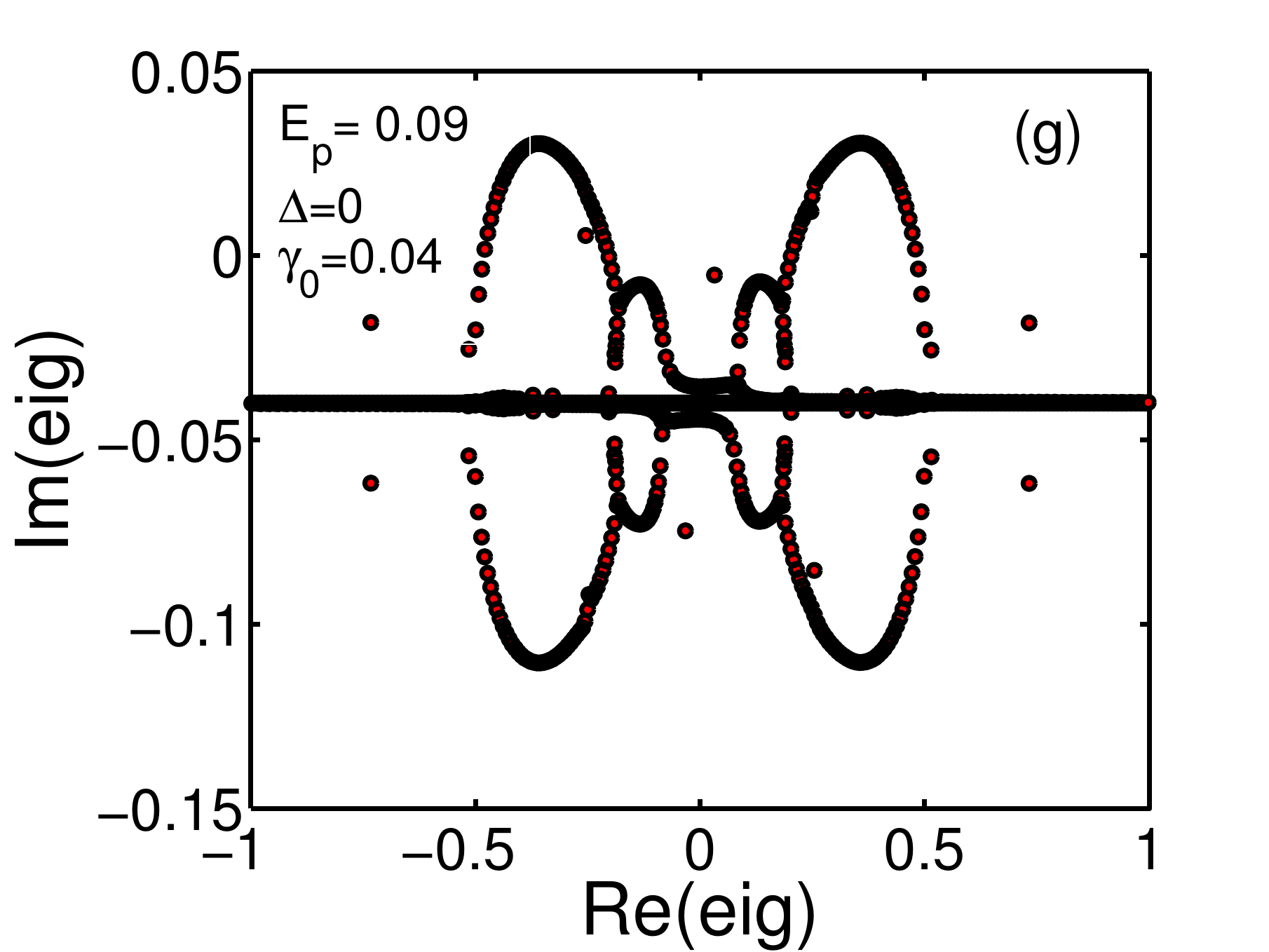}
\includegraphics[height=5cm]{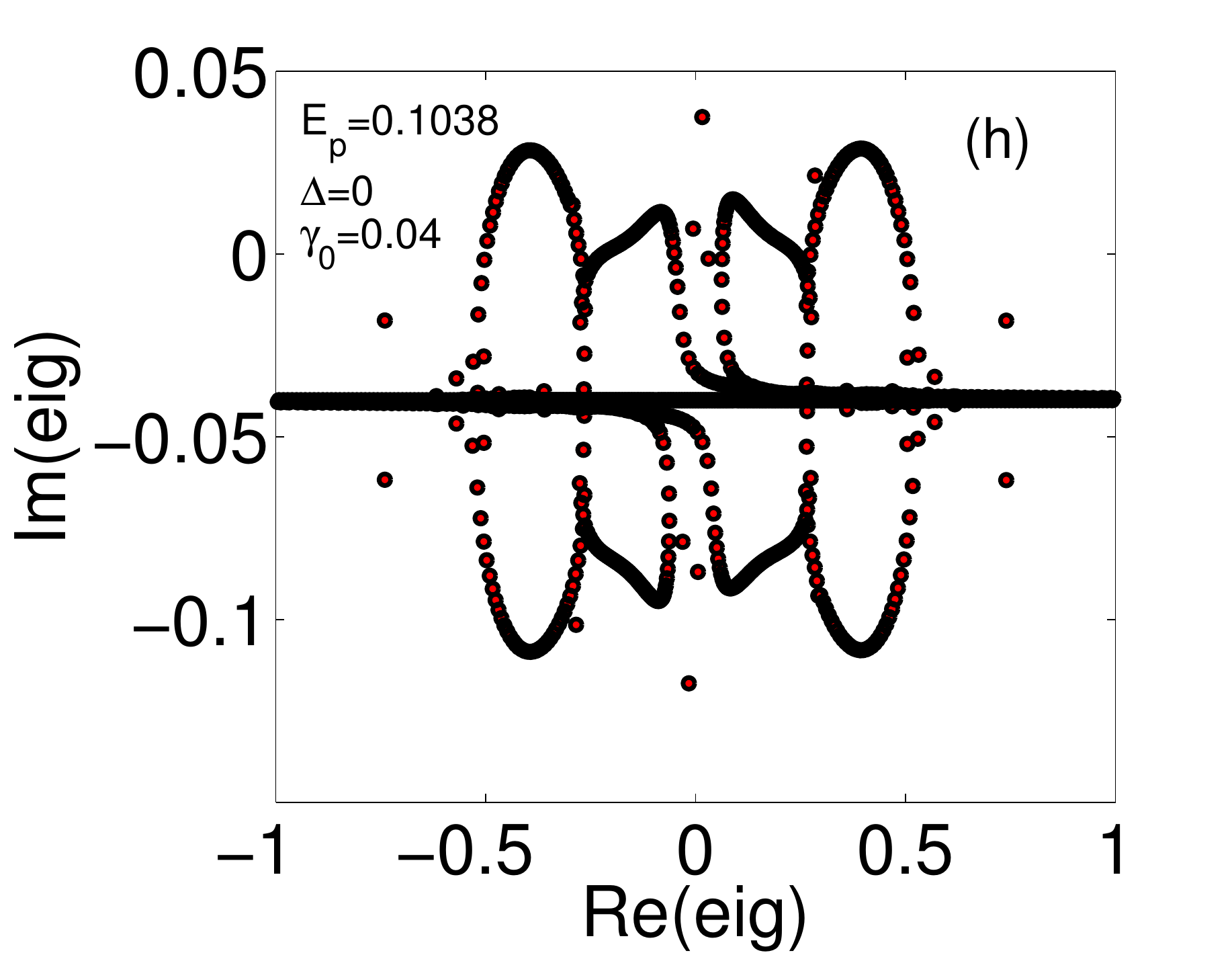}
}
\caption{Full eigenvalue spectrum of the soliton branch (Fig. \ref{Fig:Stability_analysis_homogeneous_type1_solitons_Delta_0} (a)) $\Delta=0, \gamma_0=0.04$: (a) at $E_p=0.068865$;(b) at $E_p=0.0712$: the solitons are stable for these pump amplitudes ($\lambda=Im(eig) < 0$); Unstable solitons (at least one $\lambda >0$): (c) at the right edge of type1 stable soliton branch at $E_p=0.073665$; (d) at $E_p=0.07781$ (left edge of unstable soliton branch); (e) at $E_p=0.08157$: right edge before kink; (f) at $E_p=0.08190$ after kink; (g) $E_p=0.09$ middle of unstable soliton branch; (h) at $E_p=0.1038$: right edge of unstable soliton branch.}
\label{Fig:Full_spectrum_soliton_branches_Delta_0_gamma_0p04}
\end{figure}

The evolution of the full eigenvalue spectrum with the pump amplitude, $E_p$, computed from the sparse matrix in Eq.~(\ref{eq:soliton_eigenvalue_problem}) for the soliton branch at $\Delta=0, \gamma_0=0.04$ is shown in Fig. \ref{Fig:Full_spectrum_soliton_branches_Delta_0_gamma_0p04}. The solitons are stable, as all eigenvalues have negative imaginary part ($\lambda < 0$) up to a pump amplitude of $E_p=0.073665$, corresponding to the right edge of the stable type 1 soliton branch (see Fig. \ref{Fig:Stability_analysis_homogeneous_type1_solitons_Delta_0}) (a-c). Above this pump amplitude the soliton branch type 2 is unstable (since eigenvalue imaginary part, $\lambda>0$ for at least one eigenvalue).

\begin{figure}
\resizebox{.9\textwidth}{!}{%
\includegraphics[height=5cm]{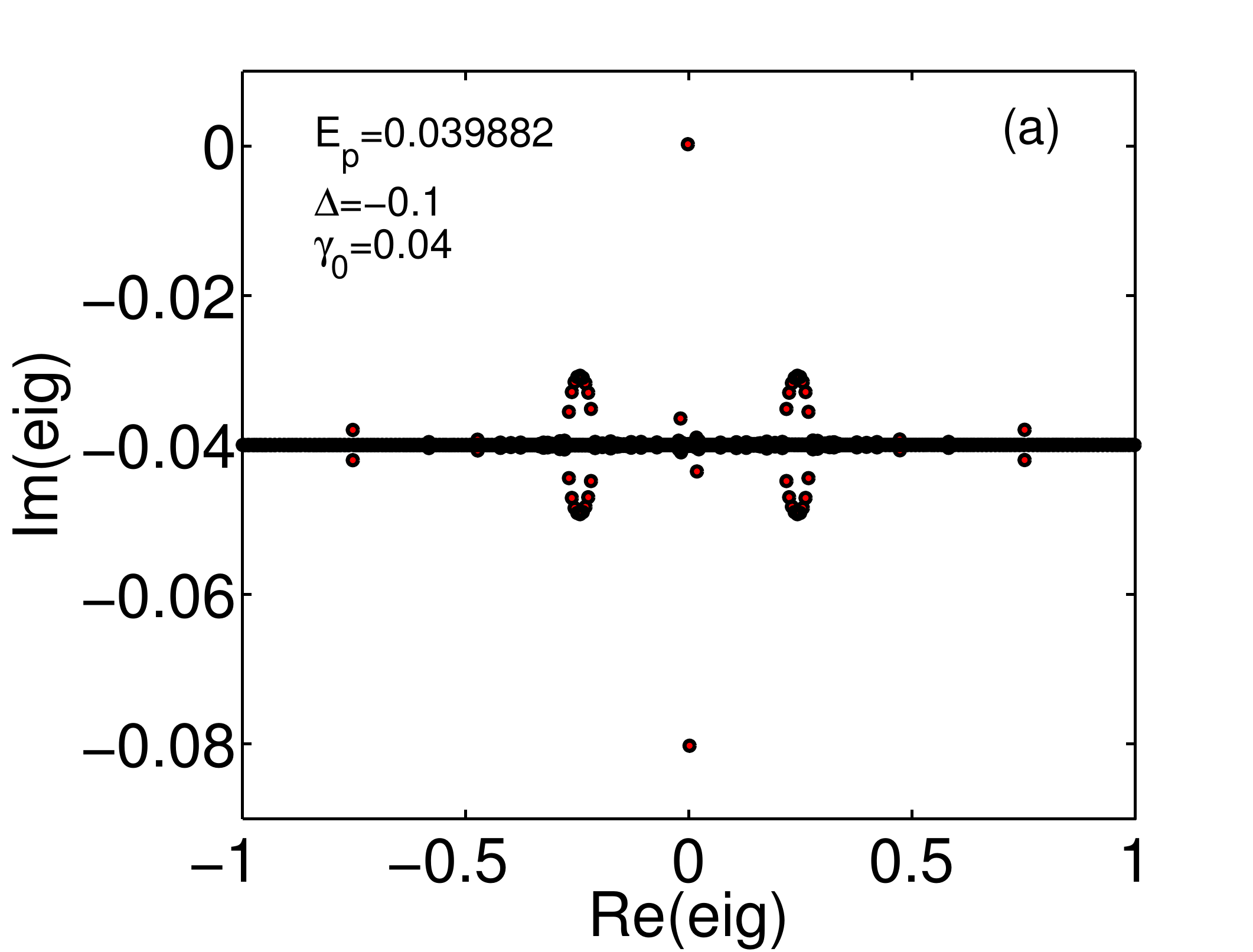}
\includegraphics[height=5cm]{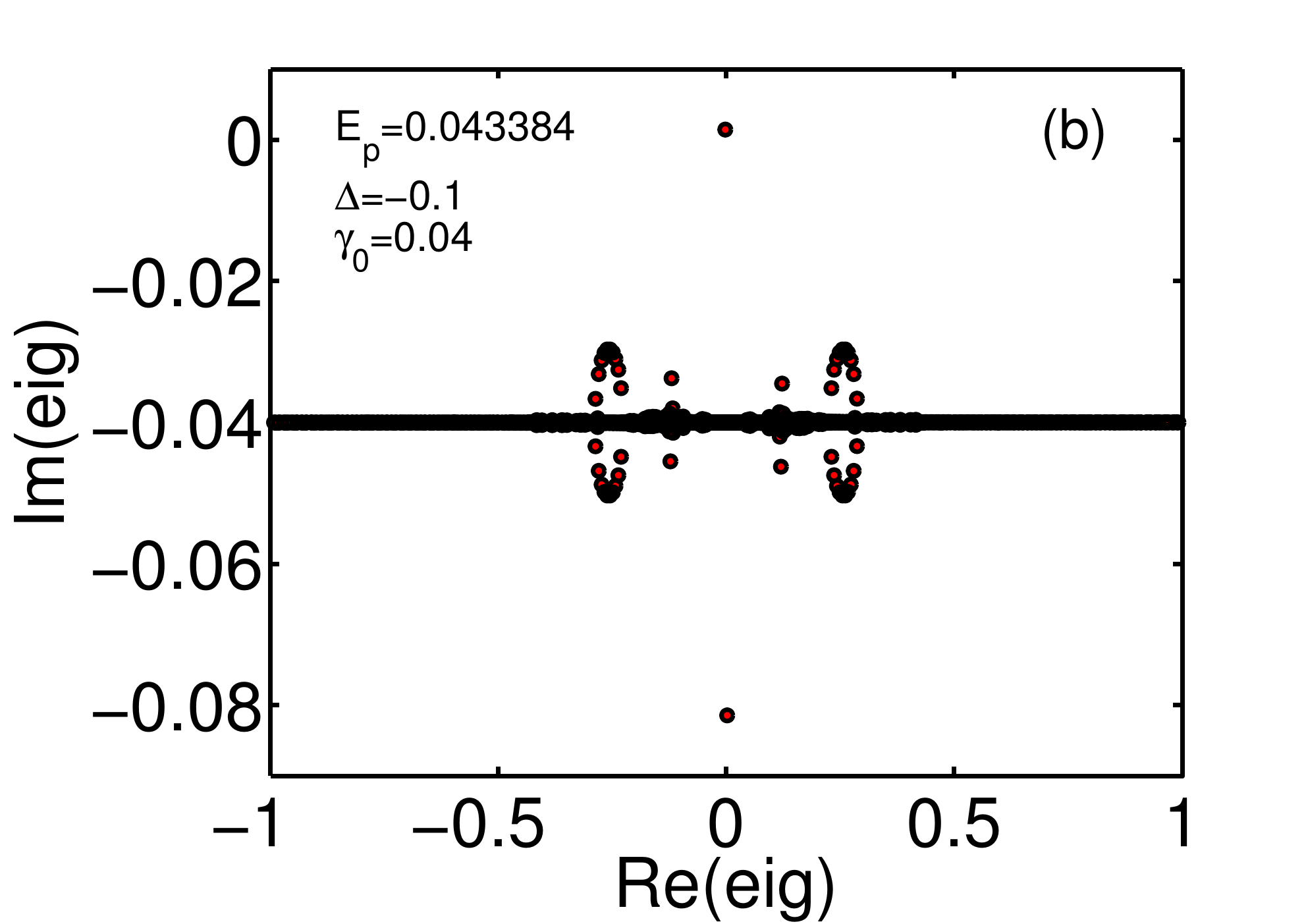}
}
\resizebox{.9\textwidth}{!}{%
\includegraphics[height=5cm]{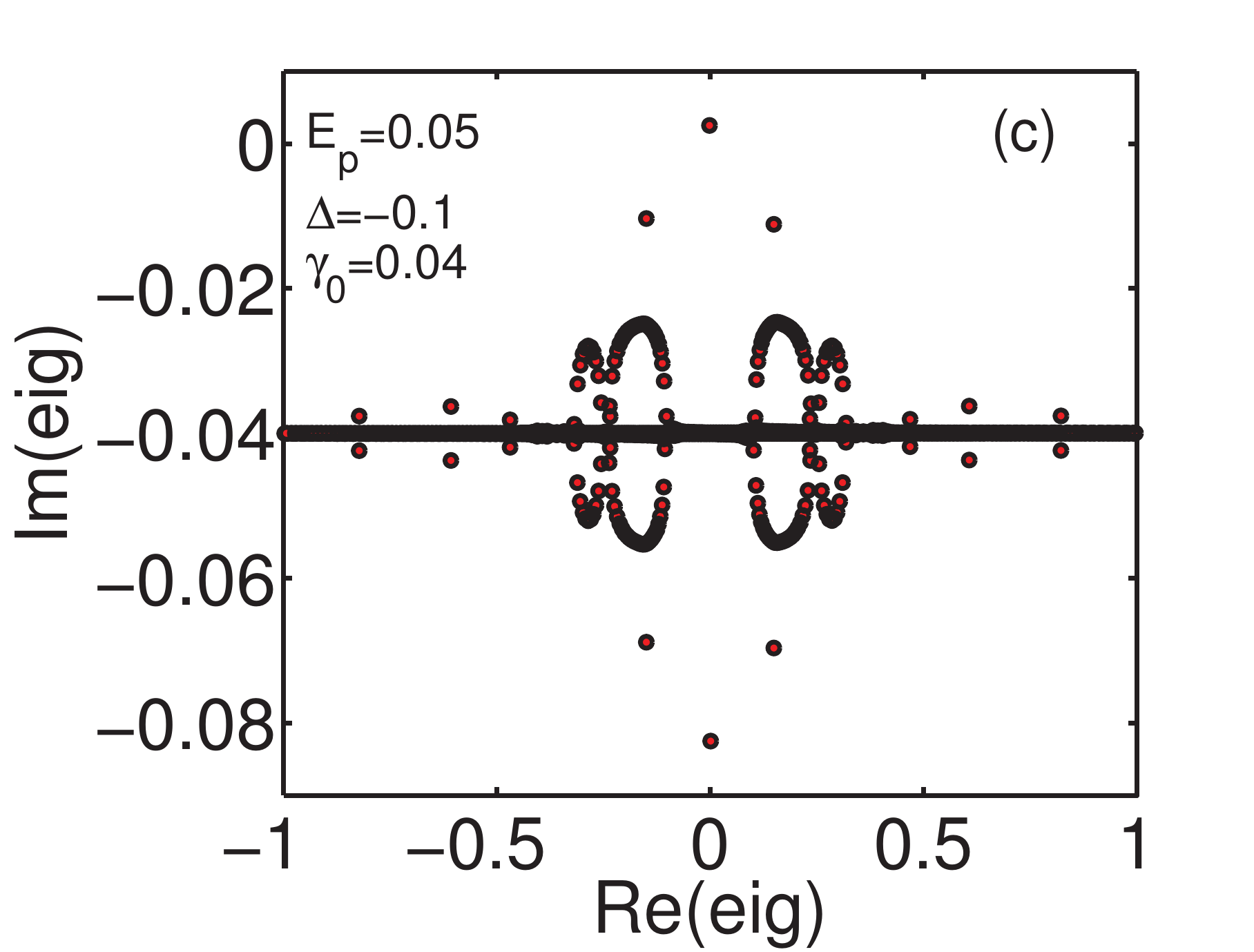}
\includegraphics[height=5cm]{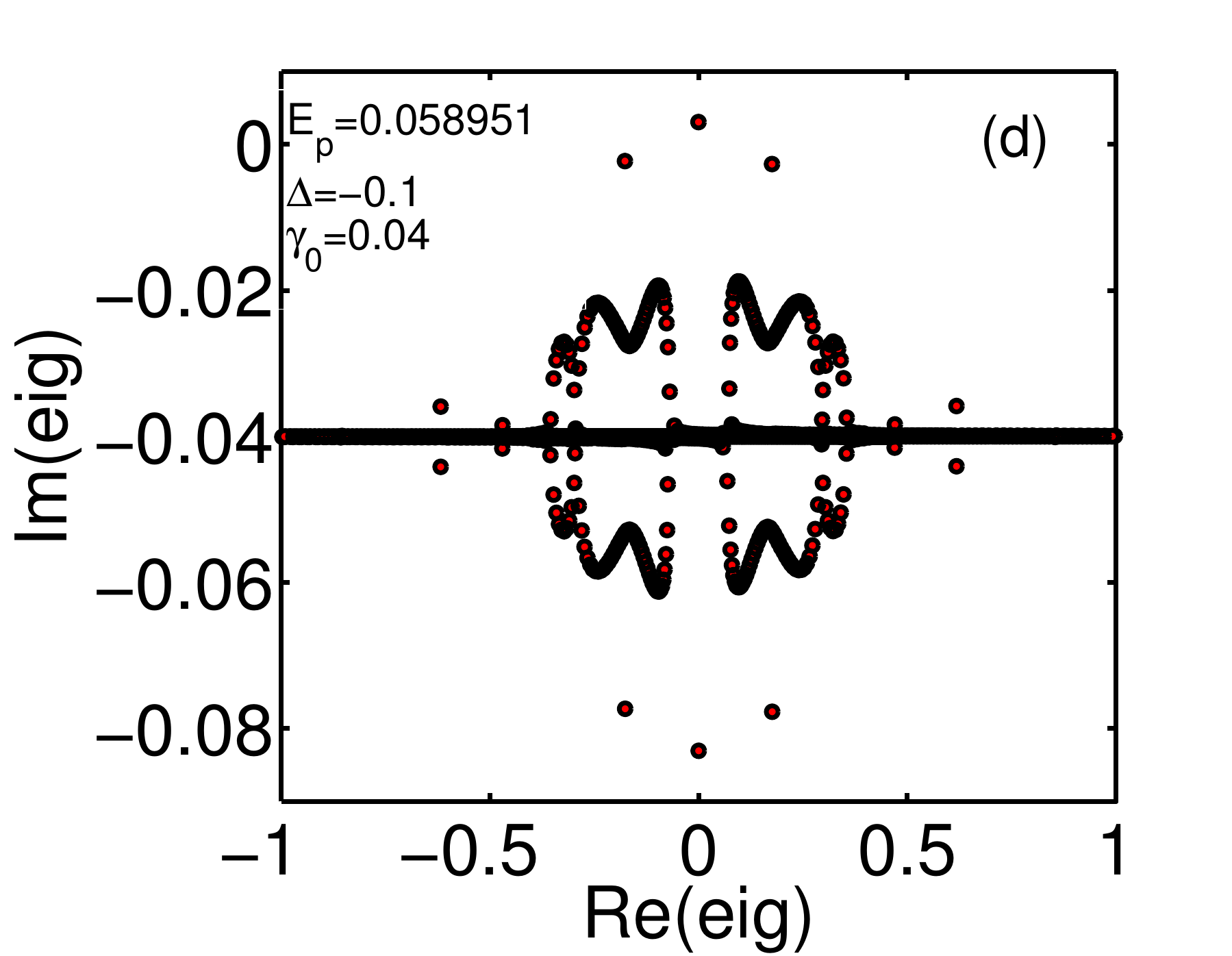}
\includegraphics[height=5cm]{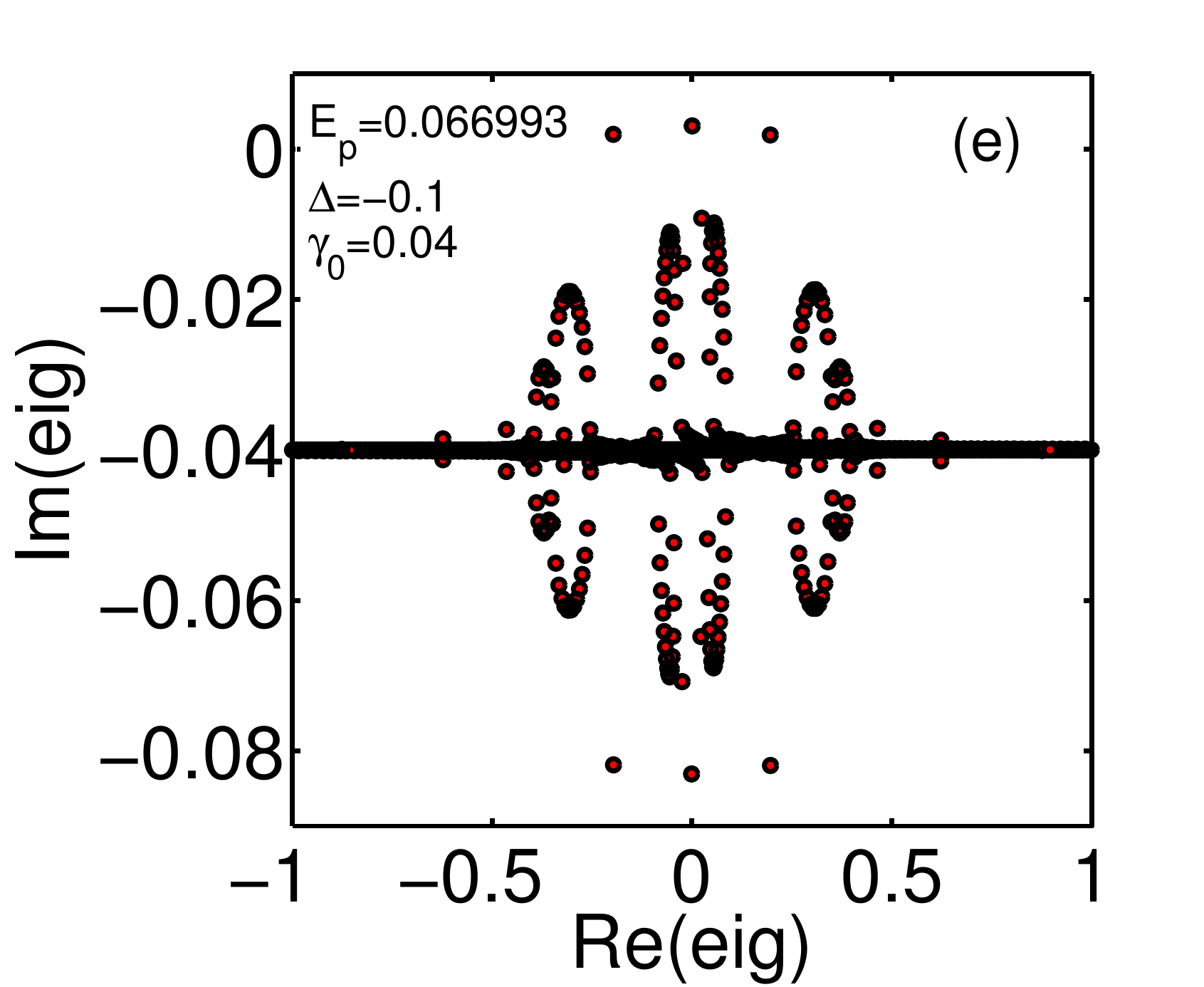}
}
\caption{Full eigenvalue spectrum for the soliton branch (Fig. \ref{Fig:Stability_analysis_homogeneous_type1_solitons_Delta_-0.1}(a)) at $\Delta=-0.1, \gamma_0=0.04$ at the: (a) left edge  $E_p=0.039882$; (b) right edge of the homogeneous stable solution background $E_p=0.043384$; (c) at $E_p=0.05$; (d) right edge of modulationally unstable background region $E_p=0.058951$; (e) right edge of soliton branch $E_p=0.066993$; the solitons are slightly unstable for all pump amplitudes ($\lambda \approx 0 > 0$).}
\label{Fig:Full_spectrum_soliton_branches_Delta_m0p1_gamma_0p04}
\end{figure}

The full eigenvalue spectrum computed from the sparse matrix in Eq.~(\ref{eq:soliton_eigenvalue_problem}) for the soliton branch at $\Delta=-0.1, \gamma_0=0.04$, for a range of pump amplitudes, starting from the left edge of the soliton branch (at $E_p=0.039882$) and sweeping the branch up to $E_p=0.066993$ is shown in Fig. \ref{Fig:Full_spectrum_soliton_branches_Delta_m0p1_gamma_0p04}. The solitons are slightly unstable towards the left edge of the soliton branch, since at least one of the eigenvalues is positive, although very close to zero, and remain slightly unstable towards the right edge.

\section{Comparison with the full model}
\label{sec4}
\begin{figure}
\includegraphics[width=8cm]{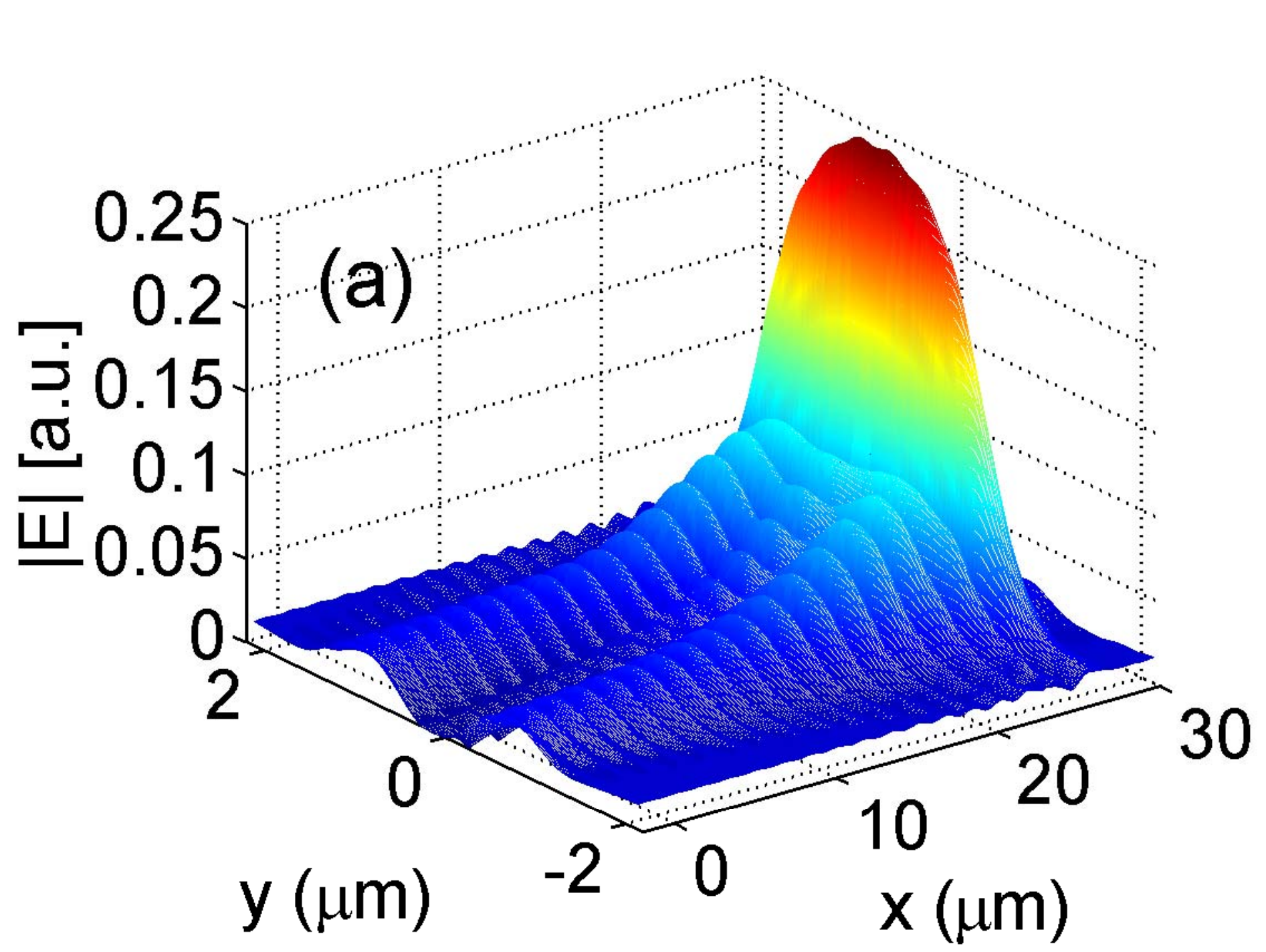}
\includegraphics[width=8cm]{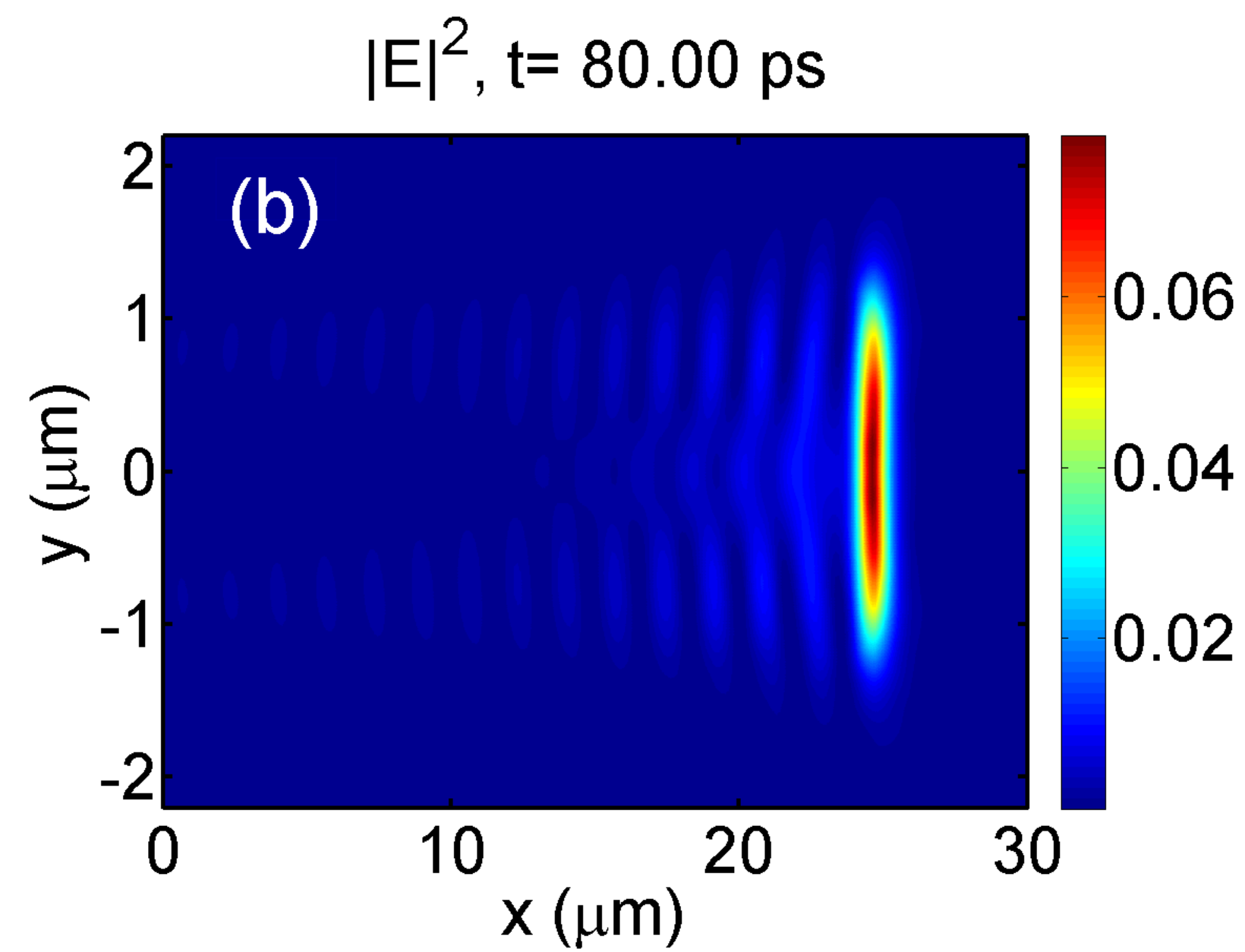}
\includegraphics[width=8cm]{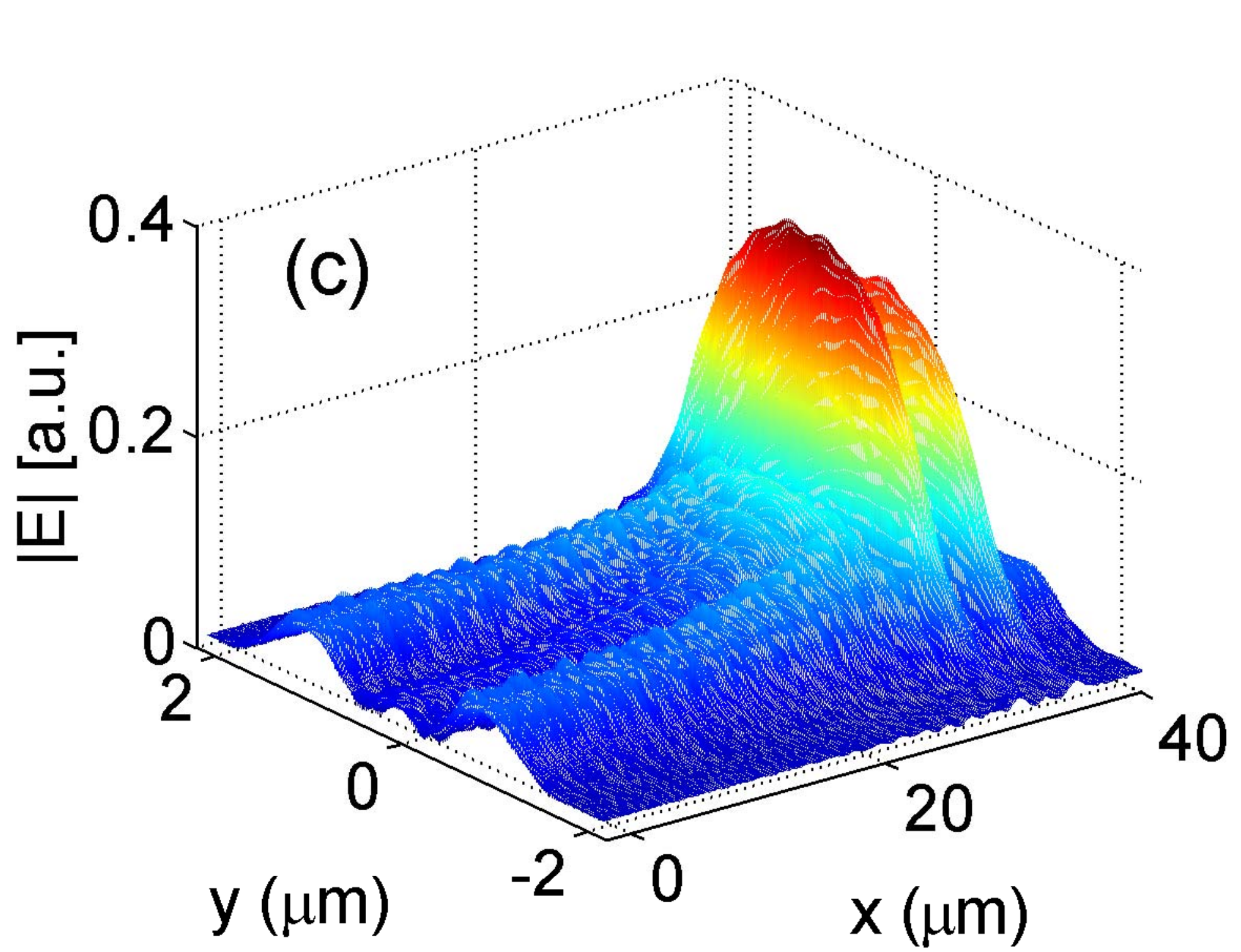}
\includegraphics[width=8cm]{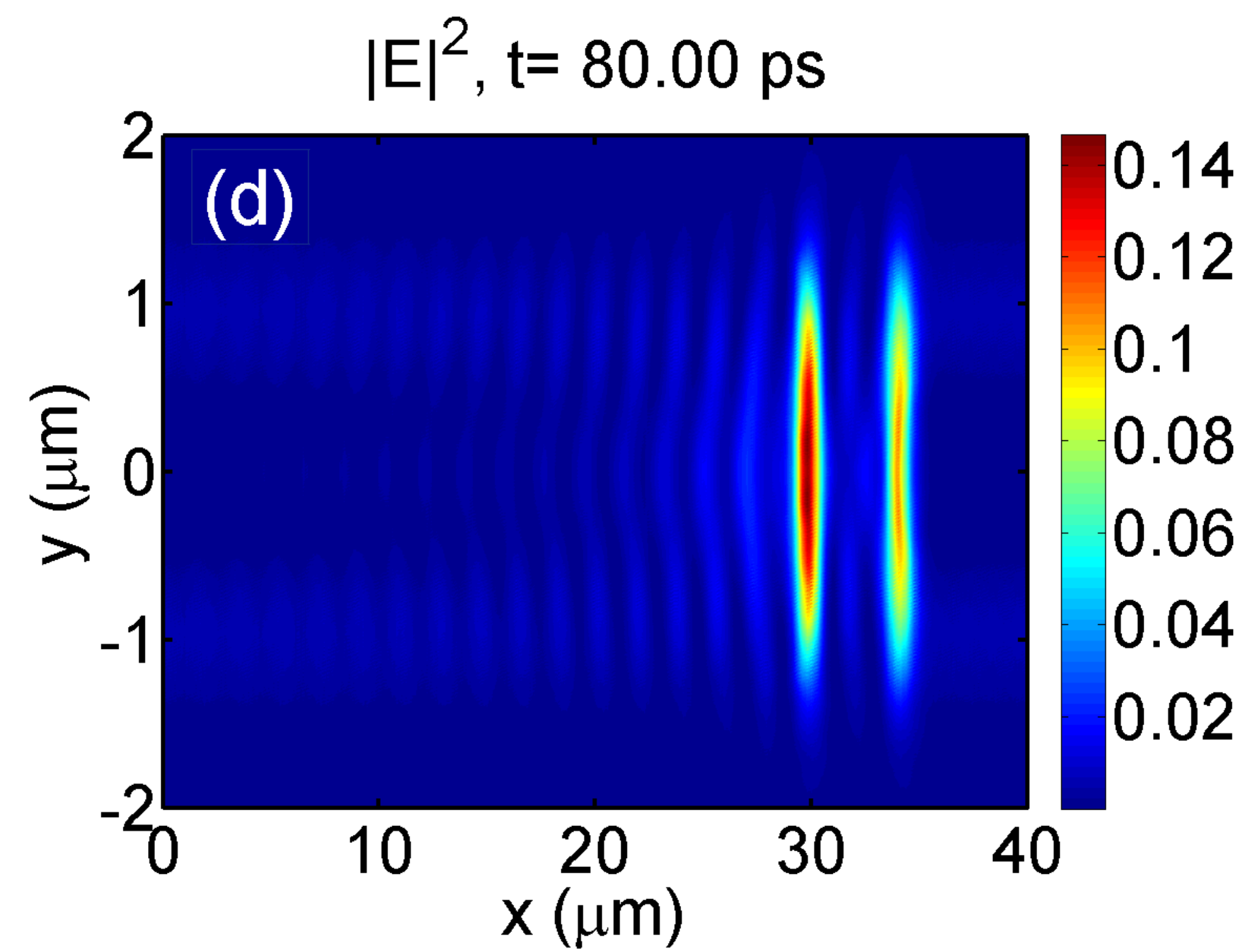}
\caption{Snapshot at $t=80 \mathrm{ps}$ of a (a-b) single-hump soliton for$E_p=0.0672$,$Es=0.34114$; (c-d) double-hump soliton for$E_p=0.0762$,$Es=0.34114$ at $\Delta=0$ }
\label{Fig:3D_solitons_full_model}
\end{figure}

In this section we compare our reduced 1D model with the full 2D model \cite{our_OL}. We solve the time-dependent full-model Eqs.~(\ref{eq:eqE}), (\ref{eq:eqPsi}) by Fourier split-step technique, taking as initial guess the lower bistability branch homogenous solution for pump amplitude  $E_p=0.0672$, applying a seed pulse with amplitude $E_s=0.34114$ and sweeping the whole soliton branch. The resulting evolved 3D soliton profiles are displayed in Fig. \ref{Fig:3D_solitons_full_model} at the left (and $E_p=0.0672$) and right ($E_p=0.0762$) edge of the upper and lower soliton branches.
We should note that our dynamical simulation predicts two types of solitons as the pump amplitude is increased, namely a single-hump soliton, shown in Fig. \ref{Fig:3D_solitons_full_model}(a,b) at $E_p=0.0672$, persisting up to a pump amplitude of $E_p=0.0692$, at which point the soliton peak splits up. The splitting between the two soliton peaks becomes larger and larger with increasing the pump amplitude, eventually resulting in a well defined double-hump soliton (Fig. \ref{Fig:3D_solitons_full_model} (c,d)).

\begin{figure}
\includegraphics[width=8cm]{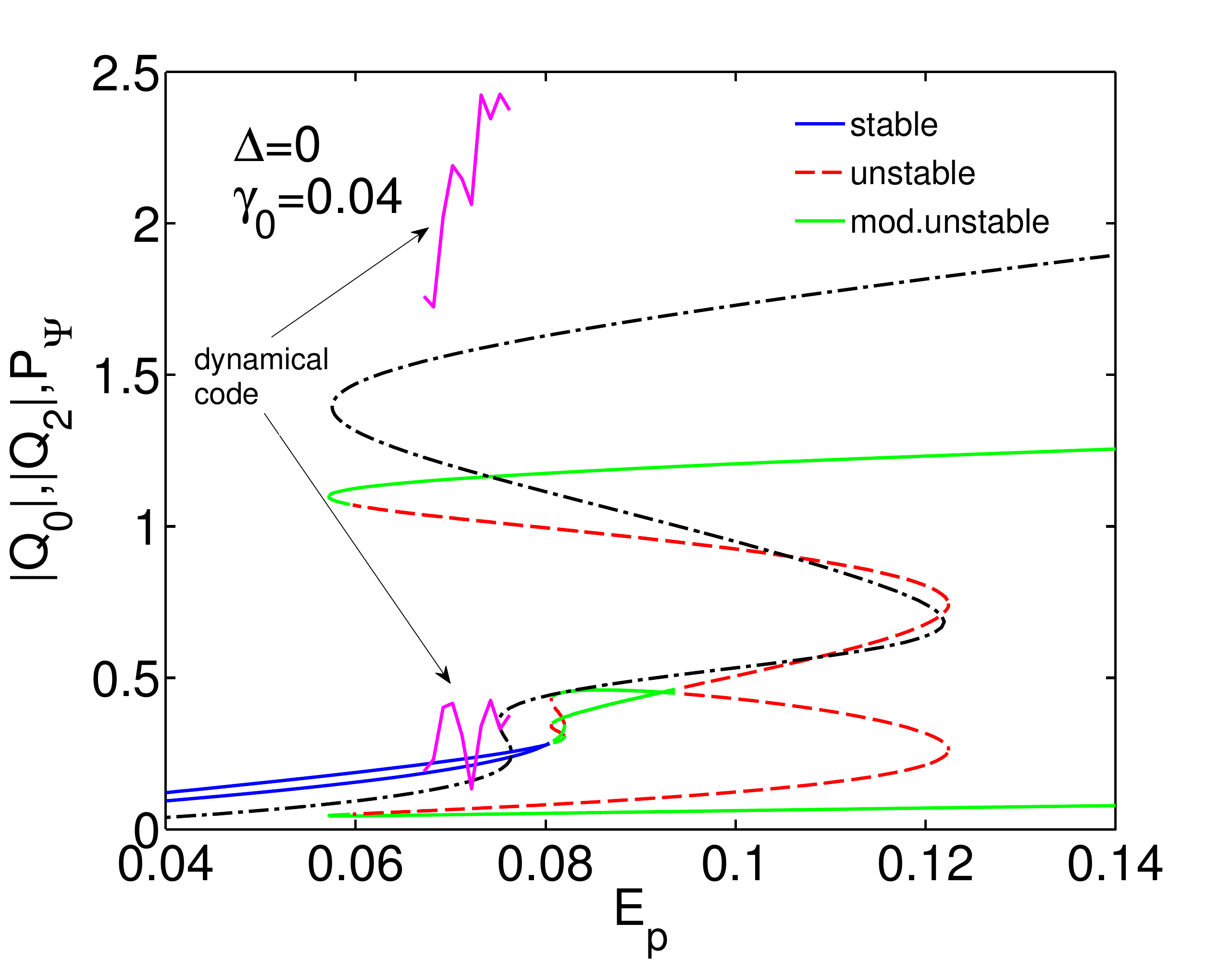}
\caption{Soliton branch inferred from dynamical computation of Eqs.~(\ref{eq:eqE}), (\ref{eq:eqPsi}) superimposed on the coupled multistability curves of the reduced model and the full model (black dash-dotted curve) at $\Delta=0, \gamma_0=0.04$: single-hump solitons persist up to the first maximum, above which double-hump solitons are formed.}
\label{Fig:soliton_branch_dynamical_code}
\end{figure}

The single/double-hump soliton branches are obtained from the time-dependent solution from the maximum through the soliton core and a slice through the soliton tail of the integrated power, $P_\Psi   = \int {\int {\left| {\Psi \left( {x,y} \right)} \right|^2 } } dxdy$, shown in Fig.\ref{Fig:soliton_branch_dynamical_code}. The transition between the single- and double-hump solitons is clearly seen from the stepwise soliton branches curves, shown in magenta. The single-hump soliton persists up to the first maximum in the soliton branch curve, above which a double-hump soliton forms.

\subsection{Reconstruction of the full model 2D soliton from stable type 1 soliton $Q_0$ and $Q_2$ profiles}
\label{sssec:4.1}
\begin{figure}
\includegraphics[width=8cm]{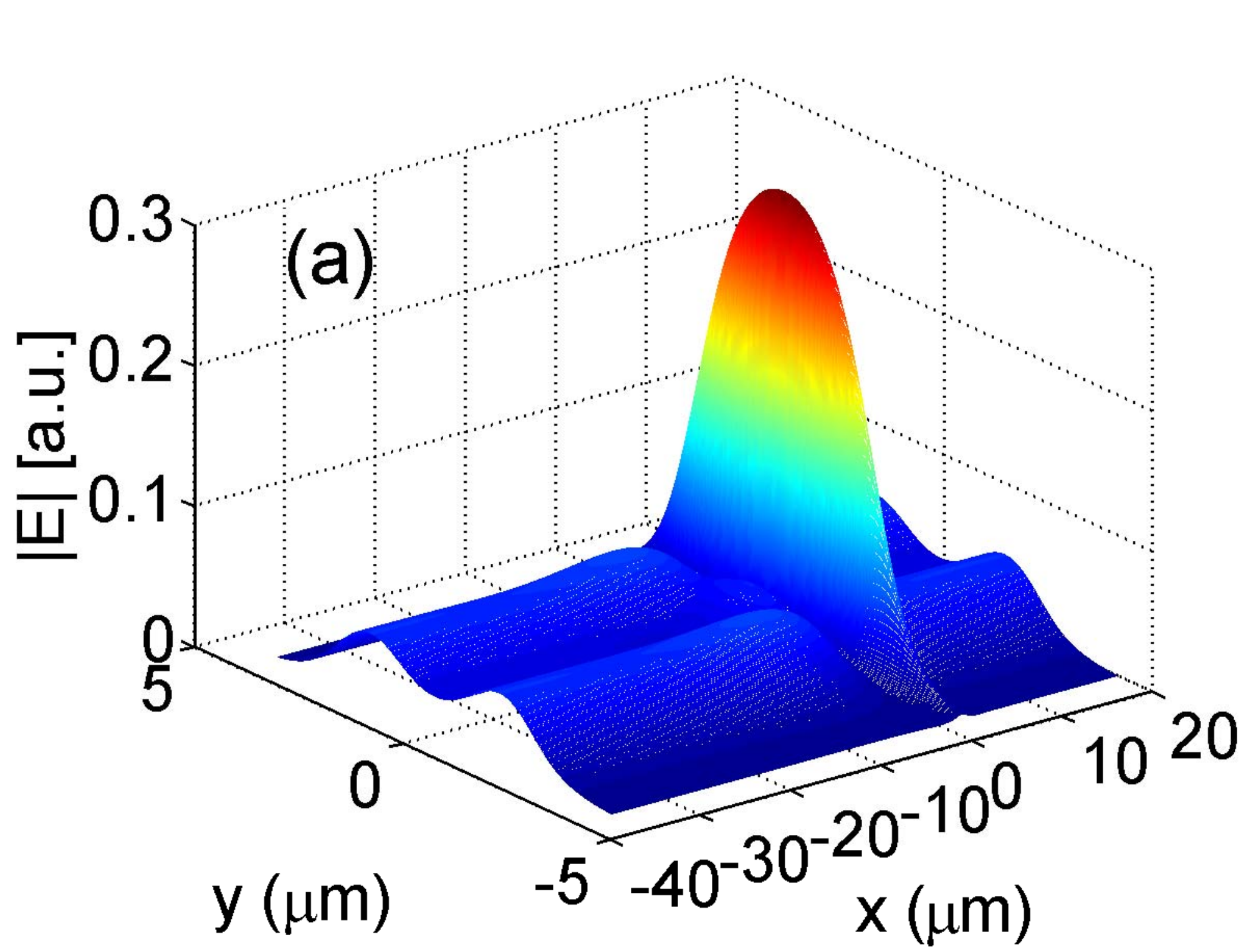}
\includegraphics[width=8cm]{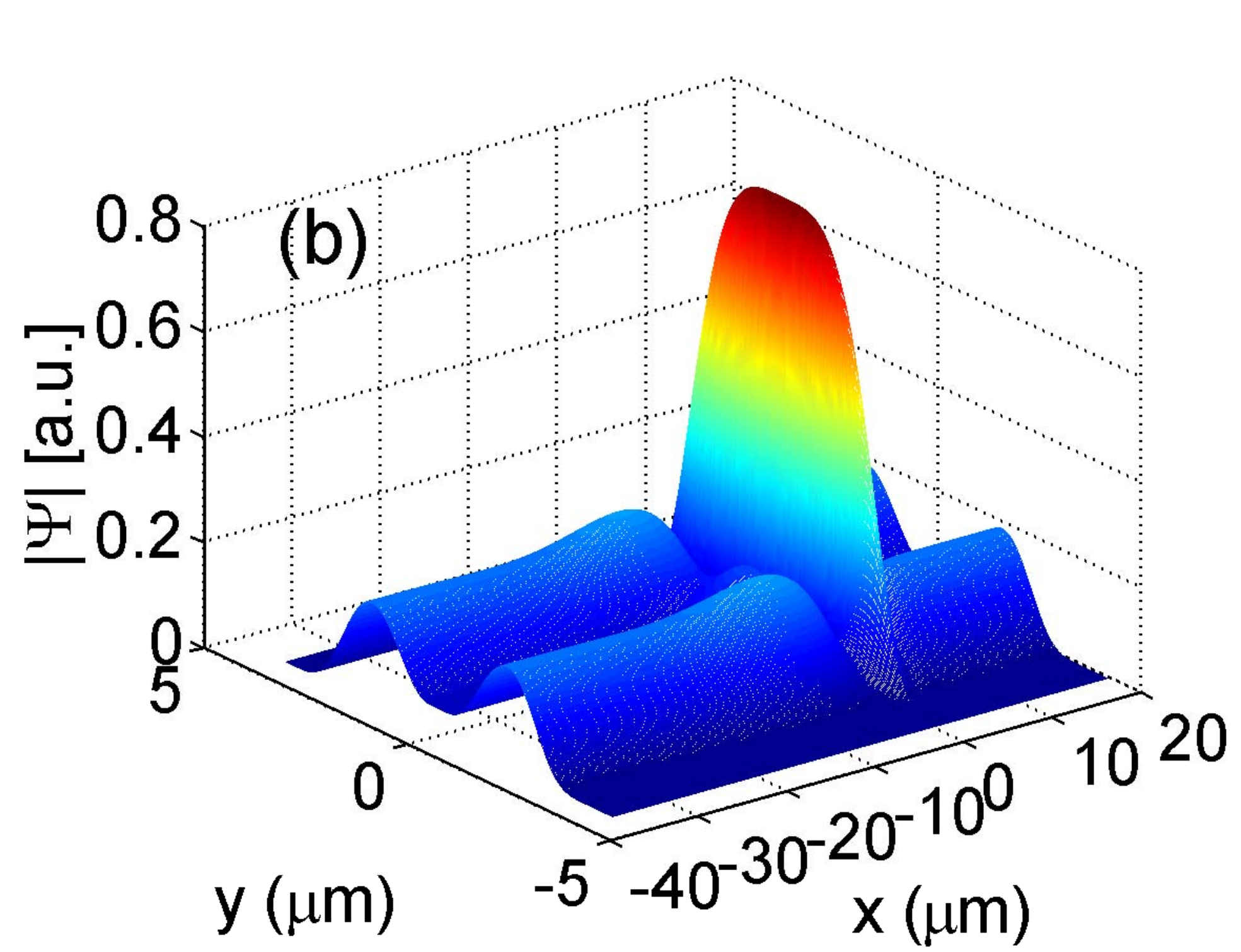}
\includegraphics[width=8cm]{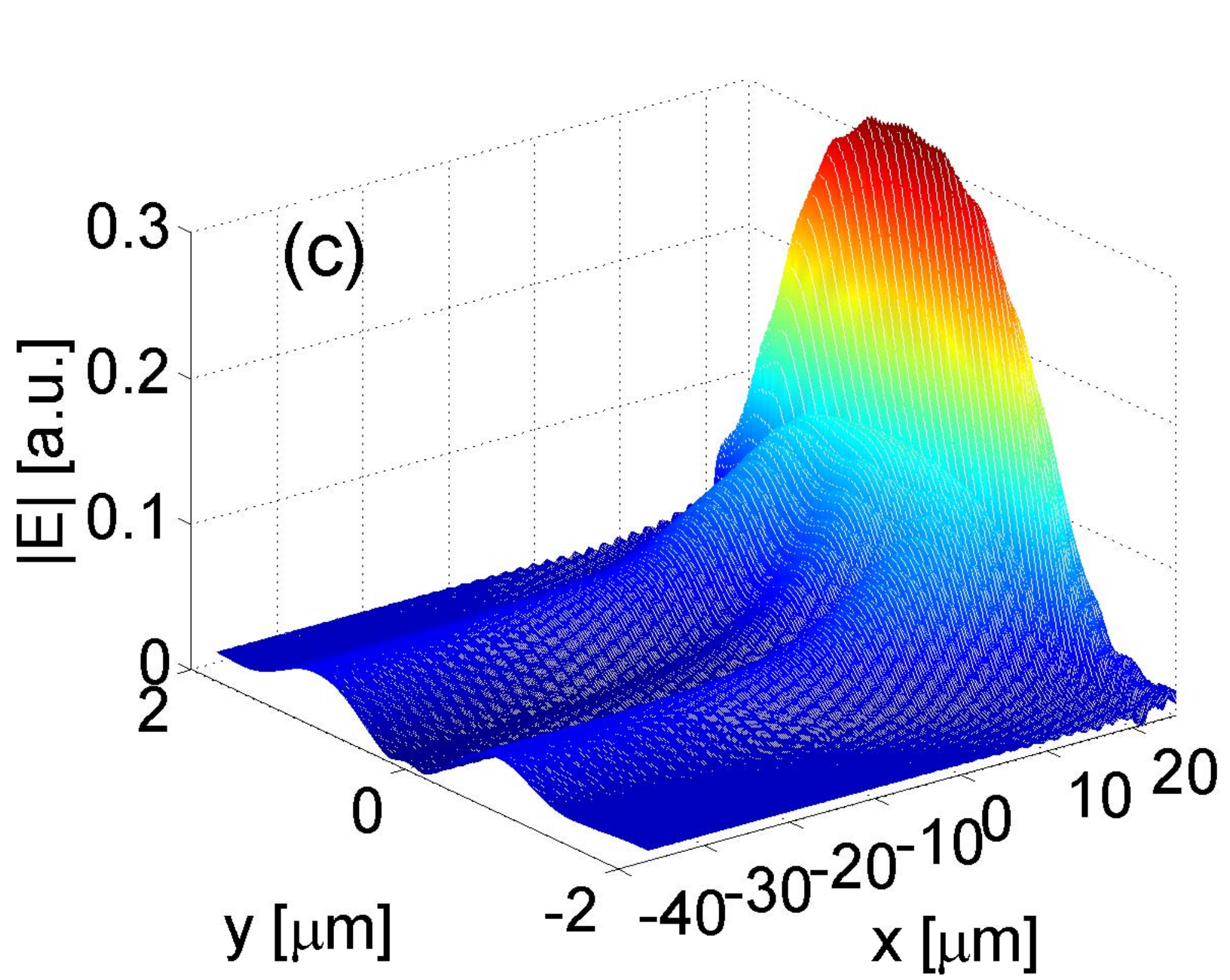}
\includegraphics[width=8cm]{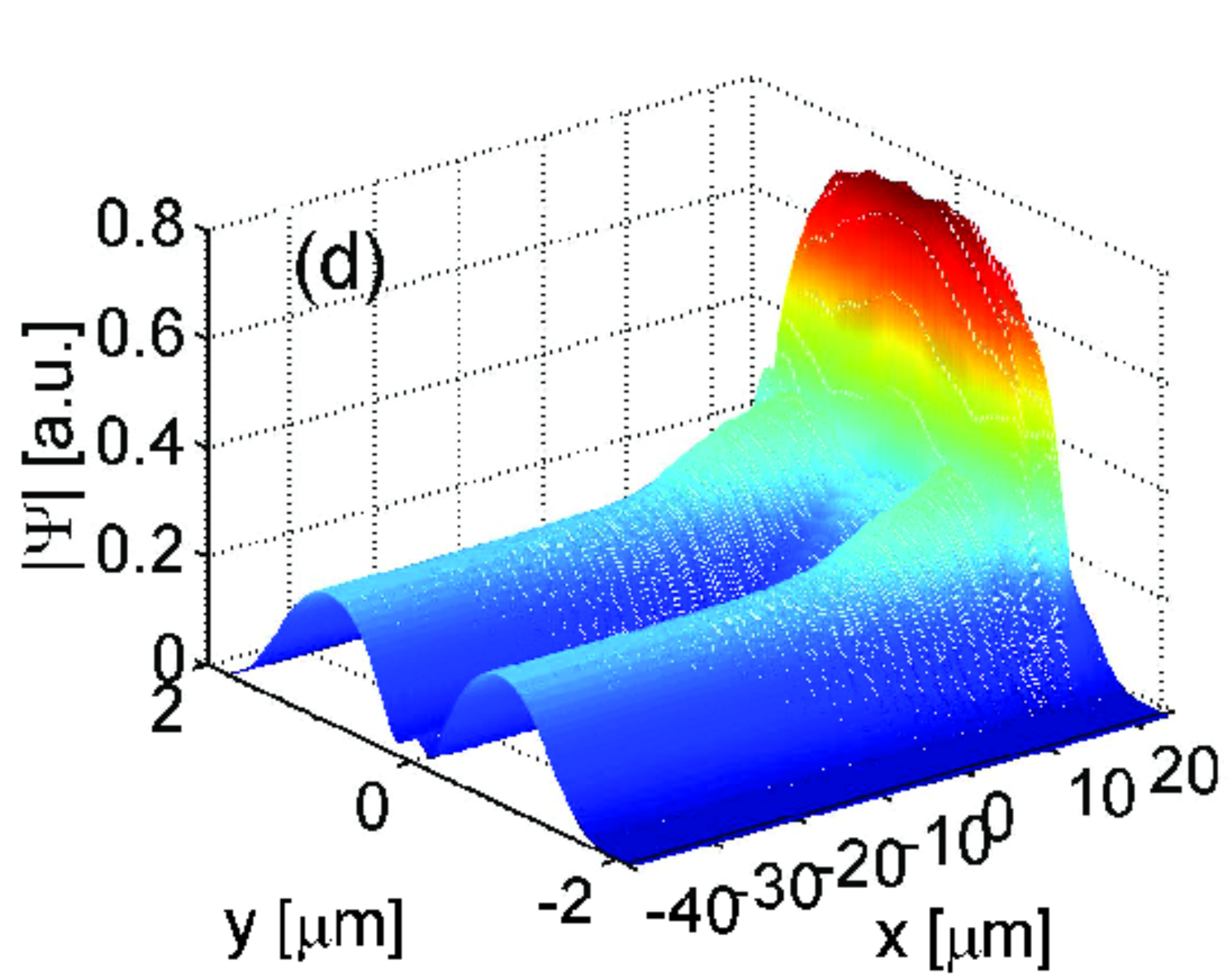}
\includegraphics[width=8cm]{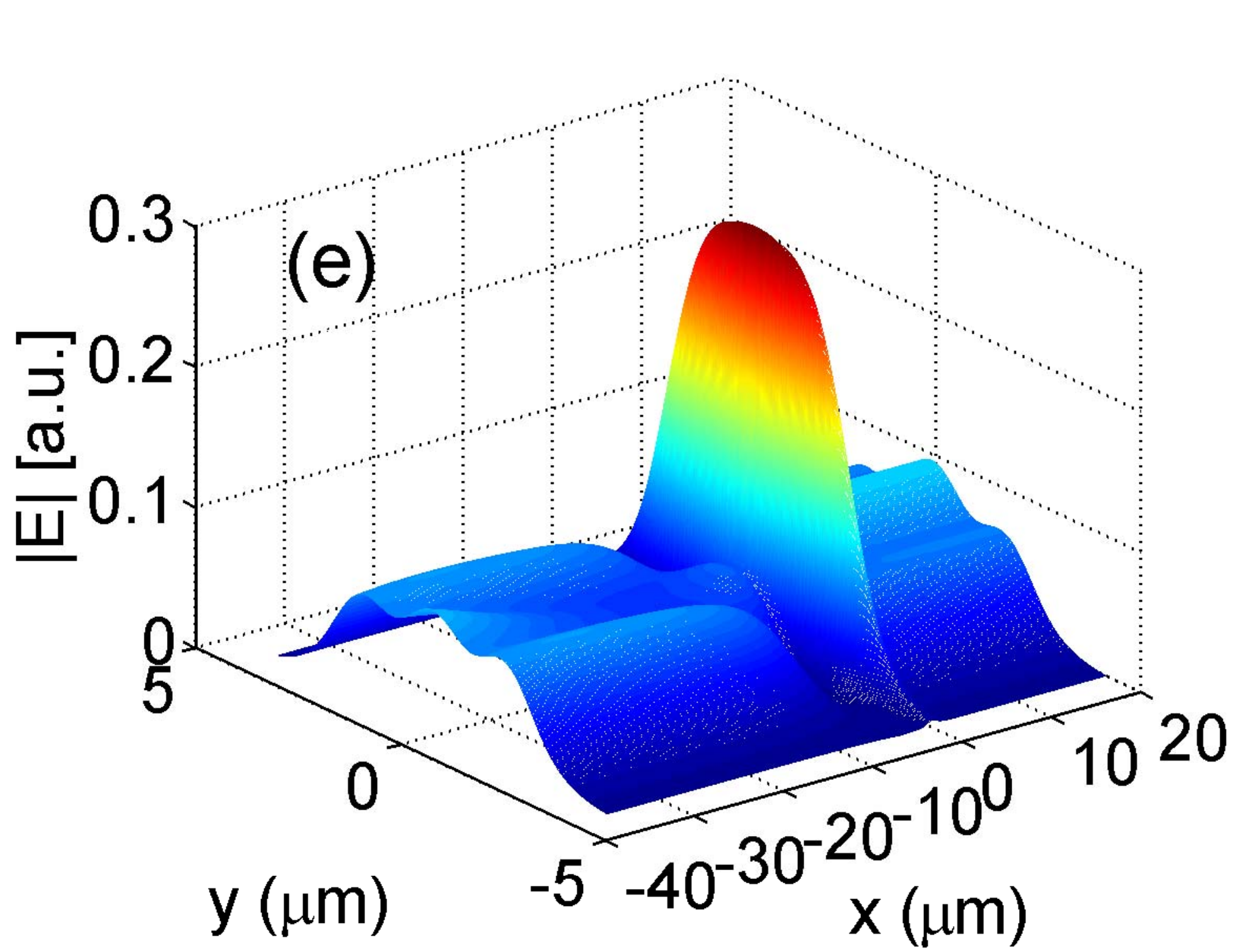}
\includegraphics[width=8cm]{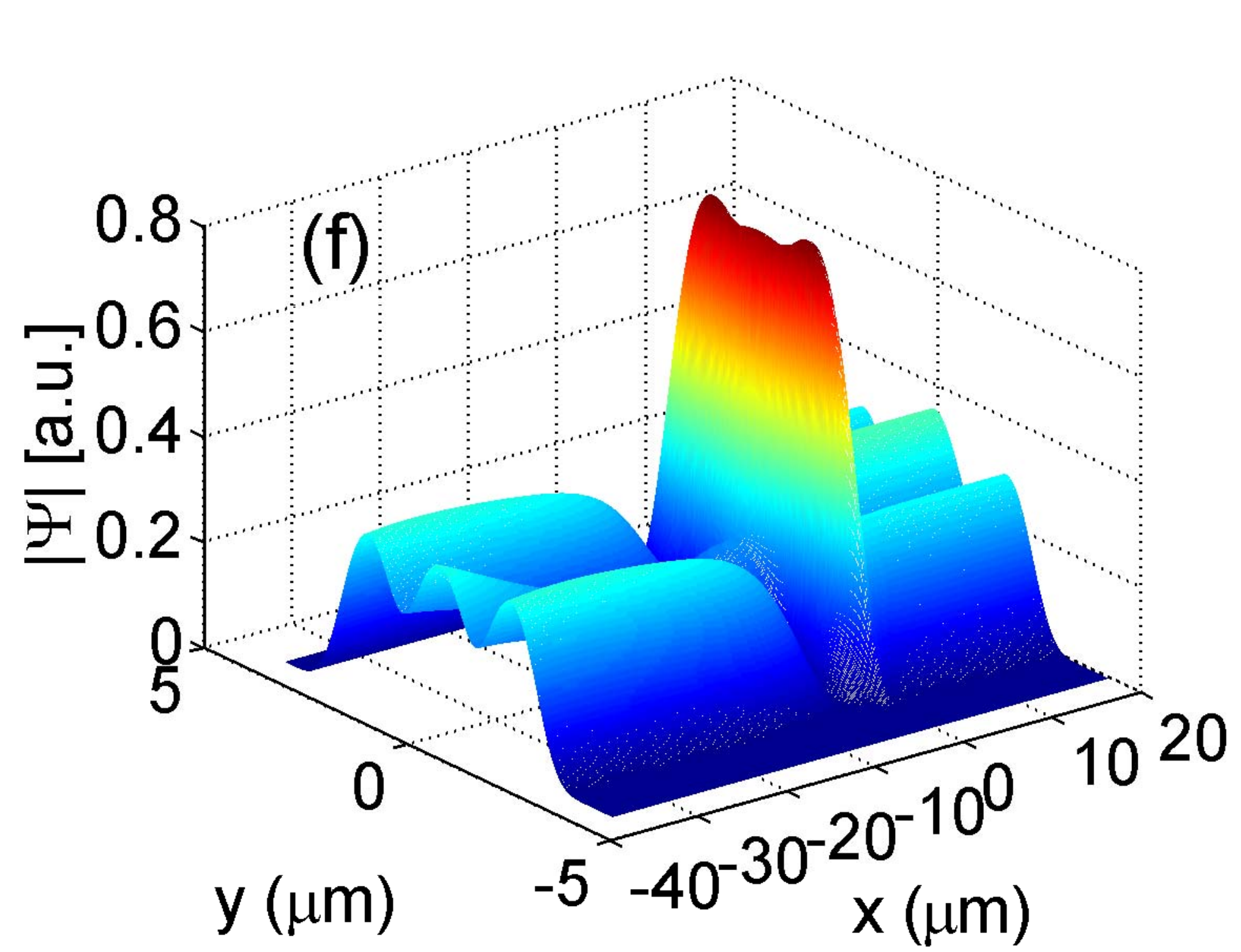}
\caption{3D plot of the reconstructed 2D soliton at $\Delta=0,\gamma_0=0.04$ (see Fig. \ref{Fig:Stability_analysis_homogeneous_type1_solitons_Delta_0}): (a) $|E|$ ; (b) $|\Psi|$ from type 1 stable soliton ($Q_0,Q_2$) profiles at $E_p=0.0715$; (c),(d) - full-model soliton solutions (Eq. \ref{eq:eqE},\ref{eq:eqPsi}) at $E_p=0.0715$; (e),(f) reconstructed $|E|$ and $|\Psi|$ 3D plots from the unstable type 2 soliton at $E_p=0.08572$.}
\label{Fig:Reconstructed_2D_soliton_Delta_0}
\end{figure}

To assess the extent to which our reduced model captures the 2D soliton dynamics, we reconstruct the 2D soliton from the obtained 1D coupled ($Q_0,Q_2$) soliton profiles, using Eqs.~(\ref{eq:0s12_solution}) and plot it in Fig.\ref{Fig:Reconstructed_2D_soliton_Delta_0}.

\begin{figure}
\includegraphics[width=8cm]{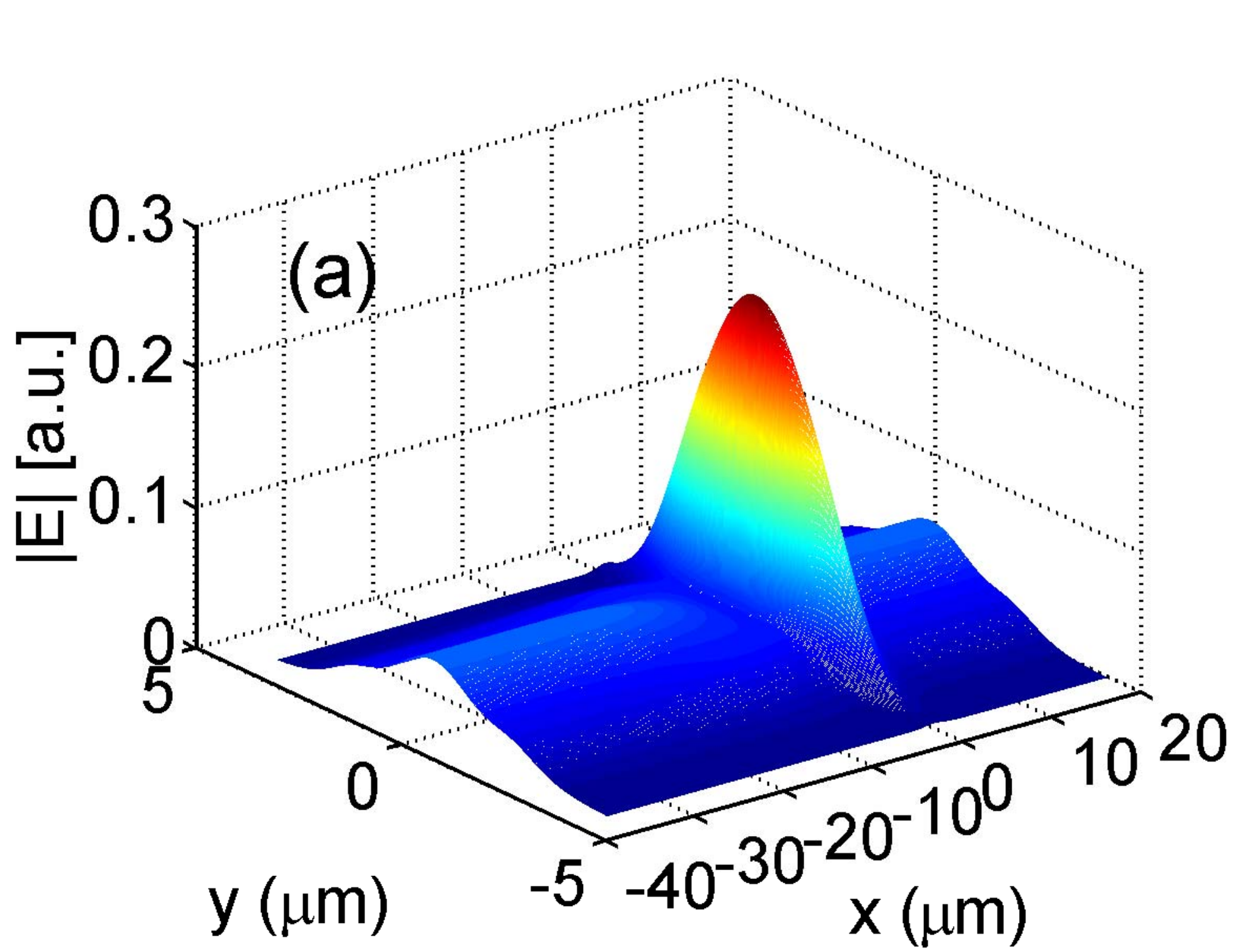}
\includegraphics[width=8cm]{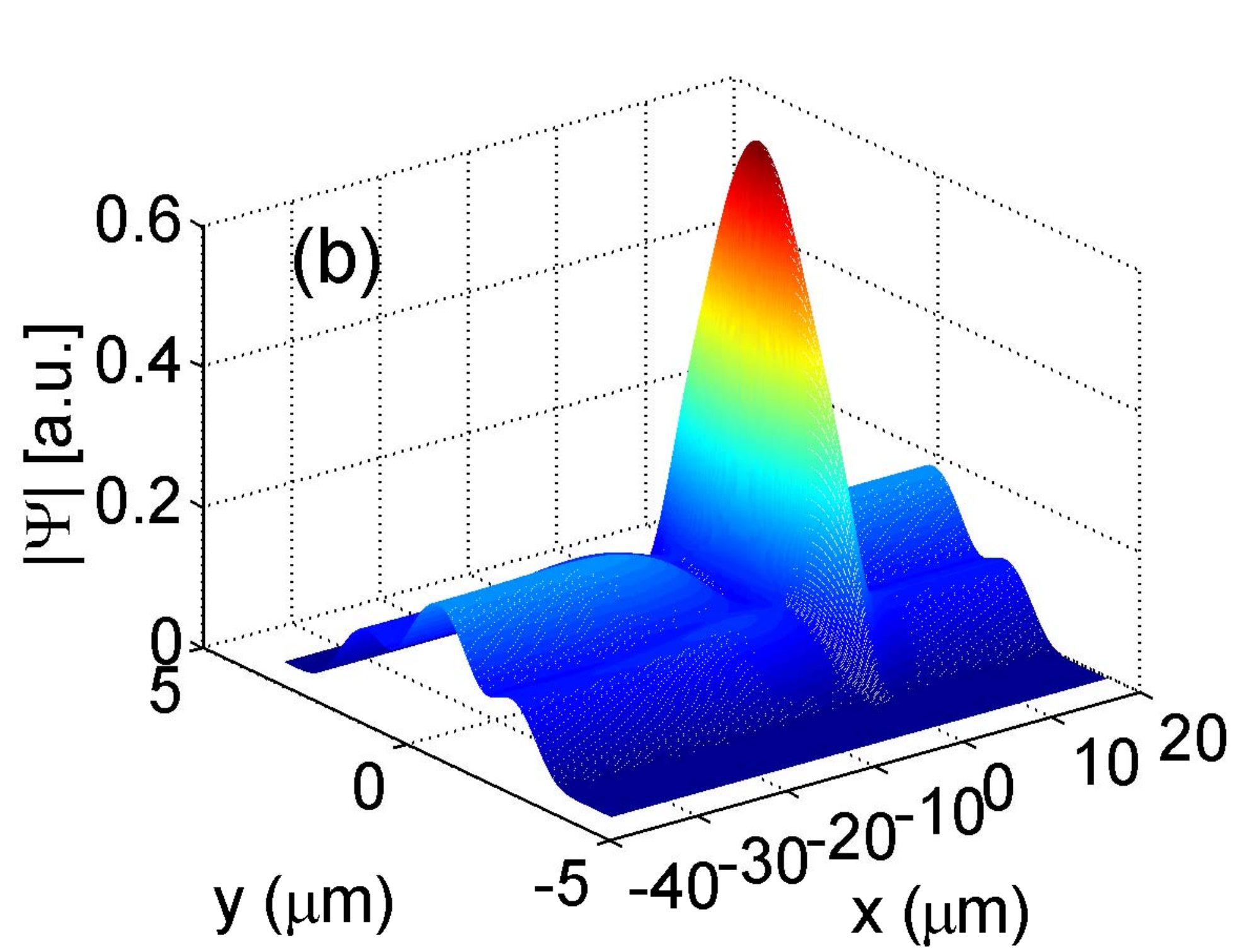}
\includegraphics[width=8cm]{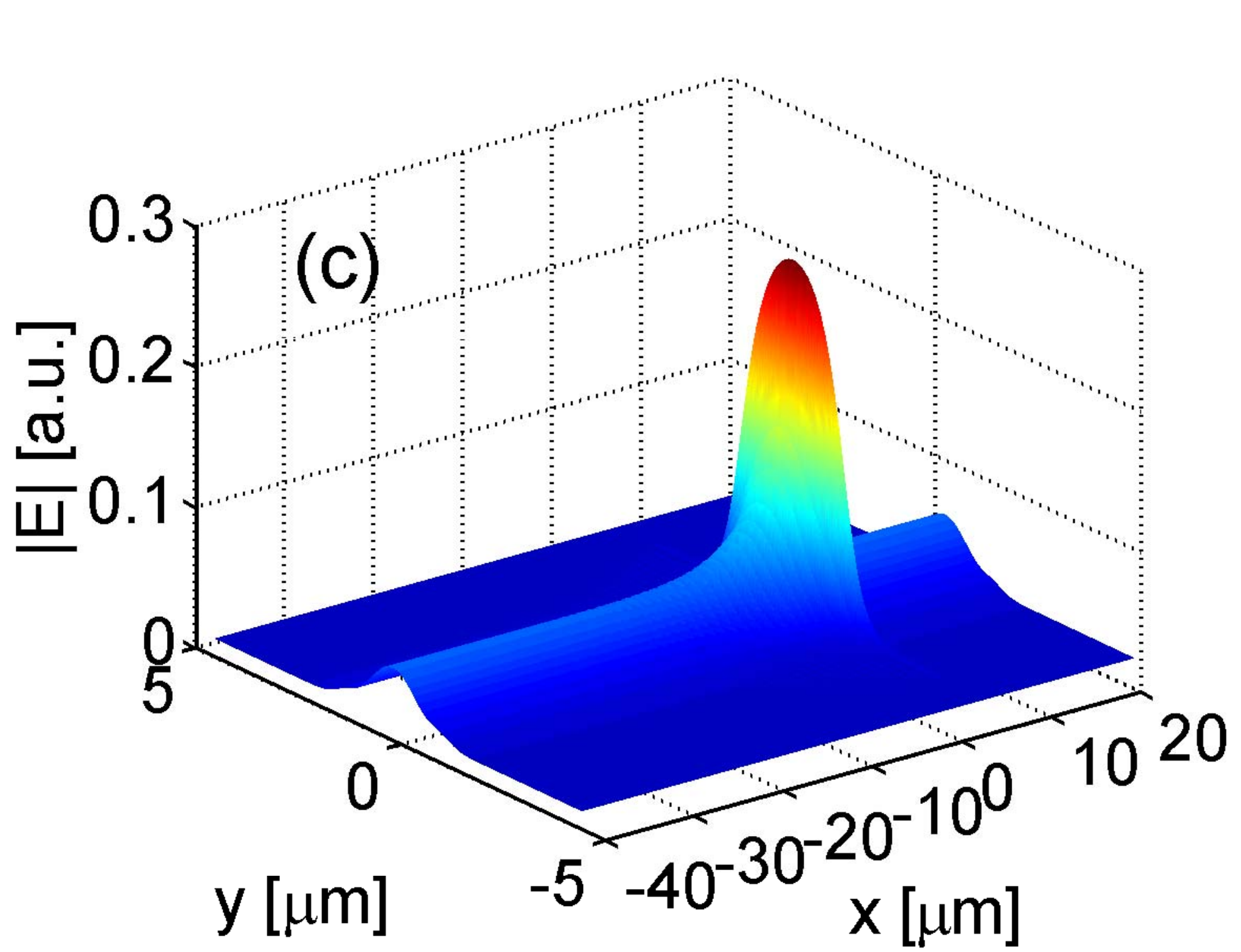}
\includegraphics[width=8cm]{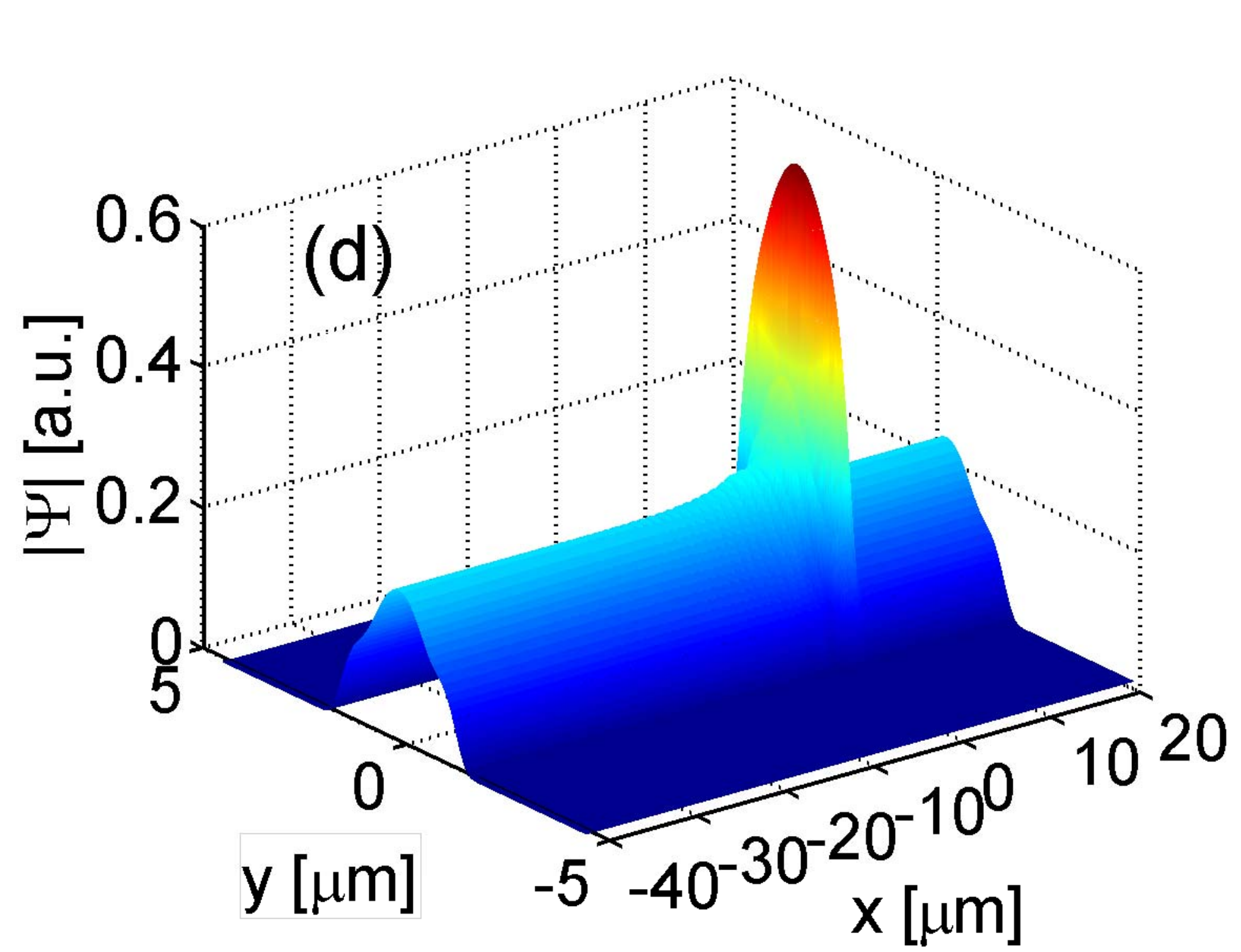}
\includegraphics[width=8cm]{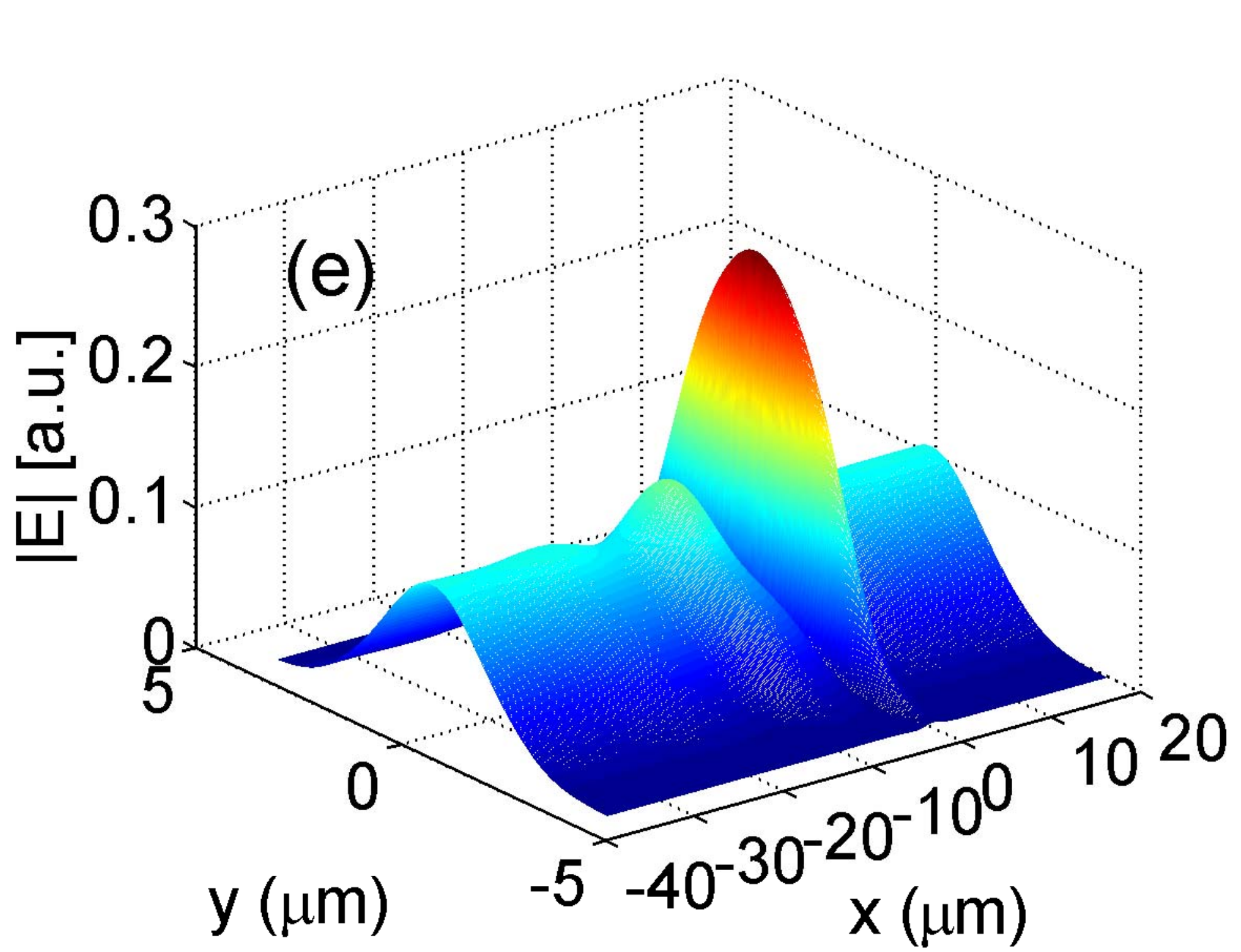}
\includegraphics[width=8cm]{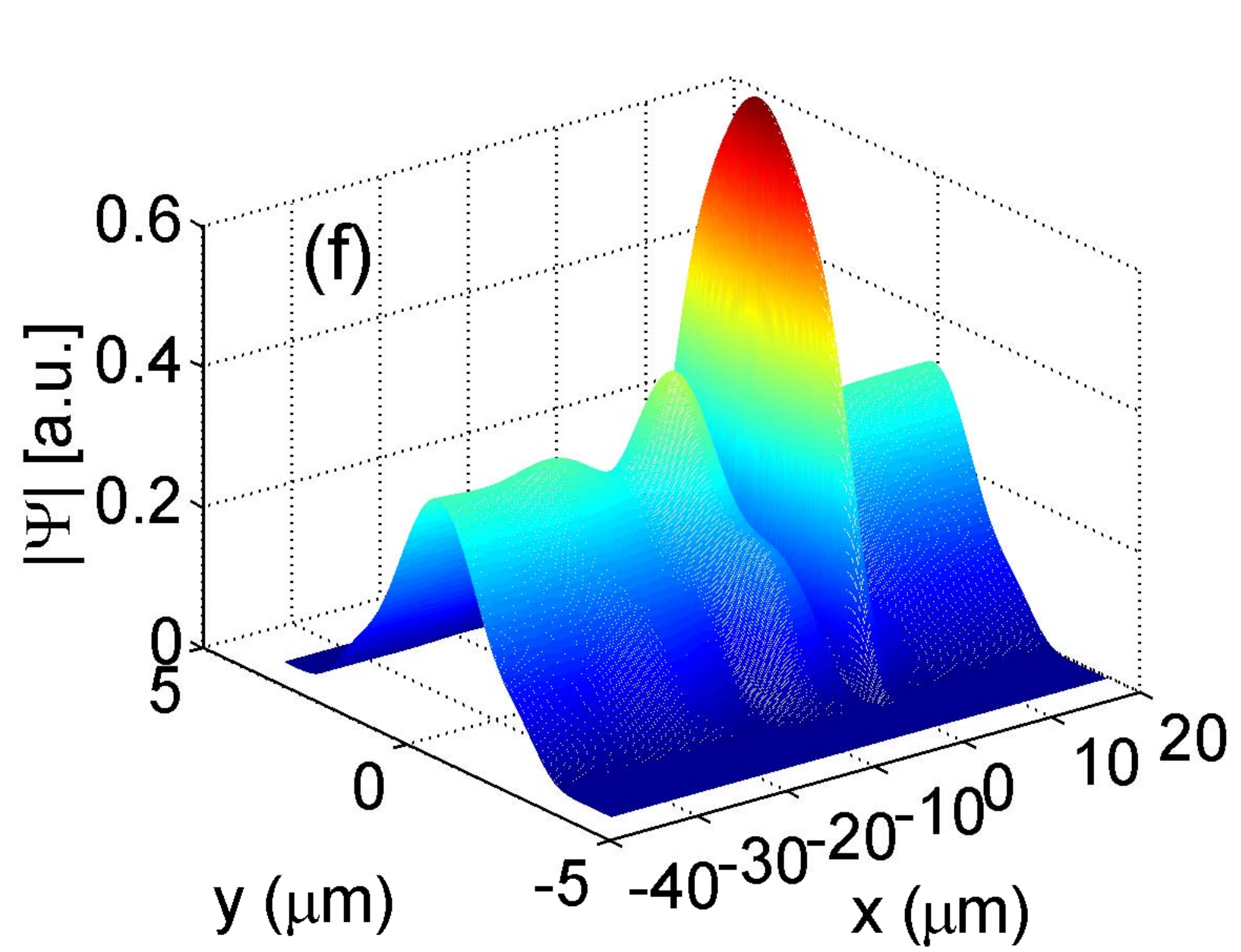}
\includegraphics[width=8cm]{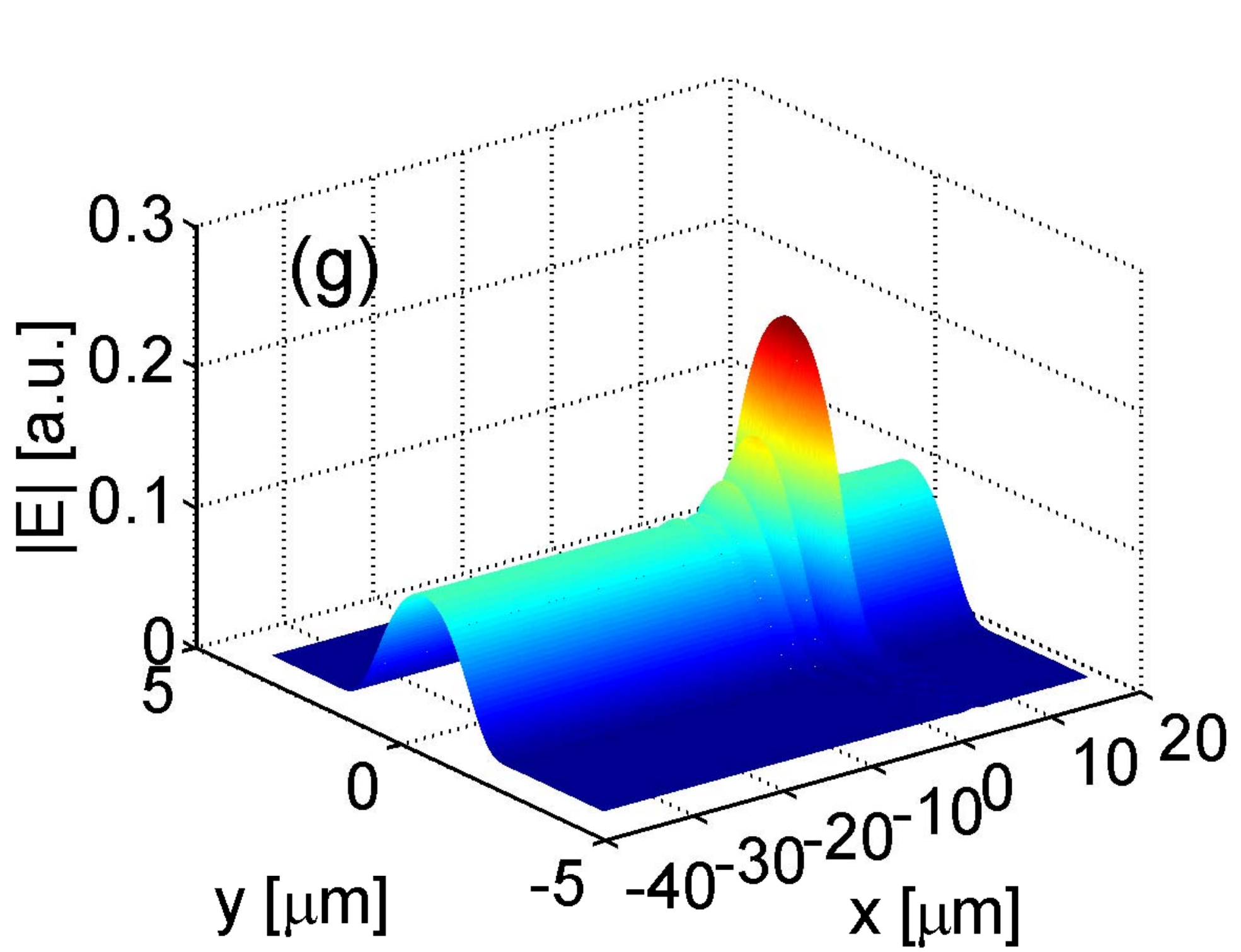}
\includegraphics[width=8cm]{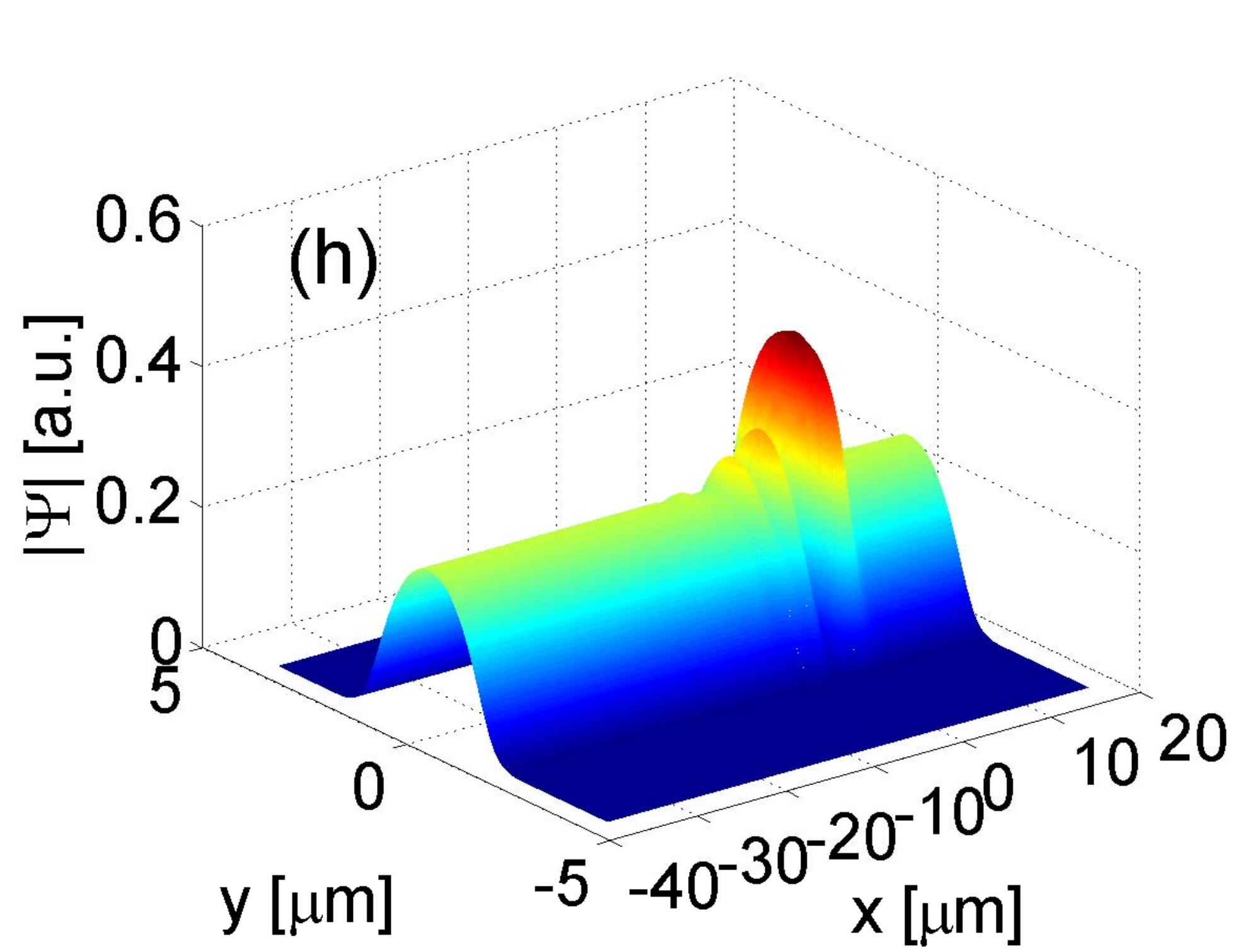}
\caption{3D plot of the reconstructed 2D soliton at $\Delta=-0.1$, $\gamma_0=0.04$ (a) $|E|$ ; (b) $|\Psi|$ from type $1/2$ stable/unstable soliton ($Q_0,Q_2$) profiles (see Fig. \ref{Fig:Stability_analysis_homogeneous_type1_solitons_Delta_-0.1}) at the (a) left edge of the soliton branch $E_p=0.03988$; (c,d) Full-model 2D soliton solutions (Eq. \ref{eq:eqE},\ref{eq:eqPsi}) for at the left edge; (e, f) Reconstructed solitons at the right edge of the soliton branch for $E_p=0.067$; (g,h) Full-model 2D solitons at the right edge of the soliton branch.}
\label{Fig:Reconstructed_2D_soliton_Delta_m0p1}
\end{figure}
We compare the reconstructed 2D solitons at $\Delta=0, \gamma_0=0.04$ for the $E$-field and $\Psi$-fields with the full-model solutions, shown in Fig. \ref{Fig:Reconstructed_2D_soliton_Delta_0} (a-d). Both solitons exhibit the characteristic two-fold split tail which can be considered as a signature of zero-detuning case (cf. Fig. 4 of \cite{our_OL}). We note that the full-model soliton (Fig.\ref{Fig:Reconstructed_2D_soliton_Delta_0} (c,d))  is more strongly-localised in a transverse direction ($y$-axis) compared to the reconstructed one, obtained from the reduced model (Fig.\ref{Fig:Reconstructed_2D_soliton_Delta_0} (a,b)). This is expected as our reduced model assumes unchanged transverse mode along $y$-direction. The unstable soliton (type 2) profiles are shown for comparison in Fig.\ref{Fig:Reconstructed_2D_soliton_Delta_0} (e,f).
The reconstructed and full-model 2D solitons for $\Delta=-0.1$ and $\gamma_0=0.04$, are displayed in Fig. \ref{Fig:Reconstructed_2D_soliton_Delta_m0p1}. Comparison between the reconstructed, Fig.\ref{Fig:Reconstructed_2D_soliton_Delta_m0p1}(a,b) and the full-model 2D solitons (c,d) at the left soliton branch edge reveals similar type of solitons with a simpler shape and a single-lobe tail. Similar to the previous case considered, the full-model solitons exhibit stronger localisation in a transverse direction to the propagation, showing again the limitations of our reduced model. The reconstructed soliton profiles at the right edge of the soliton branch in Fig. \ref{Fig:Reconstructed_2D_soliton_Delta_m0p1} (e,f) are quite similar to the full-model ones (g,h), both exhibiting tail oscillations and a stronger transverse localisation in the case of the full-model solitons.

\subsection{Projection of full-model solutions}
\label{sssec:4.2}
\begin{figure}
\resizebox{.9\textwidth}{!}{%
\includegraphics[height=5cm]{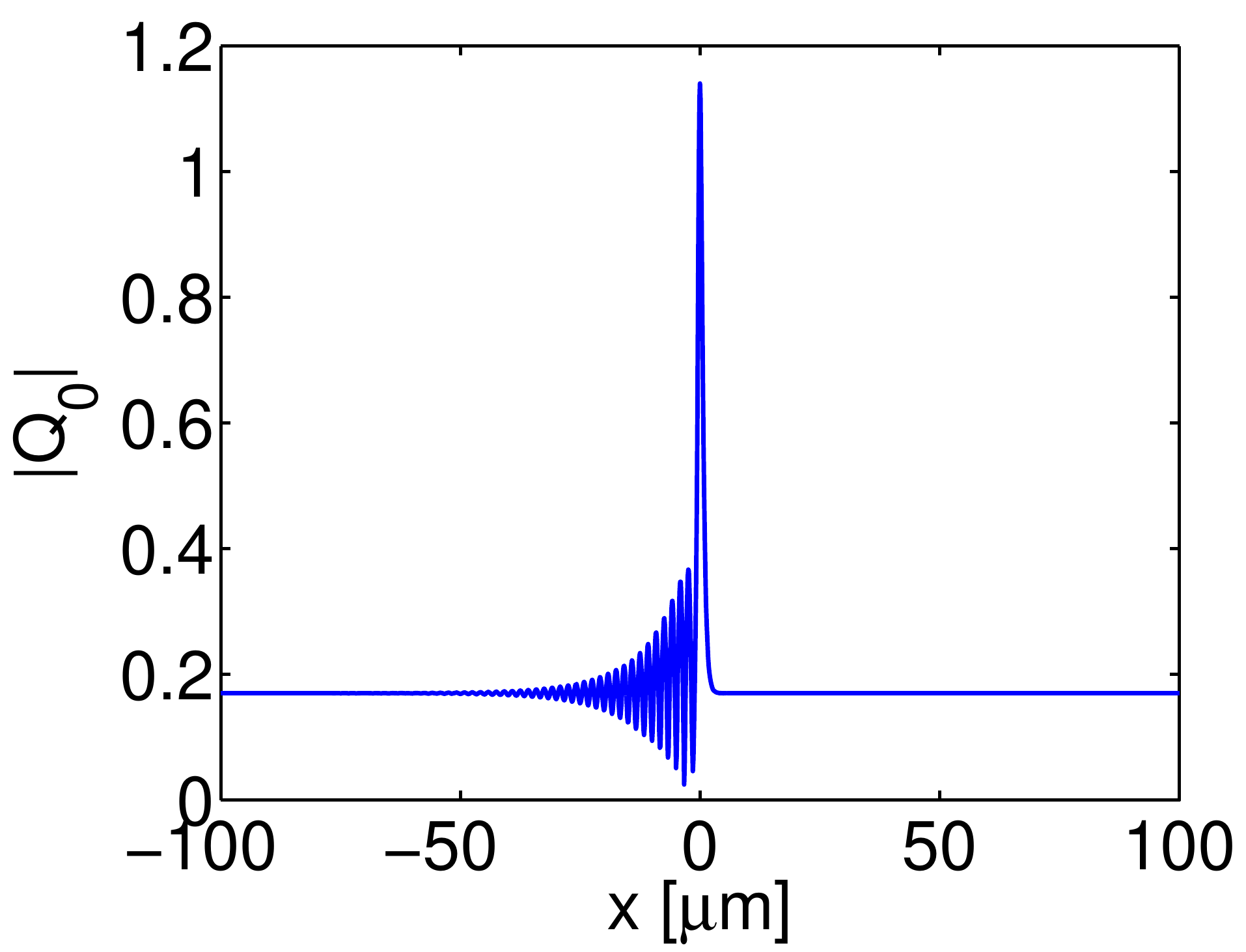}
\quad
\includegraphics[height=5cm]{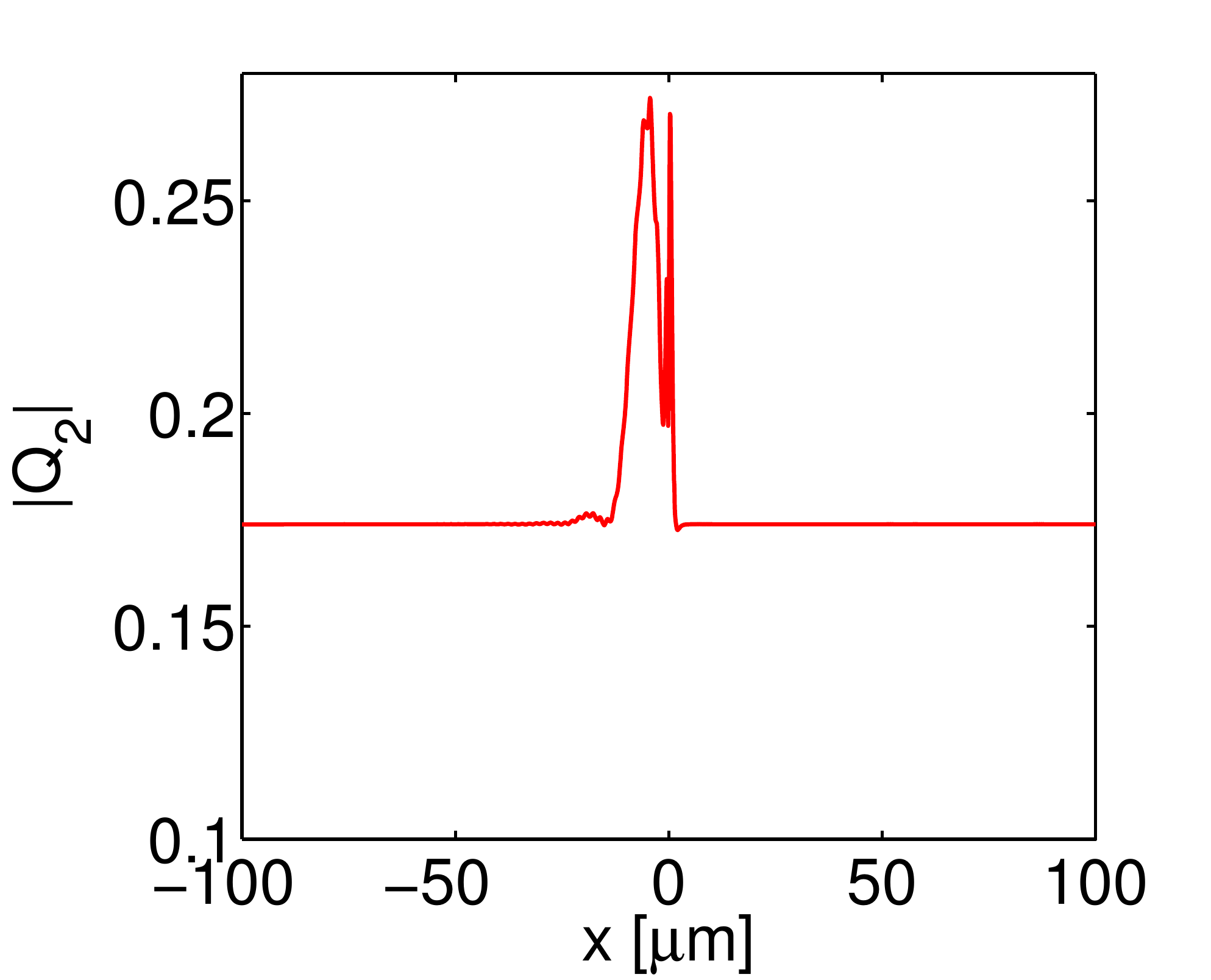}
\quad
\includegraphics[height=5cm]{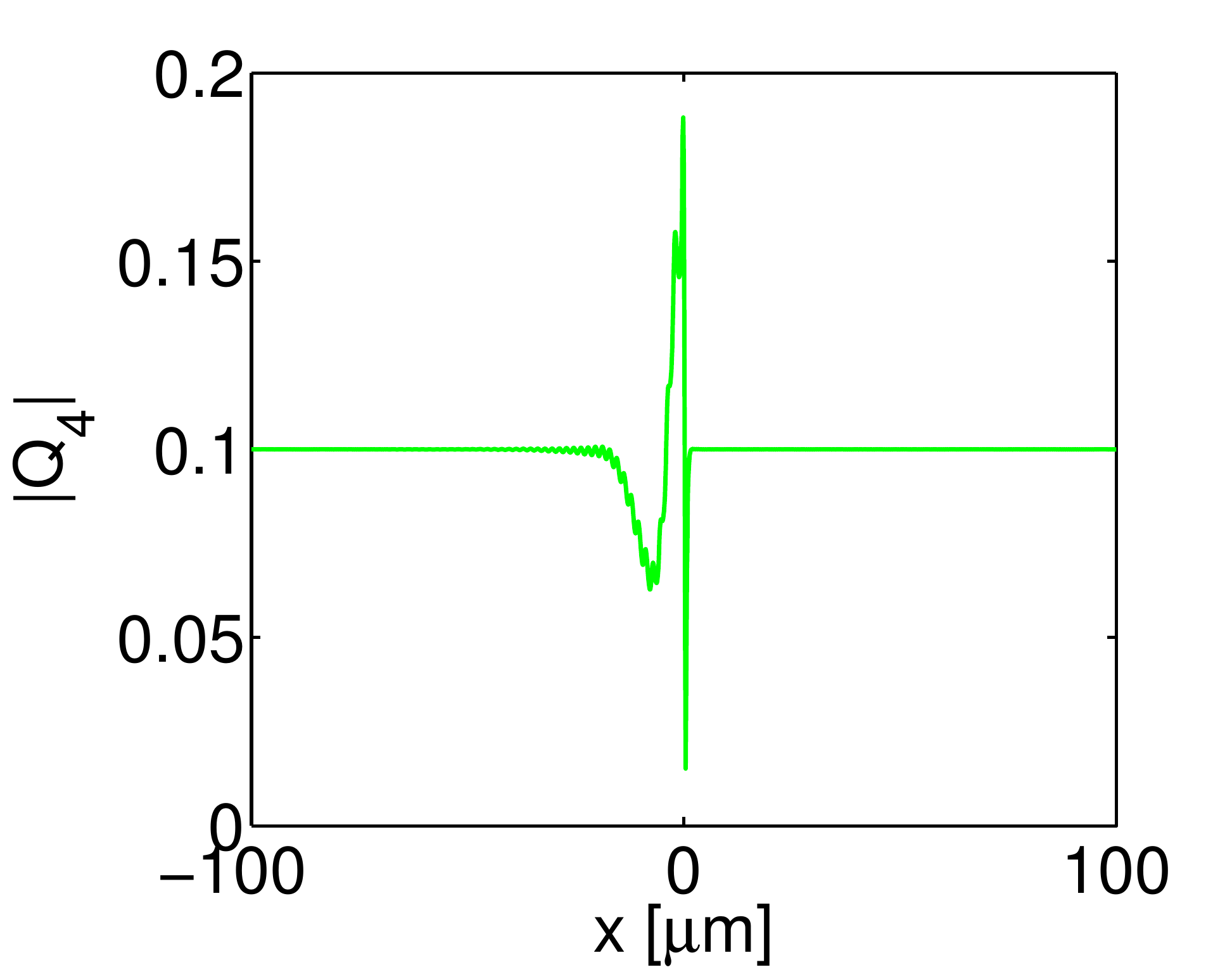}
}
\resizebox{.9\textwidth}{!}{%
\includegraphics[height=5cm]{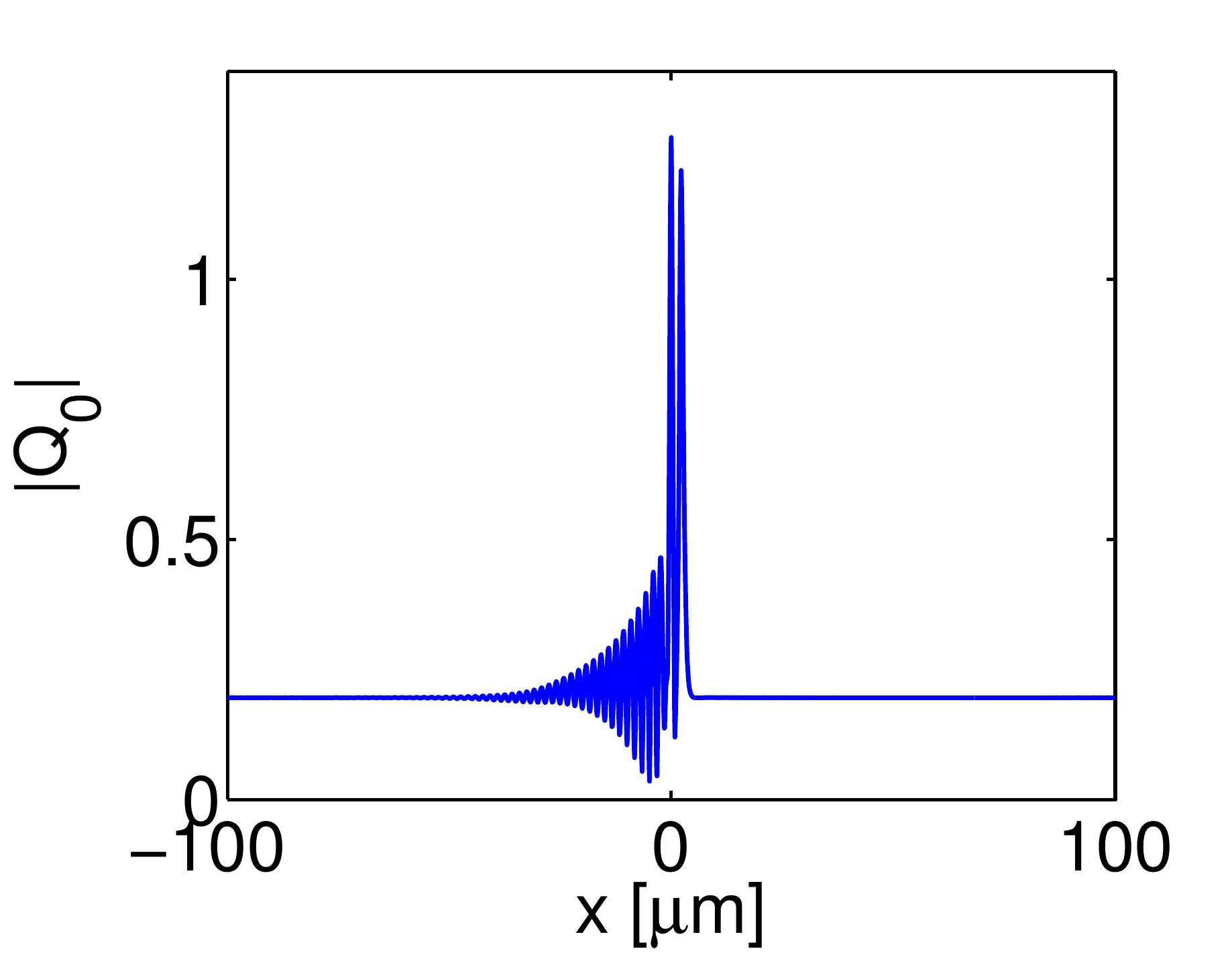}
\quad
\includegraphics[height=5cm]{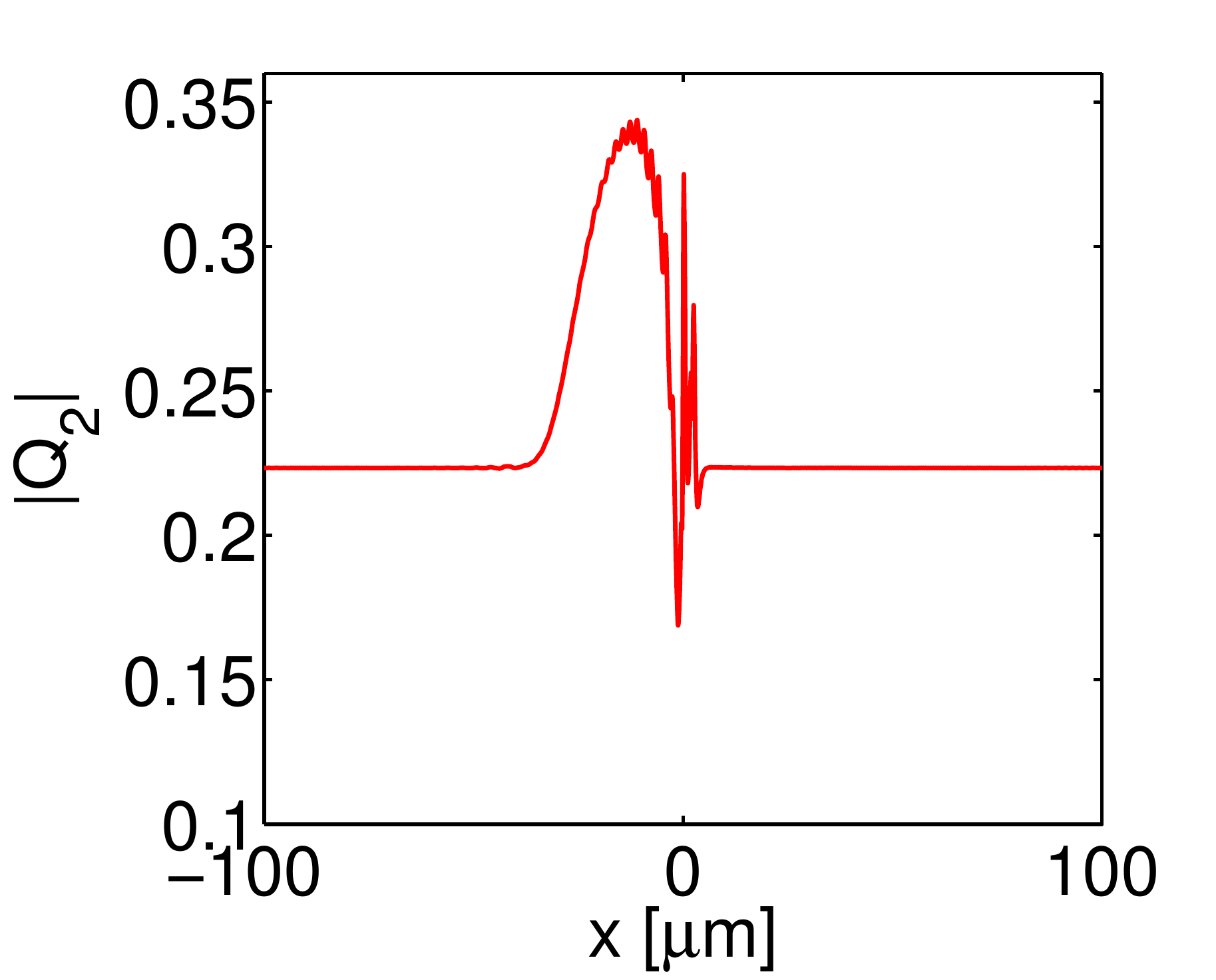}
\quad
\includegraphics[height=5cm]{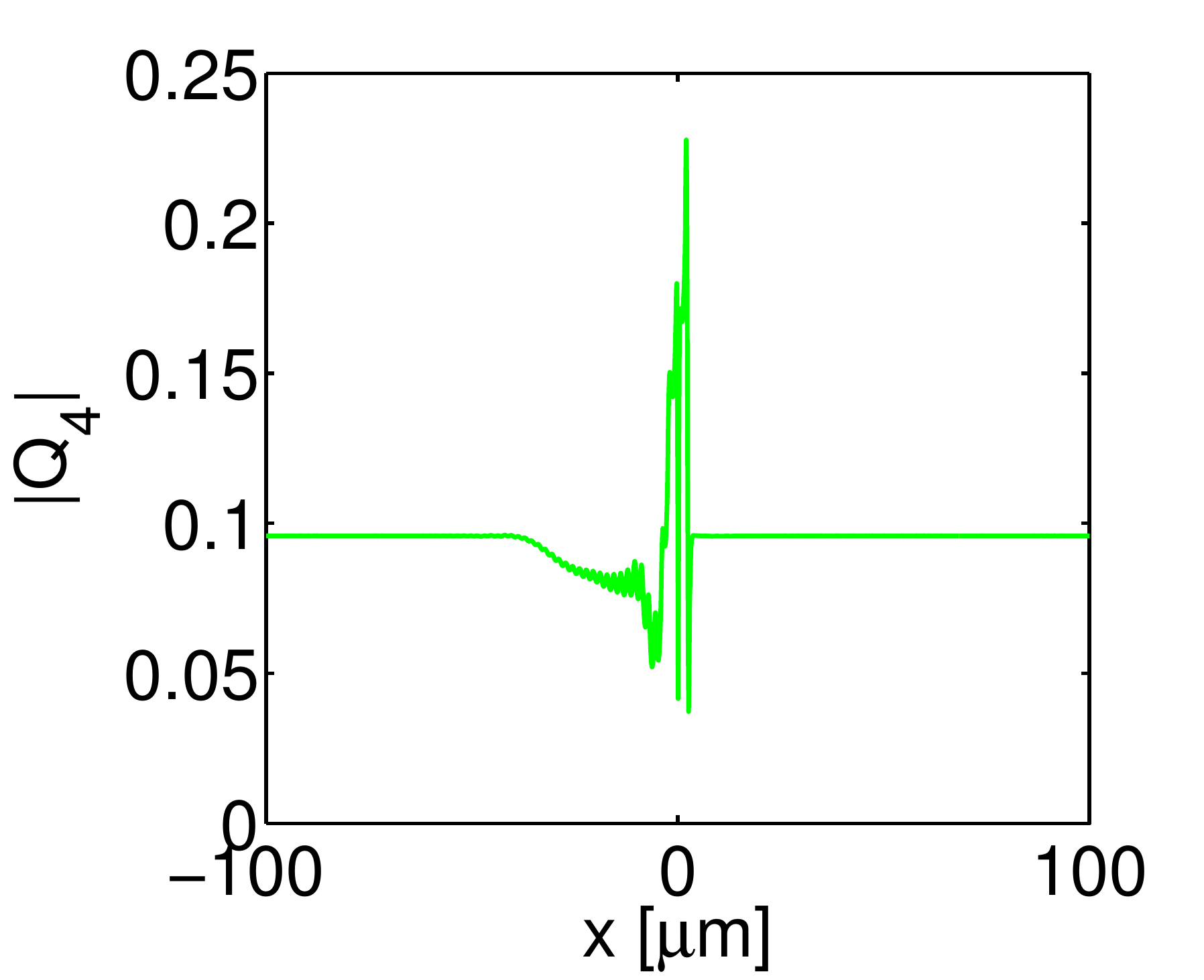}
}
\caption{Reconstructed $Q_0$,$Q_2$ and $Q_4$ soliton components from final evolved full-model ($E,\Psi$) solutions at $\Delta=0, \gamma_0=0.04$, using inverse transform (Eq.\ref{eq:modal_decomposition}): upper row - $E_p=0.0672$, corresponding to a single-hump soliton; lower row - $E_p=0.0732$, corresponding to a double-hump soliton.}
\label{Fig:full_model_projections}
\end{figure}

In the previous section we compared the reconstructed 2D soliton profiles from our reduced model with the full-model 2D dynamical solution. To complete our comparison both ways, we compare the projections of the final full-model dynamically evolved profile, as computed from Eqs.~(\ref{eq:eqE}), (\ref{eq:eqPsi}) onto mode $0, 2, 4$, using the inverse transformation (Eqs.\ref{eq:modal_decomposition}) thereby reconstructing $Q_0$, $Q_2$ and $Q_4$ soliton components. The reconstructed components for $\Delta=0, \gamma=0.04$ are shown in Fig.\ref{Fig:full_model_projections} for pump amplitudes $Ep=0.0672, 0.0732$.

Note that all soliton components at a pump amplitude $E_p=0.0672$ exhibit a single peak, and thus correspond to a single-hump soliton (cf. \ref{Fig:3D_solitons_full_model}(a,b)). By contrast, the soliton components at $E_p=0.0732$ exhibit double peaks, as expected for double-hump solitons (cf. \ref{Fig:3D_solitons_full_model}(c,d)). Note that the $Q_4$ soliton component is small, compared to $Q_2$, which justifies our modal expansion method.

The reconstructed soliton branches for $\Delta=0, \gamma_0=0.04$, computed as $max|Q_j|, j=0,2,4$ from the full model, using the inverse transformation (Eqs.\ref{eq:modal_decomposition}), are shown in Fig.\ref{Fig:Stability_analysis_homogeneous_type1_solitons_Delta_0}(a) with connected (by a magenta line) open-circles, superimposed on the homogeneous solution background of the reduced model and the soliton branches, inferred from the reduced model. A comparison between the soliton branches obtained by our reduced model and the ones, obtained by projection of the fully evolved 2D soliton from the dynamical model shows remarkable agreement between the $Q_2$ components both in amplitude and domain of soliton existence (excluding the points corresponding to double-hump solitons on the soliton projection branch).The $Q_0$ components match in domains of soliton existence but differ in amplitude. We attribute this difference to transverse localisation effects absent in the reduced model.

\subsection{Domains of soliton existence}
\label{sssec:4.3}
\begin{figure}
\resizebox{.9\textwidth}{!}{%
\includegraphics[width=8cm]{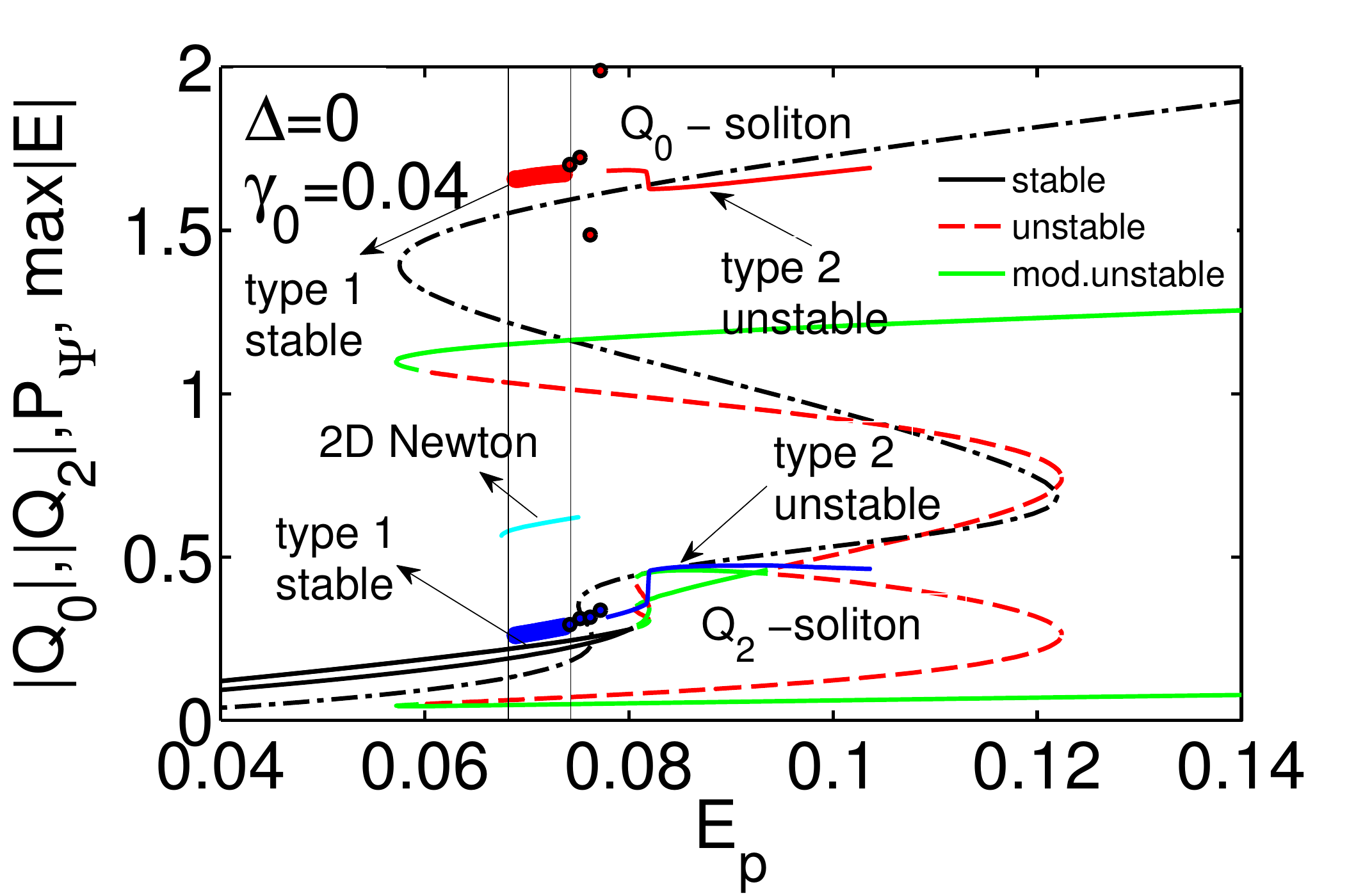}
}
\caption{Soliton branches of type 1 stable and type 2 unstable solitons at $\Delta=0, \gamma_0=0.04$ superimposed on the coupled multistability ($Q_0,Q_2$) curves of the reduced model and the multistability curve ($P_{\Psi}=\int{|\Psi|^2 dy}$ vs $E_p$) full model (black dash-dotted curve). The domain of stable soliton existence is indicated by a rectangle. The soliton branch ($max|E|$) computed by 2D Newton-Raphson method for the full model is shown for comparison. The points correspond to soliton solutions that cannot be connected to the stable branch, since they represent multi-hump soliton solutions.}
\label{Fig:soliton_branches_2D_Newton}
\end{figure}

Finally we perform a comparison between the domain of stable type 1 soliton existence for the zero-detining case ($\Delta=0, \gamma_0=0.04$) with the soliton branch computed by a 2D Newton-Raphson method (cf. shaded area in Fig. 2(c) \cite{our_OL}). Both soliton branches are superimposed on the homogeneous solutions and the full-model multistability curve (black dash-dotted line) in Fig.\ref{Fig:soliton_branches_2D_Newton}.The 2D Newton-inferred soliton branch is computed as a $max|E|$ and is shown in cyan.
The full-model 2D Newton-inferred soliton branch and our reduced model branches are in excellent agreement, thus confirming the soliton existence domain computed in \cite{our_OL}.

The open circle points in the gap between the stable (type 1)and unstable (type 2) solitons are computed by solving the time-dependent equations for the coupled ($Q_0, Q_2$) soliton (Eqs.(\ref{eq:cm3_Q0}), (\ref{eq:cm3_Q2})). We should note that although the dynamical model converges to these solutions, we were unable to connect them by Newton method to either the unstable or stable soliton branches. As these solutions happen to be in a range of pump amplitudes where the multi-humped solitons have been predicted by the full model, we attribute these solutions to multi-hump solitons. This bifurcation behaviour deserves further investigation, but will be a subject of a further study.

Our reduced model reproduces remarkably well the soliton existence domain, computed from the full model by 2D Newton method \cite{our_OL}.

\section{Conclusion}
We have developed a 1D reduced model, based on modal expansion of our full 2D mean-field model polariton solutions in a microcavity wire. We demonstrated that by considering just a two coupled modes, the fundamental and the second-order microcavity wire modes, we can correctly reproduce the onset of spatial multi-stability upon variation of pump detuning. Furthermore, we show that such a simple 1D coupled-mode model is sufficient to reproduce the stable soliton existence domain of the full model and thus provide guidance for experiments. Our simplified 1D model can be used as a framework for analysis of the complex spatio-temporal dynamics of multi-mode polariton solitons in microcavity wires and of effects, such as e.g. non-monotonous wire width dependence of the soliton existence domains, which become entangled and cannot be understood on the basis of the full mean-field model. Similar to single-mode solitons, coupled-mode solitons could provide a powerful concept for description of the complex non-linear polariton dynamics in confined microcavity structures. Understanding the complex dynamical nonlinear soliton formation phenomena within the coupled-mode picture would be of great importance for practical realisation of the future integrated polaritonic devices, based on structured microcavities, with microcavity wires acting as basic functional components.
We acknowledge that in order to fully elucidate the nonlinear polariton dynamics, further work is needed to investigate conditions of formation and stability of multi-hump solitons in this confined system which will be a subject of a future study.

We acknowledge helpful discussions with D. V. Skryabin. GS acknowledges funding through the Leverhulme Trust Research Project Grant RPG-2012-481.


\begin{thebibliography}{99}

\bibitem{Wright}
L. G. Wright, D. N. Christodoulides and F. W. Wise, Nat. Photonics, \textbf{9}, 306 (2015)

\bibitem{Russell}
A. Efimov, A. J. Taylor, F. G. Omenetto, J. C. Knight, W. J. Wadsworth, and P. St. Russell, Opt. Express \textbf{11}, 910 (2003).

\bibitem{Modotto}
D. Modotto, C. De Angelis, M. A. Maga\~na-Cervantes, R. M. De La Rue, R. Morandotti, St. Linden, H. M. van Driel, J. St. Aitchison, J. Opt. Soc. Am. B \textbf{22}, (2005)

\bibitem{Karpinski}
M. Jachura, M. Karpinski,C. Radzewicz, and K. Banaszek, Opt. Express \textbf{22}, 8624 (2014)

\bibitem{Hasegawa}
A. Hasegawa, Opt. Lett. \textbf{5}, 416 (1980)

\bibitem{Crosignani@DiPorto}
B. Crosignani and P. D. Porto, Opt. Lett. \textbf{6}, 329 (1981)

\bibitem{Crosignani&Cutolo&DiPorto}
B. Crisignani, A. Cutolo, P. D. Porto, J. Opt. Soc. Am. \textbf{72}, 1136 (1982)

\bibitem{Poletti}
F. Poletti and P. Horak, J. Opt. Soc. Am. B 25, 1645 (2008)

\bibitem{Horak}
P. Horak and F. Poletti, in "Recent progress in Otical Fibre Research," M. Yasin, ed. (2012), chap. 1, pp.3-24

\bibitem{Mafi}
A. Mafi, J. of Light. Techn. \textbf{30}, 2803 (2012)

\bibitem{Agrawal}
S. Buch and G. P Agrawal, Opt. Lett. \textbf{40}, 225 (2014)

\bibitem{Renninger}
W. H. Renninger and F. W Wise, Nat. Commun. \textbf{4}. 1719 (2013)

\bibitem{Wright&Wise}
L. G. Wright, W. H. Renninger, D. N. Christodoulides, and F. W. Wise, Opt. Express, \textbf{23}, 3492 (2015)

\bibitem{Christodoulides}
D. N. Christodoulides and R. I. Joseph, "Vector solitons in birefringent nonlinear dispersive media," Opt. Lett. \textbf{13}, 53 (1988)

\bibitem{Haelterman1}
M. Haelterman, A. P. Sheppard, and A. W. Snyder, "Bound-vector solitary waves in isotropic nonlinear dispersive media," Opt. Lett. \textbf{18}, 1406 (1993)


\bibitem{Haelterman2}
M Haelterman, A. P. Sheppard, and A. W. Snyder, "Bimodal counterpropagating spatial solitary waves," \oc \textbf{103} 145--152 (1993)

\bibitem{Snyder}
A.W. Snyder, S. J. Hewlett, and D. J. Mitchell, "Dynamic spatial solitons," Phys. Rev. Lett. \textbf{72}, 1012 (1994)


\bibitem{our_OL}
G. Slavcheva, A. V. Gorbach, A. Pimenov, A. G. Vladimirov and D. V. Skryabin, Opt. Lett. \textbf{40}, 1787 (2015)

\bibitem{Wertz_APL}
E. Wertz, L. Ferrier, D. D. Solnyshkov, P. Senellart, D. Bajoni et al., Appl. Phys. Lett., \textbf{95}, 051108-1--051108-3 (2009)

\bibitem{Wertz_Nature}
E. Wertz, L. Ferrier, D. D. Solnyshkov, R. Johne, D. Sanvitto, A. Lema\^{\i}tre, I. Sagnes, R. Grousson, A. V. Kavokin, P. Senellart, G. Malpuech and J. Bloch, Nature Physics, \textbf{6}, 860--864 (2010)

\bibitem{Kavokin}
A. Kavokin, J. Baumberg, G. Malpuech, and F. Laussy, {\it Microcavities} (Oxford University Press, Oxford, 2007).

\bibitem{Sich_NaturePhotonics}
M. Sich, D. N. Krizhanovskii, M. S. Skolnick, A. V. Gorbach, R. Hartley, D. V. Skryabin, E. A. Cerda-M\'{e}ndez, K. Biermann, R. Hey and P. V. Santos, Nature Photonics, \textbf{6}, 50--55 (2012)

\bibitem{Amo1}
A. Amo, D. Sanvitto, F. P. Laussy, D. Ballarini, E. del Valle, M. D. Martin, A. Lema\^{\i}tre, J. Bloch,
D. N. Krizhanovskii, M. S. Skolnick, C. Tejedor, and L. Vi\~{n}a, "Collective fluid dynamics of a polariton condensate in
a semiconductor microcavity," Nature \textbf{457}, 291--296 (2009)

\bibitem{Amo2}
A. Amo, S. Pigeon, D. Sanvitto, V. G. Sala, R. Hivet, I. Carusotto, F. Pisanello, G. Leménager, R. Houdré, E Giacobino, C. Ciuti, A. Bramati, "Polariton Superfluids Reveal Quantum Hydrodynamic Solitons," Science \textbf{332}, 1167 -- 1169 (2011)

\bibitem{Philippe}
V. Ardizzone, P. Lewandowski, M. H. Luk, Y. C. Tse, N. H. Kwong, A. L\"{u}cke,
M. Abbarchi, E. Baudin, E. Galopin, J. Bloch, Aristide Lemaitre,
P. T. Leung, Ph. Roussignol, R. Binder, J. Tignon, and S. Schumacher, "Formation and control of Turing patterns
in a coherent quantum fluid,", Scientific Reports, \textbf{3}:3016, 1--6 (2013)

\bibitem{Kwong}
N.H. Kwong, R. Takayama, I. Rumyantsev, M. J. Kuwata-Gonokami, and R. Binder, "Third-order exciton-correlation and nonlinear cavity-polariton effects in semiconductor microcavities," Phys. Rev. B \textbf{64}, 045316 (2001)

\bibitem{Lagoudakis}
K. G. Lagoudakis, M. Wouters, M. Richard, A. Baas, I. Carusotto, R. Andr\'{e}, Le Si Dang, and B. Deveaud-Pl\'{e}dran "Quantized vortices in an exciton- polariton condensate," Nature Phys. \textbf{4}, 706 (2008)

\bibitem{Deveaud}
C. Ouellet-Plamondon, G. Sallen, F. Morier-Genoud, D. Y. Oberli, M. T. Portella-Oberli, and B. Deveaud,
Phys. Rev. B \textbf{93}, 085313 (2016)

\bibitem{JOSAB}
G. Slavcheva, A. V. Gorbach, and A. Pimenov,"Polariton solitons and multi-stability in tapered microcavity wires', submitted JOSA B (2016)

\bibitem{Nardin}
G. Nardin, G. Grosso, Y. L\'{e}ger, B. Pietka, F. Mirier-Genoud, and B. Deveaud-Pl\'{e}dran, Nat. Phys. \textbf{7}, 635 (2011)


\bibitem{Stolen}
R. H. Stolen, IEEE J. of Quant. Electron, \textbf{11}, 100 (1975)

\bibitem{Stolen&Ashkin}
R. H. Stolen, J. E. Bjorkholm, and A. Ashkin, Appl. Phys. Lett., \textbf{24}, 308 (1974)

\end{thebibliography}
\end{document}